\documentclass[a4paper,11pt]{article}
\pdfoutput=1 % if you are submitting a pdflatex (i.e. if you have
\usepackage{jheppub} % for details on the use of the package, please
\usepackage[utf8]{inputenc}
\usepackage{setspace}
\usepackage{dsfont}
\usepackage{subfigure}
\usepackage{hyperref}
\definecolor{refkey}{gray}{0.45}
\definecolor{labelkey}{RGB}{155,48,48}
\usepackage{mathtools}
\usepackage{physics}
\usepackage{simplewick}
\usepackage[obeyFinal]{todonotes}
\usepackage[normalem]{ulem}
\usepackage {cancel}

\def\beq{\begin{eqnarray}}\def\eeq{\end{eqnarray}}
\def\be{\begin{equation}}\def\ee{\end{equation}}

\def\mes[#1]{d^{3}{#1}}

\def\del{\partial}

\newcommand{\half}{\frac{1}{2}}

\def\del{\partial}

\def\order{\ensuremath{\mathcal{O}}}

\def\deltabar{{\mathchar '26\mkern -10mu\delta}}

\reversemarginpar
\definecolor{UI_blue}{RGB}{32, 64, 151}
\definecolor{UI_red}{RGB}{187, 62, 24}
\definecolor{UI_blue2}{RGB}{0, 84, 147}
\definecolor{UI_red2}{RGB}{159, 32, 66}
\definecolor{UI_gray}{RGB}{169, 169, 169}
\definecolor{UI_sepia}{RGB}{112, 66, 20}
\definecolor{UI_bittersweet}{RGB}{254, 111, 94}
\definecolor{UI_emerald}{RGB}{80, 200, 120}
\definecolor{UI_olivegreen}{RGB}{181, 179, 92}
\definecolor{UI_cadetblue}{RGB}{95, 158, 160}
\definecolor{UI_fuchsia}{RGB}{255, 0, 255}
\definecolor{UI_midnightblue}{RGB}{25, 25, 112}
\definecolor{UI_royalblue}{RGB}{0,35, 102}
\definecolor{UI_periwinkle}{RGB}{204, 204, 255}
\definecolor{UI_redorange}{RGB}{255, 83, 73}
\definecolor{UI_brickred}{RGB}{203,65,84}	
\definecolor{UI_forestgreen}{RGB}{34, 139, 34}
\definecolor{UI_tan}{RGB}{210,180,140}	
\definecolor{UI_burlywood}{RGB}{222,184,135}
\definecolor{UI_burlywood}{RGB}{192,64,0}
\definecolor{UI_darkorchid}{RGB}{153,50,204}

\usepackage{titlesec}
\usepackage{verbatim}
\usepackage{fancyhdr}
\usepackage{enumerate}
\hypersetup{colorlinks=true,citecolor=blue,urlcolor=black}
\usepackage{slashed, enumitem}
\usepackage{array}
\newcolumntype{P}[1]{>{\centering\arraybackslash}p{#1}}
\usepackage{anyfontsize}
	\vspace{-11cm}
	
	\author[a]{Onkar Parrikar,}
	\author[b]{Sunil Kumar Sake}
	
	\affiliation[a]{\it Department of Theoretical Physics,
		Tata Institute of Fundamental Research,\\  Colaba, Mumbai, India, 400005\\}
		\affiliation[b]{\it Yukawa Institute for Theoretical Physics, Kyoto University, Kyoto 606-8502, Japan}

	\emailAdd{parrikar@theory.tifr.res.in}
	\emailAdd{sunilsake1@gmail.com}
	
	\vspace{1cm}

	\abstract{The notion of time in general relativity must arise from an internal clock, i.e., a degree of freedom in the gravitational theory internal to the system that can serve the role of a physical clock. One such internal notion of time is the York time, corresponding to constant extrinsic curvature slicing of spacetime. We study the Hartle-Hawking wavefunction of asymptotically $AdS_2$ JT gravity as a function of York time. Using both canonical quantization and the JT gravity path integral, we explicitly calculate this wavefunction and show that it satisfies a Schrodinger equation with respect to York time. We find the corresponding York Hamiltonian, which turns out to be manifestly Hermitian. Our analysis cleanly avoids operator ordering ambiguities. The dependence of the wavefunction on York time should be thought of as emerging from a unitary basis transformation of the gravitational length basis states used to compute the wavefunction, and not from time evolution of the state in the dual boundary theory. }
\title{York time in JT gravity}

\preprint{\parbox{3cm}{YITP-25-04\\TIFR/TH/25-8}}
\begin{document}
	\maketitle
	\flushbottom
	
	\vskip 10pt
	
	\parskip = 10pt
\section{Introduction}
\label{introd}
In general relativity, the notion of time is not an external parameter, but must be specified intrinsically in terms of the spacetime metric \cite{kuchar, Anderson:2010xm}. When a time-like asymtotic boundary is available, there is a clear notion of boundary time, which one may regard as an ``external'' clock. The corresponding Hamiltonian, i.e., the ADM Hamiltonian, then ends up being a boundary term. In AdS/CFT\cite{Maldacena:1997re,Witten:1998qj}, the ADM Hamiltonian is the bulk avatar of the microscopic Hamiltonian of the dual quantum mechanical system. However, in spacetimes like de Sitter\cite{Nanda:2023wne,Alonso-Monsalve:2024oii,Held:2024rmg, Chakraborty:2023yed, Chakraborty:2023los} with no timelike boundaries, one does not have recourse to a boundary notion of time and one is forced to consider an ``internal'' notion of time. Of course, one could consider an internal notion of time even in asymptotically AdS spacetimes, where we consider bulk time evolution at a fixed boundary time (i.e. we hold fixed the time at which the bulk Cauchy slices are anchored at asymptotic infinity). For instance, consider foliating the bulk spacetime with Cauchy slices, all anchored at the same boundary time but with different values of the trace of the extrinsic curvature -- this particular notion of internal time is called the \emph{York time}\cite{York1,York2,York3, Witten:2022xxp,Park:2015xoa}. The corresponding time evolution is generated, not by the boundary Hamiltonian, but some other operator that we might call the \emph{York Hamiltonian}, and it is interesting consider whether this operator might also find a microscopic realization in the dual boundary theory.

The goal of this paper is to study York time and the corresponding York Hamiltonian in the simple setting of Jackiw-Teitelboim \cite{JACKIW1985343,Teitelboim:1983ux} gravity in $AdS_2$, where the boundary dual is well-understood. At a technical level, the main object of interest in this paper is a slight generalization of the Hartle-Hawking (HH) wavefunction in JT gravity. The standard HH wavefunction can be interpreted as computing the overlap of the thermo-field double state in the boundary quantum theory \cite{Maldacena:2001kr} with certain special states which correspond to inserting fixed-length boundary conditions on a geodesic in the bulk gravitational path integral\cite{Yang:2018gdb}. From the perspective of York time, we can think of this as the wavefunction at $k=0$, i.e., at zero York time. In this paper, we explicitly derive the HH wavefunction with fixed-length boundary conditions on a Cauchy slice of constant, but non-zero extrinsic curvature. We interpret this wavefunction as the time evolution of the standard HH wavefunction in York time, but where the boundary time is held fixed. We will show that in JT gravity, the HH wavefunction satisfies a Schrodinger equation with respect to York time, and we will compute the corresponding York Hamiltonian. We show that this Hamiltonian is manifestly Hermitian, and does not suffer from any operator ordering ambiguities. We note that operator ordering ambiguities are routinely encountered in the study of the WdW equation in gravity \cite{Isham:1992ms,dewittct,HAWKING1986185} (see also \cite{Nanda:2023wne} for discussion in the two dimensional de Sitter context), but our analyis avoids such issues. 

Of course, we do not expect bulk time evolution (at fixed boundary time) to change the physical state in the dual boundary theory; indeed, bulk time is after all a coordinate choice. Nevertheless, the Hartle-Hawking wavefunction satisfies a non-trivial Schrodinger equation under York time evolution. How should we interpret this evolution in the microscopic theory? Our point of view is that the gravitational length basis states at $k=0$ form a particular basis for the gravitational Hilbert space, and that the fixed length states at a different value of $k$ correspond to a different choice of basis, related to the $k=0$ basis by a unitary transformation. While the HH state in the boundary theory is fixed, the HH \emph{wavefunction} does depend on what basis is being used to express it, and thus its York time dependence should be thought of as emerging from a unitary change of basis relative to the fixed-length states at $k=0$. From this perspective, what we think of as a gauge choice on the gravity side -- namely, a choice of the value of the time coordinate $k$ -- corresponds to another ``gauge'' choice -- namely, a choice of basis -- in the microscopic quantum theory. Of course, an arbitrary change of basis will generically take a nice, semi-classical wavefunction to an arbitrarily complicated, non-classical wavefunction, so the change of basis should be one which preserves the semi-classicality of the original wavefunction. Given the JT gravity/SSS matrix model duality\cite{Saad:2019lba}, it is conceivable that the York Hamiltonian can be given a microscopic interpretation in the dual quantum mechanics. This would need a microscopic interpretation of the fixed length states of JT gravity (see, for instance, \cite{Iliesiu:2024cnh}).

In AdS/CFT, we often talk about the emergence of a bulk spatial direction; in some sense, here we are interested in the emergence of bulk \emph{time} from the boundary theory. This idea has also been explored recently in \cite{Caputa:2020fbc,Araujo-Regado:2022gvw, Soni:2024aop}, where it was referred to as ``Cauchy slice holography''. One difference between these papers and our work is that in Cauchy slice holography, one imposes Dirichlet boundary conditions on the Cauchy slices,\footnote{Dirichlet boundary conditions in Euclidean gravity can sometimes be problematic \cite{Witten:2018lgb} because the local Einstein equations are elliptical in nature and the Dirichlet boundary conditions do not preserve the elliptic regularity of the problem. On the other hand, boundary conditions where the conformal class of the metric and the trace of the extrinsic curvature are fixed are better behaved.} where as in our case, York time slicing corresponds to part Neumann, part Dirichlet boundary conditions. The interpretation is different as well -- from our point of view, the emergence of York time corresponds to a semi-classicality-preserving unitary transformation of the gravitational length basis states. In Cauchy slice holography, the flow corresponds to a $T\bar{T}$-deformation \cite{Zamolodchikov:2004ce, Smirnov:2016lqw, Cavaglia:2016oda, McGough:2016lol, Hartman:2018tkw} of the boundary Euclidean path integral.\footnote{A Hilbert space interpretation of the $T\bar{T}$ deformation is plagued by issues such as energy eigenvalues becoming complex. However, see \cite{Kruthoff:2020hsi, Guica:2022gts, Soni:2024aop} for some progress towards a Hilbert space approach to the theory.} Loosely speaking, the two approaches should be related by a Legendre transform, but we do not have a more precise connection.

The rest of this paper is organized as follows: in section \ref{jtgbc}, we review some basics of JT gravity and discuss the appropriate boundary conditions that are of interest to us in this work. In section \ref{clsol}, we present the classical calculation of the generalized Hartle-Hawking (HH) wavefunction at non-zero York time, along the lines of \cite{Harlow:2018tqv}. In section \ref{qsols}, we compute the quantum wavefunction, along the lines of \cite{Yang:2018gdb}, and present some consistency checks. In section \ref{concl}, we end with some conclusions and open questions. Additional material about the details of the gravitational Gauss law constraints and how they are satisfied by the generalized HH wavefunction is relegated to the appendices.

\section{Preliminaries }
\label{jtgbc}
We will begin with a quick review of two-dimensional JT gravity with a negative cosmological constant, with special emphasis on various choices of boundary conditions which will be important in this work. The action for this theory in Euclidean signature is given by 
\begin{align}
	S_{\text{JT}}&=S_{\text{bulk}}+S_{\text{bdy}},\nonumber\\
	S_{\text{bulk}}&=-\frac{1}{16\pi G}\int \,d^2x\sqrt{g}\, \phi\, (R+2),\label{buljt}
\end{align}
where $G$ is the two-dimensional Newton's constant, $\phi$ is the dilaton, and $R$ is the Ricci scalar of the two dimensional spacetime. The term $S_\text{bdy}$ in the action is the boundary term that is required for a well-defined variational principle. For the standard Dirichlet boundary conditions where the dilaton and the induced metric are held fixed on the asymptotic boundary, this boundary term is the usual Gibbons-Hawking-York term given by 
\begin{align}
	S_{\text{GHY}}=-\frac{1}{8\pi G}\int_\del dx \, \sqrt{\gamma}\, \phi\, K\label{GHYyork},
\end{align}
where $\gamma$ is the induced metric on the asymptotic boundary. However, for more general boundary conditions (i.e., different from the Dirichlet case) the structure of boundary terms changes. For instance, in JT gravity, the canonical momentum conjugate to the dilaton is the extrinsic curvature. The analog of Neumann boundary conditions for the dilaton and Dirichlet boundary conditions for the metric -- where the extrinsic curvature and the induced metric on the boundary are held fixed -- does not require the addition of any boundary term, not even the GHY term (see appendix \ref{varpri}; a more detailed analysis of the boundary conditions in JT gravity for smooth boundaries can be found in \cite{Goel:2018ubv}).

The equations of motion for JT gravity are given by:
\begin{align}
	R+2=0,\label{eom1}\\
	\nabla_{\mu}\nabla_{\nu}\phi-g_{\mu\nu}\nabla^2\phi -g_{\mu\nu}\phi=0.\label{eom2}
\end{align}
The first of these equations fixes the metric to be $AdS_2$ everywhere:
\begin{align}
	ds^2=(r^2-r_s^2)dt^2+\frac{dr^2}{(r^2-r_s^2)}.\label{liele}
\end{align}
A solution for $\phi$, which is independent of $t$, is given by 
\begin{align}
	\phi=\phi_b r.\label{phphbr}
\end{align}
The same solution can also be expressed in Poincare coordinates as follows: 
\begin{align}
	ds^2&=\frac{dx^2+dy^2}{y^2},\nonumber\\
	\phi&=\phi_b \left(1+(x^2+y^2)\over 2y\right).
	\label{poinco}
	\end{align}
The coordinate transformations are given by 
\begin{align}
	&	x^\pm=\tan(\zeta^{\pm}),\quad \zeta^{\pm}=\frac{1}{2}(\tilde{t}\pm i\tilde{r}^*),\quad \tilde{r}^*=-\frac{1}{2} \log \left(\frac{\tilde{r}-1}{\tilde{r}+1}\right),\nonumber\\
	&	x=\frac{x^++x^-}{2}\,= \frac{\sqrt{\tilde{r}^2-1}\sin \tilde{t}}{\tilde{r}+\sqrt{\tilde{r}^2-1}\cos \tilde{t}},\nonumber\\
	& y=\frac{x^+ - x^-}{2i}=\frac{1}{\tilde{r}+\sqrt{\tilde{r}^2-1}\cos \tilde{t}},\nonumber\\
	\label{poltopoin}
\end{align}
where $\tilde{r}=\frac{r}{r_s},\tilde{t}=r_s t$.

\subsection{Phase space of JT gravity}
\label{canqun}
For completeness, we now give a brief review of the phase space of JT gravity. Consider the bulk action of JT gravity  given by equation \eqref{buljt}. We pick some coordinates $(t,x^a)$. For the purposes of this paper, we can essentially think of foliating spacetime with constant extrinsic curvature slices (anchored at some fixed boundary points), so the role of the time coordinate $t$ will be played the extrinsic curvature. Without loss of generality, we can always write the metric in these coordinates in the form: 
\begin{align}
	ds^2=N^2dt^2+h_{ab}(dx^a+N^a dt)(dx^b+N^b dt),\label{dsadm}
\end{align}
where $N$ is the lapse function, $N^a$ is the shift vector and $h_{ab}$ is the induced metric on constant $t$ slices. The bulk term in the action can be simplified by using the Gauss-Codazzi equation which, in general dimension,  reads:
\begin{align}
	R^{(d)}=R^{(d-1)}+K^2-K_{ab}K^{ab}+2(\nabla_\beta(\mathfrak{t}^\alpha\nabla_\alpha \mathfrak{t}^\beta)-\nabla_\alpha(\mathfrak{t}^\alpha\nabla_\beta \mathfrak{t}^\beta)),\label{gcod}
\end{align}
where $R^{(d)}$ and $R^{(d-1)}$ are the scalar Ricci curvatures corresponding to the full and the induced metric respectively, $\mathfrak{t}^\alpha$ is the normal vector to a constant $t$-slice, $K_{ab}$ is the extrinsic curvature tensor of a constant $t$ slice, and $K$ is the trace of the extrinsic curvature, $K=h^{ab}K_{ab}$. In the present case where the bulk theory is two-dimensional, the extrinsic curvature has only one component, while the Ricci scalar of the induced metric vanishes. Using these facts and simplifying equation \eqref{gcod}, we get
\begin{align}
	R=2(\nabla_\beta(\mathfrak{t}^\alpha\nabla_\alpha \mathfrak{t}^\beta)-\nabla_\alpha(\mathfrak{t}^\alpha\nabla_\beta \mathfrak{t}^\beta)),\label{rin2d}
\end{align}
where $R$, of course, is the two-dimensional Ricci scalar. Substituting equation \eqref{rin2d} in the bulk term of the action in equation \eqref{buljt} and doing an integration by parts leads to 
\begin{align}
	S_{\text{bulk}}=\int \sqrt{g}(\mathfrak{t}^\alpha(\nabla_\alpha \mathfrak{t}^\beta)( \nabla_\beta\phi)+\phi\nabla_\alpha(K\mathfrak{t}^\alpha)-\phi)-\int_\del \phi n_\beta \mathfrak{t}^\alpha \nabla_\alpha \mathfrak{t}^\beta,\label{sbul2d}
\end{align}
where $n$ is the normal vector to the boundary of the spacetime manifold.\footnote{Henceforth, we will set $8\pi G=1$.} In general, the boundary could have multiple components; for instance, in the calculation of the HH wavefunction, the boundary consists of two segments -- the asymptotic boundary and the constant time Cauchy slice.  For the portion of the boundary corresponding to a constant time slice, so that $n^\mu$ coincides with $\mathfrak{t}^\mu$, the boundary term vanishes. Furthermore, noting that 
\begin{align}
&	\mathfrak{t}_\mu=(\mathfrak{t}_t,\mathfrak{t}_x) =(N,0),\nonumber\\
&	\mathfrak{t}^\mu=(\mathfrak{t}^t,\mathfrak{t}^x)=N^{-1}\left(1,-{N_\perp}\right),\label{nt}
\end{align}
we can manipulate the first term in equation \eqref{sbul2d} to get
\begin{align}
	\mathfrak{t}^\alpha \nabla_\alpha \mathfrak{t}^\beta\nabla_\beta\phi=\frac{1}{N}h^{\beta\alpha}\nabla_\beta N\nabla_\alpha\phi\label{normman}.
\end{align}
With this, we find that equation \eqref{sbul2d} becomes
\begin{align}
		S_{\text{bulk}}&=\int dt\int dx\sqrt{h}(h^{\beta\alpha}\nabla_\beta N\nabla_\alpha\phi+N\phi\nabla_\alpha(K\mathfrak{t}^\alpha)-N\phi ) -\int_\del \phi \,n_\beta \mathfrak{t}^\alpha \nabla_\alpha \mathfrak{t}^\beta.
	%	&=\int dt\int dx\sqrt{h}(h^{\beta\alpha}\nabla_\beta N\nabla_\alpha\phi-\frac{\phi}{\sqrt{h}}\del_t(K\sqrt{h})+N\phi)\nonumber\\
		 \label{sbul1}
\end{align}
From here, we can read off the conjugate momenta as
\begin{align}
	\pi_\phi&\equiv \frac{\del S_{\phi, h}}{\del \dot{\phi}}=-\sqrt{h}K,\nonumber\\
	\pi_{h}^{ab}&\equiv \frac{\del S_{\phi, h}}{\del \dot{h}_{ab}}=-\frac{\sqrt{h}}{2}h^{ab}\mathfrak{t}^\alpha\nabla_\alpha\phi.\label{dilconmo2}
\end{align}
where dot denote derivative with respect to $t$.

Naively, the phase space consists of $(\phi,\pi_{\phi},h_{ab},\pi_h^{ab})$,\footnote{Note that the induced metric $h_{ab}$ is just one number, $h_{xx}$, and similarly $\pi_h^{ab}$ is simply $\pi_h^{xx}$.} all of which are functions of $x$. The variables $N(x)$ and $N_i(x)$ are Lagrange multipliers and do not have any canonical structure. Their role is to enforce the Gauss law constraints of JT gravity:

	{
		\begin{align}
			\frac{\delta S}{\delta N}\equiv\mathcal{H}&=\sqrt{h}(-\phi -D^2\phi +K \mathfrak{t}^\alpha \nabla_\alpha\phi),\nonumber\\
			\frac{\delta S}{\delta N_\perp}\equiv\mathcal{P}&=\sqrt{h}\left(K\del_x\phi-\del_x(\mathfrak{t}^\alpha\nabla_\alpha\phi)  \right).
			\label{delsn1}
		\end{align}
	}

To see the actual dimensionality of the physical phase space, we can resort to foliating spacetime with slices of constant extrinsic curvature between two fixed end points on the asymptotic boundary, and use the value of the extrinsic curvature as our time coordinate. Furthermore, we can choose the spatial coordinate to be the proper length along these slices to fix the induced metric to unity everywhere on a slice. The only physical information in the induced metric, then, is the (renormalized) length of the slice. Finally, the Gauss law constraints allow us to solve for the dilaton and its normal derivative, up to a constant of integration. This constant corresponds to the canonical momentum of the length of constant time slices. Thus, the physical phase space of JT gravity is two-dimensional.

\subsection{Corner terms}

As alluded to above, we will be interested in a more general class of boundary conditions where the boundary is not necessarily smooth, i.e., where part of the boundary has one set of boundary conditions and the remaining boundary has an alternate set of boundary conditions. For instance,  such a boundary condition arises in the computation of the Hartle-Hawking wavefunction. With non-smooth boundary conditions, one must be careful to satisfy the variational principle at the points where the boundary conditions jump. To this end, let us analyze the variation of the bulk term of the action more closely. The most general variation of the bulk term in the action, for non-smooth boundaries is given by (see appendix \ref{varpri} for details):
\begin{align}
	\delta S_{\text{bulk}}&=-\half \int d^2x \left[\sqrt{g}\,\delta\phi(R+2)+\delta g_{\mu\nu}(	\nabla_{\mu}\nabla_{\nu}\phi-g_{\mu\nu}\nabla^2\phi -g_{\mu\nu}\phi)\right] -\delta S_{\del},\nonumber\\
	\delta S_{\del}&=\half \int_\del \sqrt{h}\,\left[ (\phi K-n\cdot \nabla\phi) h_{ab} \delta h^{ab}-	2\phi \, \delta K \right]+\sum_{j\in\, \text{Corners}}\phi_j \delta\theta_j.\label{totvarbucb}
\end{align}
Here, $\delta S_{\del}$ is the full set of the boundary terms obtained from the variation of the bulk action upon integration by parts, $n^\mu$ is the normal vector to the boundary, $h_{ab}$ is the induced metric on the boundary, and $K_{ab}$ is the extrinsic curvature of the boundary. For a general co-dimension one surface $x^{\mu}(y^a)$ with the normal vector $n^{\mu}$, the extrinsic curvature in terms of the normal vector is defined as 
\begin{align}
	K_{ab}=\frac{\partial x^{\mu}}{\partial y^a}\frac{\partial x^{\nu}}{\partial y^b}\nabla_{\mu} n_{\nu} \label{kmunu}.
\end{align}

\begin{figure}
    \centering

\tikzset{every picture/.style={line width=0.75pt}} %set default line width to 0.75pt        

\begin{tikzpicture}[x=0.75pt,y=0.75pt,yscale=-1,xscale=1]

%Curve Lines [id:da7950702797911976] 
\draw    (254,122) .. controls (294,92) and (313,137) .. (354,122) ;
%Shape: Free Drawing [id:dp0603115431395399] 
\draw  [line width=0.75] [line join = round][line cap = round] (255,120) .. controls (255,120.33) and (255,120.67) .. (255,121) ;
%Straight Lines [id:da9823994674778809] 
\draw    (254,122) -- (226.67,139.9) ;
\draw [shift={(225,141)}, rotate = 326.77] [color={rgb, 255:red, 0; green, 0; blue, 0 }  ][line width=0.75]    (10.93,-3.29) .. controls (6.95,-1.4) and (3.31,-0.3) .. (0,0) .. controls (3.31,0.3) and (6.95,1.4) .. (10.93,3.29)   ;
%Straight Lines [id:da41211847196991325] 
\draw    (254,122) -- (254,93) ;
\draw [shift={(254,91)}, rotate = 90] [color={rgb, 255:red, 0; green, 0; blue, 0 }  ][line width=0.75]    (10.93,-3.29) .. controls (6.95,-1.4) and (3.31,-0.3) .. (0,0) .. controls (3.31,0.3) and (6.95,1.4) .. (10.93,3.29)   ;
%Shape: Free Drawing [id:dp1176367689620863] 
\draw  [line width=0.75] [line join = round][line cap = round] (255,120) .. controls (255,125) and (255.31,130.01) .. (255,135) .. controls (254.68,140.04) and (249.53,144.13) .. (251,150) .. controls (252.25,155) and (255.09,154.09) .. (258,157) .. controls (261.31,160.31) and (267.33,166.78) .. (271,168) .. controls (277.82,170.27) and (285.32,169.66) .. (292,173) .. controls (295.33,174.67) and (296.32,180.64) .. (298,184) .. controls (299.04,186.08) and (306.24,188.5) .. (310,189) .. controls (323.93,190.86) and (324.32,181) .. (333,181) ;
%Shape: Free Drawing [id:dp8943977132531115] 
\draw  [dash pattern={on 0.84pt off 2.51pt}][line width=0.75] [line join = round][line cap = round] (332,181) .. controls (339.44,181) and (344.07,181.48) .. (350,180) .. controls (352.09,179.48) and (354.31,174.15) .. (357,172) .. controls (361.53,168.38) and (371,164.31) .. (371,156) ;
%Shape: Free Drawing [id:dp3374380653314517] 
\draw  [dash pattern={on 0.84pt off 2.51pt}][line width=0.75] [line join = round][line cap = round] (351,122) .. controls (358.69,122) and (363.76,109) .. (377,109) ;
%Shape: Free Drawing [id:dp4647985559287511] 
\draw  [line width=0.75] [line join = round][line cap = round] (254,113) .. controls (248.12,113) and (245,121.99) .. (245,128) ;

% Text Node
\draw (229,102.4) node [anchor=north west][inner sep=0.75pt]    {$\theta_j $};
% Text Node
\draw (248,69.4) node [anchor=north west][inner sep=0.75pt]    {$r_{j}$};
% Text Node
\draw (213,138.4) node [anchor=north west][inner sep=0.75pt]    {$\tilde{r}_{j}$};

\end{tikzpicture}
\caption{Corner angle between two boundaries}
    \label{cornerang}
\end{figure}
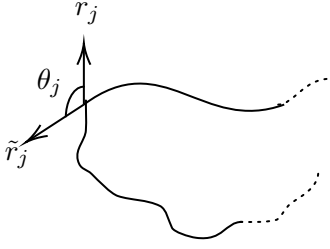

In our context, since the boundary is one dimensional, the extrinsic curvature has only one component. The extra terms in equation \eqref{totvarbucb} compared to the smooth boundary case are the contributions coming from the ``corners'', i.e., points where the boundary conditions change abruptly. The quantity $\theta_j$ in equation \eqref{totvarbucb}  is the angle between the two boundaries at the $j^{\text{th}}$ corner. The corner angle can be expressed in terms of the tangent vectors to the boundaries at the corner as
\begin{align}
\cos\theta_j=r_j\cdot\tilde{r}_j\label{rrj}
\end{align}
where $r_j,\tilde{r}_j$ are the ``outward" drawn (unit normalized) tangent vectors to the boundaries that meet at the $j^{\text{th}}$ corner, see fig.\ref{cornerang}.  

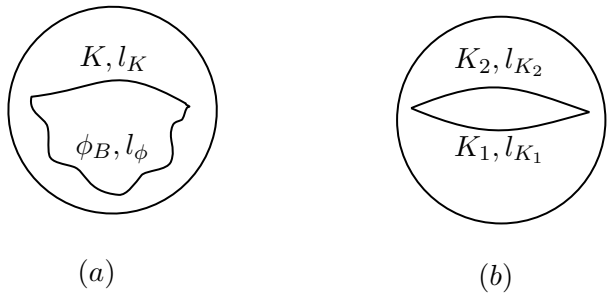
\begin{figure}
	
	\centering
	\label{fig:bcfig}
	
	\tikzset{every picture/.style={line width=0.75pt}} %set default line width to 0.75pt        
	
	\begin{tikzpicture}[x=0.75pt,y=0.75pt,yscale=-1,xscale=1]
		
		%Shape: Circle [id:dp5917095101757305] 
		\draw   (177.18,113.97) .. controls (183.28,85.9) and (210.97,68.09) .. (239.03,74.18) .. controls (267.1,80.28) and (284.91,107.97) .. (278.82,136.03) .. controls (272.72,164.1) and (245.03,181.91) .. (216.97,175.82) .. controls (188.9,169.72) and (171.09,142.03) .. (177.18,113.97) -- cycle ;
		%Shape: Free Drawing [id:dp8057048901321816] 
		\draw  [line width=0.75] [line join = round][line cap = round] (187.51,118.26) .. controls (185.66,126.78) and (194.29,128.57) .. (195.5,133.29) .. controls (196.71,138.03) and (195.29,143.31) .. (196.44,147.82) .. controls (197.38,151.5) and (204.04,152.01) .. (208.51,153.51) .. controls (215.01,155.7) and (214.46,166.13) .. (231.03,167.61) .. controls (235.16,167.98) and (239.4,161.49) .. (242.37,157.8) .. controls (244.54,155.1) and (255.41,155.59) .. (257.19,150.78) .. controls (259.9,143.49) and (255.12,139.36) .. (258.08,132.55) .. controls (259.06,130.32) and (262.16,130.64) .. (262.84,129.49) .. controls (263.87,127.75) and (264.06,125.59) .. (265.09,123.84) .. controls (265.33,123.44) and (266.65,123.37) .. (266.28,123.08) .. controls (265.7,122.61) and (264.38,122.41) .. (264.54,121.68) ;
		%Curve Lines [id:da14942571002860272] 
		\draw    (187.51,118.26) .. controls (224.26,113.95) and (227.23,100.27) .. (264.54,121.68) ;
		%Shape: Circle [id:dp9980339136030475] 
		\draw   (370,130) .. controls (370,101.28) and (393.28,78) .. (422,78) .. controls (450.72,78) and (474,101.28) .. (474,130) .. controls (474,158.72) and (450.72,182) .. (422,182) .. controls (393.28,182) and (370,158.72) .. (370,130) -- cycle ;
		%Curve Lines [id:da8173119888007423] 
		\draw    (377,124) .. controls (418,107) and (429,113) .. (466,126) ;
		%Curve Lines [id:da9663039539271185] 
		\draw    (377,124) .. controls (417,141) and (430,136) .. (466,126) ;
		
		% Text Node
		\draw (208,135.4) node [anchor=north west][inner sep=0.75pt]    {$\phi _{B} ,l_{\phi }$};
		% Text Node
		\draw (210,91.4) node [anchor=north west][inner sep=0.75pt]    {$K,l_{K}$};
		% Text Node
		\draw (397,92.4) node [anchor=north west][inner sep=0.75pt]    {$K_{2} ,l_{K_{2}}$};
		% Text Node
		\draw (397,136.4) node [anchor=north west][inner sep=0.75pt]    {$K_{1} ,l_{K_{1}}$};
		% Text Node
		\draw (209,198.4) node [anchor=north west][inner sep=0.75pt]    {$( a)$};
		% Text Node
		\draw (409,200.4) node [anchor=north west][inner sep=0.75pt]    {$( b)$};

	\end{tikzpicture}
	\caption{ (a) Part of the boundary with dilaton and metric fixed and part of the boundary with extrinsic curvature and metric fixed. (b)Two boundaries on each of which the extrinsic curvature and metric are fixed but to different values.}
\end{figure}

For boundaries which are smooth, there would be no corner terms. However, in several instances (see figure \ref{fig:bcfig}), one is interested in boundary conditions which change abruptly at some point along the boundary. For instance, this is the case for the HH wavefunction. In this case, corner terms arise in the variation of the action, and in order to cancel these corner contributions, one must add the following corner terms to the action:
\begin{equation}\label{cornerterms}
S_{\text{corner}}=\sum_{i \in \text{corners}}\phi_i\theta_i,
\end{equation}
where $\phi_i$ is the dilaton at the $i$th corner and $\theta_i$ is the $i$th corner angle.  These terms are the 2D analog of the Hayward-corner terms in higher dimensional setups \cite{Hayward:1993my}. 
{
For wavefunctions we shall compute in what follows, we also need to add a `counterterm' for the corner terms, 

\begin{align}
    S_{\text{ct,corner}}=-\frac{\pi}{2}\sum_{i \in \text{corners}}\phi_i\label{ctforcor}
\end{align}

This term is important for the net corner term contribution to vanish when we compute overlap of two wavefunctions, as will be shown in appendix \ref{conv2wf}. In most of the paper, we shall not keep track of the contribution of this to the wavefunction except in appendix \ref{conv2wf}, where we compute the overlap of wavefunctions. 
}

Summarizing, for a well-defined variational principle, the bulk action must be supplemented by some boundary terms. On portions of the boundary with Dirichlet boundary conditions (i.e., fixed dilaton and induced metric), we must add the GHY boundary term. No boundary terms are needed on portions of the boundary with York boundary conditions (i.e., fixed extrinsic curvature and induced metric). Finally, at points on the boundary where boundary conditions change abruptly, we must add the corner terms mentioned in equation \eqref{cornerterms}. 
%is given as follows: 

\section{Classical calculation of the HH wavefunction}
\label{clsol}
In this section, we discuss the classical calculation of the HH wavefunction at non-zero values of York time. The Hartle-Hawking wavefunction is given by the gravitational path integral with part of the boundary being an asymptotic boundary where the dilaton and length are fixed (and taken to infinity in the standard way), and the rest being a York boundary where we fix the extrinsic curvature and renormalized length. Previous calculations \cite{Yang:2018gdb,Iliesiu:2024cnh,Miyaji:2024ity,Miyaji:2025yvm} fixed the extrinsic curvature on the York boundary to zero, but here we will fix it to some non-zero constant $k$. We will interpret this as the HH wavefunction at a particular value of York time $k$; changing $k$  corresponds to York time evolution. If we interpret the insertion of a fixed-length boundary in the gravitational path integral at some fixed value of extrinsic curvature as taking the overlap with a fixed-length state $|\ell,k\rangle$, then the generalized HH wavefunction can be interpreted as the overlap $\langle \ell, k| \Psi_{\beta}\rangle$, where $\Psi_{\beta}$ is the thermo-field double state at inverse temperature $\beta$.

Let $\del_{A}$ and $\del_K$ stand for the asymptotic boundary and York boundary respectively. To be more specific, the boundary conditions for the fields are specified as:
\begin{align}
	&	\phi\big\vert_{\del_{A}}\equiv q=\phi_B\sim \frac{\phi_b}{\epsilon},\ell\big\vert_{\del_{A}}\equiv \ell_\phi\sim \frac{u}{\epsilon},\nonumber\\
	&	K\big\vert_{\del_K}=k, \ell\big\vert_{\del_K}\equiv \ell_k\sim \ell_{\text{ren.}} + \frac{2}{\sqrt{1-k^2}}\log (\frac{\sqrt{1-k^2}}{\epsilon}),\label{bdcond}
\end{align}
where $\epsilon\to 0$ and $\phi_b, \beta, k, \ell_{\text{ren.}}$ are all finite. Here, $\ell_{\text{ren.}}$ can be thought of as the ``renormalized length'' of the spatial slice of extrinsic curvature $k${\footnote{The factor of $1-k^2$ inside the logarithm turns out to simplify various formulae as we will see in section \ref{qsols}. The significance of this particular renormalization is unclear to us.}}. It is also convenient to introduce another variable $L$ defined in terms of the renormalized length as  
\beq \label{bdcond2}
\ell_{\text{ren.}} =\frac{2}{\sqrt{1-k^2}}\log (L),
\eeq 
which will appear in some of our formulae below. 
 
The full action, including the boundary terms appropriate for the  boundary conditions in eq.\eqref{bdcond}, is given by 
\begin{align}
	S=-\half\int d^2x \sqrt{g} \,\phi (R+2)-\int_{\del_A} dx\, \sqrt{\gamma}\phi (K-1)+ \sum_{j\in \text{corners}}\phi_j \,\theta_j,\label{sjt}
\end{align}
where we have added an  additional counter term to render the on-shell value of the action finite. The dilaton equation sets $R=-2$; the solution for the metric in polar coordinates is given by equation \eqref{liele}. The dilaton is then given by $\phi = \phi_b r.$

%The equation of motion obtained by varying the dilaton  is given by 
%\begin{align}
%	R+2=0\label{r2}
%\end{align}

Let $r=r_c$ be the location of the asymptotic boundary. From the boundary condition in equation \eqref{bdcond}, we see that
\begin{align}
	r_c\sim\frac{1}{\epsilon}\gg 1\label{rcepsilon}.
\end{align}
Let us now find the trajectory of the York boundary with constant extrinsic curvature. Let $\lambda$ be a proper length coordinate along this curve in terms of which, the trajectory is given by $(t(\lambda),r(\lambda))$. The extrinsic curvature in terms of the functions $t(\lambda)$ and  $r(\lambda)$ is given by 
\begin{align}
	K&={\nabla_\mu n^\mu},\nonumber\\
	&=\frac{\sqrt{r^2-1} \left(\left(r^2-1\right) r' t ''+t ' \left(r''+r \left(-r r''+3 r'^2+\left(r^2-1\right)^2 t '^2\right)\right)\right)}{\left(r'^2+\left(r^2-1\right)^2 t '^2\right)^{3/2}},\label{adsexkgbgc}
\end{align}
where in the above equation primes denote derivatives with respect to $\lambda$. 
Since, $\lambda$ is taken to be a proper length coordinate along the curve, we have the following relation between the functions $t(\lambda)$ and $r(\lambda)$:
\begin{align}
	(r^2-r_s^2)t'^2+\frac{r'^2}{r^2-r_s^2}=1.\label{prptim}
\end{align}
Further differentiating this condition with respect to $\lambda$, we can solve for $t''(\lambda)$ in terms of $r(\lambda)$:
\begin{align}
	t''(\lambda)=\frac{r'(\lambda ) \left(r''(\lambda ) \left(r_s^2-r(\lambda )^2\right)+r(\lambda ) \left(2 r'(\lambda )^2-r(\lambda )^2+r_s^2\right)\right)}{\left(r_s^2-r(\lambda )^2\right)^2 \sqrt{-r'(\lambda )^2+r(\lambda )^2-r_s^2}}.\label{tdpinrdp}
\end{align}
Substituting the above results for $t''(\lambda)$ and $t'(\lambda)$ in eq.\eqref{adsexkgbgc}, we get
\begin{align}
	K=\frac{r(\lambda )-r''(\lambda )}{\sqrt{-r'(\lambda )^2+r(\lambda )^2-r_s^2}}.\label{rrdpink}
\end{align}
We need to solve the above equation with extrinsic curvature $K(\lambda)$ set to a constant (i.e., $\lambda$-independent) value $k$ to obtain the function $r(\lambda)$. The solutions for $(r(\lambda), t(\lambda))$ are then given by:
\begin{align}
	\tilde{r}\equiv	\frac{r(\lambda)}{r_s}&=\frac{\sqrt{\tilde{J}^2+1} \cosh \left(\sqrt{1-k^2} \lambda \right)+\tilde{J} k}{\sqrt{1-k^2}},\nonumber\\
	\tilde{t}\equiv	r_st(\lambda)&= \arctan(\frac{\sinh(\sqrt{1-k^2}\lambda)}{\tilde{J}\cosh(\sqrt{1-k^2}\lambda)+k\sqrt{1+\tilde{J}^2}}),
	\label{rtslasol}
\end{align}
where $\tilde{J}$ and $r_s$ are constants. 
In the limit $k\rightarrow 0$, the above equations become
\begin{align}
	\frac{r(\lambda)}{r_s}&={\sqrt{\tilde{J}^2+1} \cosh \left(\lambda \right)},\nonumber\\
	r_st(\lambda)&= \arctan(\frac{1}{\tilde{J}}{\tanh(\lambda)}),
	\label{rtslasolkzer}
\end{align}
which agree with the corresponding expressions in equation (3.27) of \cite{Harlow:2018tqv} with the identification $\tilde{J}=\frac{J}{r_s}$. 

Our next task is to impose the boundary conditions at the two end points of the above curve and compute the on-shell action for the corresponding solution. The boundary conditions will end up determining the constants $r_s$ and $\tilde{J}$ in terms of the boundary quantities, $\phi_b, \beta, L$ and $k$. The constant-$K$ curve meets the asymptotic boundary at two points. Since the asymptotic boundary is located at $r=r_c$, these intersection points also have $r=r_c$. Let $\lambda_1, \lambda_2$ be the values of the proper length coordinate $\lambda$ at these intersection points. We then have
\begin{align}
	\frac{\sqrt{\tilde{J}^2+1} \cosh \left(\sqrt{1-k^2} \lambda_2 \right)+\tilde{J} k}{\sqrt{1-k^2}}=\frac{r_c}{r_s}=\frac{\sqrt{\tilde{J}^2+1} \cosh \left(\sqrt{1-k^2} \lambda_1 \right)+\tilde{J} k}{\sqrt{1-k^2}}\label{l2l1con},
\end{align}
which implies that 
\begin{align}
	\lambda_2=-\lambda_1.\label{l2l1rel}
\end{align}
Further noting that  $r_c\gg 1$, 
\begin{align}
	\sqrt{1-k^2}\lambda_2\simeq \log(2\frac{r_c}{r_s}\sqrt{\frac{1-k^2}{1+\tilde{J}^2}}).\label{lamd2}
\end{align}
The length of the constant $K$-boundary is given by 
\begin{align}
	\ell_{\del_K}=\int_{\lambda_1}^{\lambda_2} d\lambda=2\lambda_2\simeq \frac{2}{\sqrt{1-k^2}}\log(2\frac{r_c}{r_s}\sqrt{\frac{1-k^2}{1+\tilde{J}^2}}).\label{lkb}
\end{align}
Comparing with equations \eqref{bdcond} and \eqref{bdcond2} gives
\begin{align}
	L=\frac{2}{r_s}\frac{1}{\sqrt{1+\tilde{J}^2}}.\label{lrsrel}
\end{align}
Now, we need to impose the condition that the length of the asymptotic boundary is fixed. Since the boundary is located at $r=r_c\gg 1$, the line element at the boundary is given by 
\begin{align}
	ds^2\simeq r_c^2dt^2\label{bdlele}
\end{align}
and so the length of the asymptotic boundary is given by 
\begin{align}
	\ell_{\phi}=\int ds=r_c\int_{t(\lambda_1)}^{t(\lambda_2)} dt\label{ldela}
\end{align}
and further noting the condition in equation \eqref{bdcond} we get
\begin{align}
	t(\lambda_2)-t(\lambda_1)=u.\label{asbcon}
\end{align}
 Noting equation \eqref{l2l1rel} and using the solution for $t(\lambda)$ in equation \eqref{rtslasolkzer}, we get
\begin{align}
	\frac{\sinh(\sqrt{1-k^2}\lambda_2)}{\tilde{J}\cosh(\sqrt{1-k^2}\lambda_2)+k\sqrt{1+\tilde{J}^2}}=\tan(\frac{u r_s}{2}).\label{jrsbeq}
\end{align}
From equation \eqref{lamd2} or \eqref{l2l1con}, we notice that $\cosh(\sqrt{1-k^2}\lambda_2)\sim \sinh(\sqrt{1-k^2}\lambda_2)\gg 1$ since $r_c\gg 1$. Thus, we can safely neglect the term $k\sqrt{1+\tilde{J}^2}$ in the denominator in equation \eqref{jrsbeq} to get
\begin{align}
	\tilde{J}\simeq \cot(\frac{u r_s}{2}).\label{jcon}
\end{align}
Using this in equation \eqref{lrsrel}, we find
\begin{align}
	{L}=\frac{2}{r_s}\sin({u r_s\over 2}),\label{lrs}
\end{align}
which is an implicit equation for $r_s$ in terms of the boundary data. 
We now have all the ingredients to evaluate the on-shell value of the action. The action is given by equation \eqref{sjt}.

The bulk term vanishes as a result of equation \eqref{eom1}. The extrinsic curvature at the asymptotic boundary is given by 
\begin{align}
	K_{\del_A}=1+\frac{1}{2}\frac{r_s^2}{r_c^2}.\label{exkrsrc}
\end{align}
So, the second term in equation \eqref{sjt} has the value
\begin{align}
	\int_{\del_A}\phi (K-1)ds=\half \phi_b r_s^2 u.\label{phikasy}
\end{align}

Finally, we have to evaluate the contribution from the corner terms in equation \eqref{sjt}. There are two corner terms, one at $\lambda=\lambda_1$ and another at $\lambda=\lambda_2$. However since the solution we are considering is symmetric in $\lambda$, the angle will be the same at these corners.  The cosine of the angle between vectors can be obtained using the tangent vectors
\begin{align}
	\cos\theta=\frac{\vec{v}_1\cdot \vec{v}_2}{
		\sqrt{\vec{v}_1\cdot\vec{v}_1\vec{v}_2\cdot\vec{v}_2}}\label{vects}
\end{align}
where $\vec{v}_1,\vec{v}_2$ are the tangent vectors between which the angle is to be computed at the corner points. The vectors are given in our case by 
\begin{align}
	\vec{v}_1=(1,0),\quad 	\vec{v}_2=(t'(\lambda),r'(\lambda)),\label{vecsval}
\end{align}
where $\vec{v}_1$ indicates the direction along the asymptotic boundary and $\vec{v}_2$ along the boundary of constant $K$. Using the line element eq.\eqref{liele} to compute the inner products and noting the relation \eqref{prptim} (which essentially means that $\vec{v}_2$ is already unit normalized), we get
\begin{align}
	\cos\theta &=\sqrt{r_c^2-r_s^2}\,t'(\lambda)\nonumber\\
	&=\frac{\tilde{J} r_s \,\sqrt{1-k^2} +k r_c}{\sqrt{r_c^2-r_s^2}}\nonumber\\
	&\simeq k+\frac{\tilde{J}r_s}{r_c}\sqrt{1-k^2},
	\label{tlcos}
\end{align}
where in obtaining the second line we used the function $t(\lambda)$ in eq.\eqref{rtslasol} and simplified the result using the condition eq.\eqref{l2l1con}. The third line follows in the limit of $r_c\gg 1$.
Combining the results in eq.\eqref{phikasy} and eq.\eqref{tlcos}, we get
\begin{align}
	-S_{\text{on-shell}}&\simeq \half \phi_b r_s^2 u - 2\phi_b r_c\arccos(k + \frac{\tilde{J}r_s}{r_c}\sqrt{1-k^2})\nonumber\\
	&\simeq \half \phi_b r_s^2u-2\phi_b r_c\left(\arccos(k )- \frac{\tilde{J}r_s}{r_c}\right)\nonumber\\
	&\simeq \half \phi_b r_s^2 u-2\phi_b r_c\arccos(k )+ 2\phi_b{r_s}\cot(\frac{u r_s}{2}),
	\label{onsheval}
\end{align}
where we used eq.\eqref{jcon} in obtaining the final expression and  $r_s$ is understood to be a function of boundary data through eq.\eqref{lrs}. Note that, we have not included the corner term contribution from eq.\eqref{ctforcor} in writing the above result, which simply adds a term proportional to $\phi_B$. The wavefunction at the classical level is thus given by 
\begin{align}
	\Psi(k,\ell)=e^{-S_{\text{on-shell}}}=\exp(\half \phi_b r_s^2u-2\phi_b r_c\arccos(k )+ 2\phi_b{r_s}\cot(\frac{u r_s}{2})).\label{psicl}
\end{align}
{
We end this section with the following remark on the conventions. {For the York boundary, the outward drawn normal (outward with respect to the bulk region enclosed by  the asymptotic and York boundaries) has extrinsic curvature positive if the enclosed bulk region encompasses the geodesic curve between the corner points and negative otherwise. Note that for the York boundary between two points, for the extrinsic curvature to be positive, the normal would have to be directed away from the corresponding geodesic between those points and towards the boundary. It is useful to look more carefully at the angle in eq.\eqref{tlcos}. In the limit $k\rightarrow 0$, we see that the angle in eq.\eqref{tlcos} becomes
\begin{align}
    \theta\simeq \frac{\pi}{2}-k-\frac{\tilde{J}r_s}{r_c}\label{kzerang}
\end{align}
Thus, the angle between the asymptotic boundary and the York boundary is less than $\frac{\pi}{2}$. So, the geodesic curve between the corner points where the boundaries meet, see fig.\ref{clasfig}(a), lies outside the bulk region. For a configuration in which the geodesic curve lies inside the bulk as in fig.\ref{clasfig}(b), the corresponding result would be the same as eq.\eqref{psicl} but with the replacement $k\rightarrow -k$,}
\begin{align}
	\Psi=e^{-S_{\text{on-shell}}}=\exp(\half \phi_b r_s^2u-2\phi_b r_c\arccos(-k )+ 2\phi_b{r_s}\cot(\frac{u r_s}{2})).\label{psicl2}
\end{align}
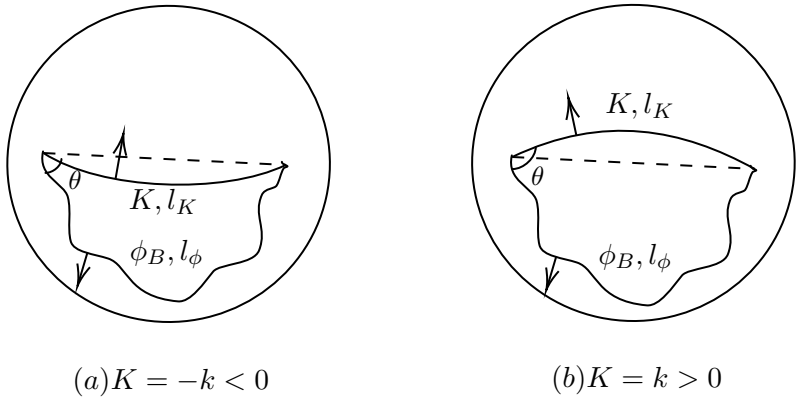
\begin{figure}
    \centering

\tikzset{every picture/.style={line width=0.75pt}} %set default line width to 0.75pt        

\begin{tikzpicture}[x=0.75pt,y=0.75pt,yscale=-1,xscale=1]

%Shape: Ellipse [id:dp6597482656017942] 
\draw   (125.84,120.63) .. controls (135.27,77.73) and (178.13,50.5) .. (221.57,59.81) .. controls (265.02,69.13) and (292.58,111.45) .. (283.15,154.36) .. controls (273.72,197.26) and (230.86,224.49) .. (187.42,215.18) .. controls (143.97,205.86) and (116.4,163.53) .. (125.84,120.63) -- cycle ;
%Shape: Free Drawing [id:dp599934342474739] 
\draw  [line width=0.75] [line join = round][line cap = round] (141.82,131.19) .. controls (138.95,144.22) and (152.32,146.95) .. (154.19,154.17) .. controls (156.06,161.42) and (153.86,169.48) .. (155.64,176.39) .. controls (157.09,182.01) and (167.41,182.79) .. (174.32,185.08) .. controls (184.39,188.43) and (183.53,204.36) .. (209.18,206.63) .. controls (215.58,207.2) and (222.14,197.27) .. (226.74,191.63) .. controls (230.09,187.51) and (246.92,188.25) .. (249.68,180.9) .. controls (253.88,169.76) and (246.47,163.44) .. (251.06,153.04) .. controls (252.57,149.63) and (257.37,150.11) .. (258.42,148.36) .. controls (260.02,145.7) and (260.3,142.39) .. (261.91,139.72) .. controls (262.28,139.1) and (264.32,139) .. (263.75,138.55) .. controls (262.85,137.84) and (260.81,137.52) .. (261.05,136.41) ;
%Curve Lines [id:da2511186128962245] 
\draw    (142,132) .. controls (171,151) and (232,153) .. (263,138) ;
%Straight Lines [id:da8076577879331953] 
\draw  [dash pattern={on 4.5pt off 4.5pt}]  (142,132) -- (263,138) ;
%Straight Lines [id:da4029283688915869] 
\draw    (164,182) -- (159.56,197.08) ;
\draw [shift={(159,199)}, rotate = 286.39] [color={rgb, 255:red, 0; green, 0; blue, 0 }  ][line width=0.75]    (10.93,-3.29) .. controls (6.95,-1.4) and (3.31,-0.3) .. (0,0) .. controls (3.31,0.3) and (6.95,1.4) .. (10.93,3.29)   ;
%Straight Lines [id:da6401959427402941] 
\draw    (178,145) -- (181.66,123.97) ;
\draw [shift={(182,122)}, rotate = 99.87] [color={rgb, 255:red, 0; green, 0; blue, 0 }  ][line width=0.75]    (10.93,-3.29) .. controls (6.95,-1.4) and (3.31,-0.3) .. (0,0) .. controls (3.31,0.3) and (6.95,1.4) .. (10.93,3.29)   ;
%Shape: Free Drawing [id:dp8313302956340525] 
\draw  [line width=0.75] [line join = round][line cap = round] (143,142) .. controls (146.83,142) and (151,138.39) .. (151,135) ;
%Shape: Ellipse [id:dp3385078484189772] 
\draw   (357.84,120.14) .. controls (367.28,77.23) and (410.14,50.01) .. (453.58,59.32) .. controls (497.02,68.63) and (524.59,110.96) .. (515.16,153.86) .. controls (505.72,196.77) and (462.86,223.99) .. (419.42,214.68) .. controls (375.98,205.37) and (348.41,163.04) .. (357.84,120.14) -- cycle ;
%Shape: Free Drawing [id:dp6008861903329619] 
\draw  [line width=0.75] [line join = round][line cap = round] (375.82,133.19) .. controls (372.95,146.22) and (386.32,148.95) .. (388.19,156.17) .. controls (390.06,163.42) and (387.86,171.48) .. (389.64,178.39) .. controls (391.09,184.01) and (401.41,184.79) .. (408.32,187.08) .. controls (418.39,190.43) and (417.53,206.36) .. (443.18,208.63) .. controls (449.58,209.2) and (456.14,199.27) .. (460.74,193.63) .. controls (464.09,189.51) and (480.92,190.25) .. (483.68,182.9) .. controls (487.88,171.76) and (480.47,165.44) .. (485.06,155.04) .. controls (486.57,151.63) and (491.37,152.11) .. (492.42,150.36) .. controls (494.02,147.7) and (494.3,144.39) .. (495.91,141.72) .. controls (496.28,141.1) and (498.32,141) .. (497.75,140.55) .. controls (496.85,139.84) and (494.81,139.52) .. (495.05,138.41) ;
%Curve Lines [id:da8637710764057177] 
\draw    (376,134) .. controls (422,112) and (461,120) .. (497,140) ;
%Straight Lines [id:da7468439962185359] 
\draw  [dash pattern={on 4.5pt off 4.5pt}]  (376,134) -- (497,140) ;
%Straight Lines [id:da38216237637224104] 
\draw    (398,184) -- (393.56,199.08) ;
\draw [shift={(393,201)}, rotate = 286.39] [color={rgb, 255:red, 0; green, 0; blue, 0 }  ][line width=0.75]    (10.93,-3.29) .. controls (6.95,-1.4) and (3.31,-0.3) .. (0,0) .. controls (3.31,0.3) and (6.95,1.4) .. (10.93,3.29)   ;
%Straight Lines [id:da9510164461170604] 
\draw    (408,123) -- (404.41,105.96) ;
\draw [shift={(404,104)}, rotate = 78.11] [color={rgb, 255:red, 0; green, 0; blue, 0 }  ][line width=0.75]    (10.93,-3.29) .. controls (6.95,-1.4) and (3.31,-0.3) .. (0,0) .. controls (3.31,0.3) and (6.95,1.4) .. (10.93,3.29)   ;
%Shape: Free Drawing [id:dp9503045669989129] 
\draw  [line width=0.75] [line join = round][line cap = round] (376.22,140.06) .. controls (381.75,140.06) and (387.78,134.32) .. (387.78,128.94) ;

% Text Node
\draw (183.59,173.94) node [anchor=north west][inner sep=0.75pt]    {$\phi _{B} ,l_{\phi }$};
% Text Node
\draw (183.22,148.68) node [anchor=north west][inner sep=0.75pt]    {$K,l_{K}$};
% Text Node
\draw (153,139.4) node [anchor=north west][inner sep=0.75pt]  [font=\small]  {$\theta $};
% Text Node
\draw (417.59,175.94) node [anchor=north west][inner sep=0.75pt]    {$\phi _{B} ,l_{\phi }$};
% Text Node
\draw (421,100.4) node [anchor=north west][inner sep=0.75pt]    {$K,l_{K}$};
% Text Node
\draw (384,137.4) node [anchor=north west][inner sep=0.75pt]  [font=\small]  {$\theta $};
% Text Node
\draw (155,238.4) node [anchor=north west][inner sep=0.75pt]    {$( a) K=-k< 0$};
% Text Node
\draw (394,237.4) node [anchor=north west][inner sep=0.75pt]    {$( b) K=k >0$};

\end{tikzpicture}
    \caption{The mixed boundary case with $\theta$ denoting the angle between the two boundaries and the arrows indicating the normals to the boundaries. The dotted line corresponds to the geodesic boundary with $K=0$.}
    \label{clasfig}
\end{figure}
This will be important to compare with results to appear later for the full quantum wavefunction.
}

\section{Quantum HH Wavefunction}
\label{qsols}

We now turn to the quantum wavefunction. We will derive the quantum wavefunction in two ways: first, from the Hamiltonian perspective (following \cite{Harlow:2018tqv}), and second from the path integral perspective following the boundary particle formalism of \cite{Yang:2018gdb,Kitaev:2018wpr}. The wavefunction we obtain in this section will reduce to the results of the previous section in the classical limit. 
\subsection{Hamiltonian approach}
\label{wfenegis}
 The analysis of this section closely follows that of \cite{Harlow:2018tqv} (see also Appendix A of \cite{Blommaert:2025avl} for a related calculation). We work in Lorentzian signature, where the metric and dilaton solution for a  black hole are given by 
\begin{align}
	ds^2&=-(r^2-r_s^2)dt^2+\frac{dr^2}{r^2-r_s^2},\nonumber\\
	\phi&=\phi_b r.
	\label{lormet}
\end{align}
However, since the $(t,r)$ coordinates do not cover the entire spacetime, it is useful to switch to global coordinates $(x,\tau)$ in terms of which the metric and dilaton are given by 
\begin{align}
	ds^2&=-(1+x^2)d\tau^2+\frac{dx^2}{(x^2+1)},\nonumber\\
	\phi&=\phi_h \sqrt{1+x^2}\cos(\tau).
	\label{loresol}
\end{align}
The relation between the  two coordinate systems above is given by 
\begin{align}
	\frac{r}{r_s}=\sqrt{1+x^2}\cos\tau,\,\quad \tanh(r_s t)=\frac{\sqrt{1+x^2}}{x}\sin\tau,\label{rrstcod}
\end{align}
along with the identification of the constants $\phi_h=\phi_b r_s$.
A straightforward evaluation then shows that the full Hamiltonian of the theory, which is the sum of the ADM Hamiltonian at the left and right boundaries, is given by 
\begin{align}
	H\equiv H_L+H_R=\frac{\phi_h^2}{\phi_b},\label{hlr}
\end{align}
where $H_{L,R}$ are the left and right Hamiltonian/ADM energies\footnote{Note that we are working in units of $8\pi G=1$. In \cite{Harlow:2018tqv}, the ADM energy is in units of $16\pi G=1$, which is the reason for the factor of $2$ difference with the result there.}. The classical phase space is two dimensional with the symplectic form on this phase space given by 
\begin{align}
	\omega= d\delta \wedge dH,\label{sym}
\end{align}
where $\delta$ is the linear combination of left and right Schwarzschild time given by 
\begin{align}
	\delta =\frac{t_L+t_R}{2}.\label{dlt}
\end{align}
The Schwarzschild times $t_L,t_R$ are related to the global time $\tau$ at the boundaries by 
\begin{align}
	\cos\tau=\sech (r_s t_L)=\sech (r_s t_R).\label{gtst}
\end{align}
We are interested in computing overlaps between energy eigenstates of the Hamiltonian with fixed-length states at constant (non-zero) extrinsic curvature. For this purpose, let us compute the length of a slice of constant extrinsic curvature.
\begin{figure}
    \centering

\tikzset{every picture/.style={line width=0.75pt}} %set default line width to 0.75pt        

\tikzset{every picture/.style={line width=0.75pt}} %set default line width to 0.75pt        

\begin{tikzpicture}[x=0.75pt,y=0.75pt,yscale=-1,xscale=1]

%Straight Lines [id:da3501227050514617] 
\draw    (305.47,102.54) -- (305.99,247.7) ;
%Straight Lines [id:da9517001930353255] 
\draw    (403.81,101.35) -- (405.5,246.74) ;
%Shape: Free Drawing [id:dp43569488268444645] 
\draw  [line width=0.75] [line join = round][line cap = round] (315.53,106.11) .. controls (315.53,112.87) and (317.07,119.61) .. (316.65,126.35) .. controls (316.34,131.29) and (313.77,141.01) .. (314.41,147.79) .. controls (314.89,152.94) and (315.9,158.11) .. (315.53,163.27) .. controls (315.09,169.36) and (312.37,171.25) .. (313.29,181.13) .. controls (313.52,183.5) and (317.62,189.48) .. (317.76,191.84) .. controls (318.56,205.51) and (316.13,210.22) .. (315.53,219.23) .. controls (315.06,226.2) and (315.53,243.28) .. (315.53,249) ;
%Shape: Free Drawing [id:dp5796299811439682] 
\draw  [line width=0.75] [line join = round][line cap = round] (395.99,106.11) .. controls (395.99,107.09) and (398.22,109.61) .. (398.23,109.68) .. controls (398.6,114.03) and (398.6,118.43) .. (398.23,122.78) .. controls (398.04,124.91) and (396.19,126.61) .. (395.99,128.73) .. controls (394.97,139.56) and (398.39,140.58) .. (397.11,150.17) .. controls (396.63,153.76) and (392.71,157.68) .. (393.76,163.27) .. controls (394.02,164.67) and (395.86,165.41) .. (395.99,166.84) .. controls (396.95,177.08) and (392.46,182.18) .. (393.76,191.84) .. controls (395.9,207.84) and (398.23,225.12) .. (398.23,243.05) ;
%Shape: Free Drawing [id:dp12090496550810648] 
\draw  [line width=3] [line join = round][line cap = round] (317,160) .. controls (317,162.8) and (313.09,160) .. (316,160) ;
%Shape: Free Drawing [id:dp31436529364996213] 
\draw  [line width=3] [line join = round][line cap = round] (394,161) .. controls (394,160.53) and (393,160.47) .. (393,160) ;
%Straight Lines [id:da2110233288183644] 
\draw    (316,160) -- (392,161) ;
%Curve Lines [id:da8026775007833316] 
\draw    (316,160) .. controls (350,140) and (368,146) .. (392,161) ;
%Curve Lines [id:da08116351852960779] 
\draw    (316,160) .. controls (349,123) and (363,122) .. (392,161) ;
%Curve Lines [id:da5126335185489354] 
\draw    (318.7,110.97) .. controls (337.9,103.48) and (345.02,95.36) .. (344.07,68.66) ;
\draw [shift={(344,67)}, rotate = 87.33] [color={rgb, 255:red, 0; green, 0; blue, 0 }  ][line width=0.75]    (10.93,-3.29) .. controls (6.95,-1.4) and (3.31,-0.3) .. (0,0) .. controls (3.31,0.3) and (6.95,1.4) .. (10.93,3.29)   ;
%Curve Lines [id:da9091986564798887] 
\draw    (396,115) .. controls (361.7,100.3) and (362.27,95.23) .. (361.08,68.65) ;
\draw [shift={(361,67)}, rotate = 87.33] [color={rgb, 255:red, 0; green, 0; blue, 0 }  ][line width=0.75]    (10.93,-3.29) .. controls (6.95,-1.4) and (3.31,-0.3) .. (0,0) .. controls (3.31,0.3) and (6.95,1.4) .. (10.93,3.29)   ;

% Text Node
\draw (340,115) node [anchor=north west][inner sep=0.75pt]   [align=left] {$k>0$};
% Text Node
\draw (342,160) node [anchor=north west][inner sep=0.75pt]   [align=left] {$k=0$};
% Text Node
\draw (327.3,38.4) node [anchor=north west][inner sep=0.75pt]  [font=\small]  {$\phi =\phi _{B}$};

\end{tikzpicture}
    \caption{Slices of constant extrinsic curvature between asymptotic boundaries ending at fixed boundary times.}
    \label{fig:yorkslices}
\end{figure}
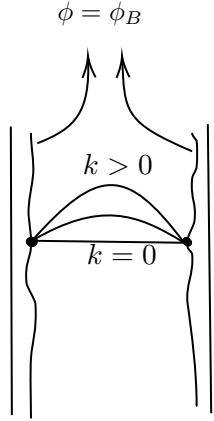
Note that, in AdS$_2$ spacetime under consideration, there always exists a unique codimension-1 spacelike slice with specified extrinsic curvature and ending on the given boundary time slice, see fig.\ref{fig:yorkslices}.
 It can be checked that such a slice can be parametrized in global coordinates as
\begin{align}
	x(\lambda)&=\frac{\sinh \left(\sqrt{k^2+1}\, \lambda \right)}{\sqrt{k^2+1}},\nonumber\\
	\tau(\lambda)&=\tau_0+\tan ^{-1}\left(k \sech\left(\lambda\sqrt{k^2+1}  \right)\right),\label{xtlambda}
\end{align}
where $\tau_0$ is a parameter that will be related to $\delta$ in eq.\eqref{dlt}. However, we do not need to explicitly evaluate it. The range of $\lambda$ is determined by the value of the dilaton through the relation for the solution for dilaton in eq.\eqref{loresol} and eq.\eqref{gtst}. The dilation on the boundary can be written as
\begin{align}
	\phi_B=\phi_h\cos(\tau)\sqrt{1+\frac{\sinh^2 \left(\sqrt{k^2+1} \lambda \right)}{1+k^2}}=\phi_h\sech(r_s\delta)\sqrt{1+\frac{\sinh^2 \left(\sqrt{k^2+1} \lambda \right)}{1+k^2}},
	\end{align}
where in obtaining the first equality we have used the first expression in eq.\eqref{xtlambda} and in obtaining the second equality we have made use of eq.\eqref{gtst} along with the restriction $t_L=t_R$  so that $\delta=t_L=t_R$. Solving for $\lambda$ we get
\begin{align}
	\lambda_\pm =\pm \frac{1}{\sqrt{k^2+1}}\sinh ^{-1}\left(\sqrt{\left(k^2+1\right) \left(\frac{\phi _B^2 \cosh ^2\left(\delta  r_s\right)}{\phi _h^2}-1\right)}\right),\label{lmpm}
\end{align}
which label the endpoints of the slice of constant extrinsic curvature. The length of the slice is given by 
\begin{align}
	\ell(k)&=\int_{\lambda_-}^{\lambda_+} d\lambda \sqrt{-(1+x^2)\tau'(\lambda)^2+\frac{x'(\lambda)^2}{(x^2+1)}}=\int_{\lambda_-}^{\lambda_+} d\lambda\nonumber\\
	&=\frac{2}{\sqrt{1+k^2}}\sinh ^{-1}\left(\sqrt{\left(1+k^2\right) \left(\frac{\phi _B^2 \cosh ^2\left(\delta  r_s\right)}{\phi _h^2}-1\right)}\right)\nonumber\\
	&\simeq \frac{2}{ \sqrt{1+k^2}}\log \left(\frac{2\phi _B}{\phi _h} \sqrt{1+k^2}  \cosh\left( \frac{\phi_h}{\phi_b}\delta \right)\right),\label{lkink1}
\end{align}
where the last approximation is in the large $\phi_B$ limit. Let us define the renormalized length as
\begin{align}
	{\ell}_{\text{ren.}}(k)\equiv \ell(k)-\frac{2}{\sqrt{1+k^2}}\log(\frac{\sqrt{1+k^2}}{\epsilon})= \frac{2}{ \sqrt{1+k^2}}\log \left(\frac{2\phi_b}{\phi _h} \cosh\left( \frac{\phi_h}{\phi_b}\delta \right)\right).\label{lkre2}
\end{align}
Note that ${\ell}_{\text{ren}}$ defined as above is related to ${\ell}_{\text{ren}}$ defined in eq.\eqref{bdcond} after taking $k\rightarrow ik$.
%\begin{align}
%    \tilde{\ell}_{\text{ren}}(k)={\ell}_{\text{ren}}(k)-\frac{2}{\sqrt{1+k^2}}\log(\phi_b)
%\end{align}
%\begin{color}{red}(OP: Why are we defining the renormalized length in two different  ways?)\end{color} 
Defining the quantity $P(k)$ as 
\begin{align}
    P(k)=\phi_h\sqrt{1+k^2}\tanh\left(\frac{\phi_h \delta}{\phi_b}\right),\label{plk}
\end{align}
we find that  the symplectic form in eq.\eqref{sym} can be equivalently written in terms of $\ell_{\text{ren.}}(k)$ and $P(k)$ as
\begin{align}
\omega =d\ell_{\text{ren.}}\wedge dP.\label{pklkw}
\end{align}
Thus, $P$ has the interpretation of the conjugate momentum for the renormalized length. The ADM Hamiltonian eq.\eqref{hlr} in terms of these variables can be written as
\begin{align}
	H(k)=\frac{P^2}{\phi_b(1+k^2)}+{4}\phi_b e^{-\ell_{\text{ren.}}\sqrt{1+k^2}}.\label{hinpklk}
\end{align}
We now promote this to a quantum operator on Hilbert space, or more precisely a one-parameter family of Hamiltonians labelled by the extrinsic curvature $k$. Let $|\ell,k\rangle$ be states with fixed renormalized length at extrinsic curvature $k$:
\begin{equation}
    \widehat{\ell}_{\text{ren.}}(k)|\ell,k\rangle = \ell |\ell,k\rangle.
\end{equation}
Our goal is to write an expression for the overlap between energy eigenstates of the ADM Hamiltonian and the fixed length states $|\ell,k\rangle$. This overlap can be computed by making the replacement $P \to -i\frac{d}{d \ell_{\text{ren.}}}$ in the above expression for the ADM Hamiltonian. Denoting the overlap by $\Psi_E(\ell_{\text{ren.}},k)$, we have
\begin{align}
	\frac{1}{\phi_b}\left(-\frac{1}{(1+k^2)}\del_{\ell_{\text{ren}}}^2+{4}\phi_b^2e^{-\ell_{\text{ren}}\sqrt{1+k^2}}\right)\Psi_E(\ell_{\text{ren}},k)=E\,\Psi_E(\ell_{\text{ren}},k).\label{eveq}
\end{align}
The solutions are given by 
%\begin{color}{red}(OP:Why is the second set of solutions not permitted?)\end{color}
\begin{align}
	\Psi_E(\ell_{\text{ren}},k)=K_{2i\sqrt{E\phi_b}}\left({4\phi_b}e^{-\half \ell_{\text{ren}}\sqrt{1+k^2}}\right).\label{psiek}
\end{align}
The HH wavefunction in the constant extrinsic curvature basis can now be obtained as
\begin{equation}
\Psi_{HH}(\ell_{\text{ren}},k) = \langle \ell_{\text{ren}}, k| \text{TFD}(\beta)\rangle= \int_0^{\infty}dE\,\rho(E)e^{-\frac{\beta}{2}E}\Psi_E(\ell_{\text{ren}},k).
\end{equation}
{We will drop the subscript `ren' from here on to keep the notation simple.
We can immediately compute the norm of the states $|\ell,k\rangle$ by noting that the energy eigenstates have the following completeness relations
\begin{align}
    \int_0^{\infty} dE\,\rho(E)\,|E\rangle \langle E| =\mathds{1}, \quad \rho(E)=\frac{2}{\pi^2}\sinh(2\pi\sqrt{E}).\label{ecomp}
\end{align}
It then follows that 
\begin{align}
    \langle \ell_{1},k|\ell_{2},k\rangle &=\int_0^{\infty} dE\, \rho(E) \,\Psi_E(\ell_{2},k)^*\Psi_E(\ell_{1},k)\nonumber\\
    &=\frac{1}{\sqrt{1+k^2}}\delta (\ell_1-\ell_2).
\end{align}
}
%{
%\color{teal}
%It turns out that the appropriate inner product for $\Psi_E(\ell_{\text{ren}},k)$ is given by \cite{Harlow:2018tqv}
%\begin{align}
%({\Psi}_{E_1},{\Psi}_{E_2})=\int d\ell_{\text{ren}} {\Psi}_{E_1}^*(\ell_\text{ren},k){\Psi}_{E_2}(\ell_\text{ren},k)\label{harlownorm}
%\end{align}
%and equipped with this inner product, the wavefunctions in eq.\eqref{psiek} can be seen to %be delta-function normalizable,
%\begin{align}
%    ({\Psi}_{E_1},{\Psi}_{E_2})=\frac{\pi^3}{2\phi_b\sinh(2\pi\sqrt{E_1\phi_b})}\frac{\delta(E_1-E_2)}{\sqrt{1+k^2}}\label{harlonorml}
%\end{align}
The above result shows that the length basis states defined above have a normalization which is York time-dependent. We will later see in subsection \ref{buphham} that this time dependence in the overlap of length states is important to keep track of in order to make the York time evolution manifestly unitary.

%{{\color{blue}SS: $\ell_{ren}$ or $\ell$ in above equations?}}

\subsection{Path integral approach}
In this section, we give a second derivation of the above HH wavefunction, this time from a path integral perspective. 
\subsubsection*{Review of boundary particle formalism}
\label{algelf}
Consider the partition function of JT gravity with a single smooth asymptotic boundary where the dilaton and metric are held fixed. We illustrate the formalism in this setup; generalization to the case of non-smooth boundary conditions is straightforward.

The key point is that the action for the JT gravity path integral, which originally is an integral over the metric and dilaton configurations in the two-dimensional spacetime, boils down to a boundary term. The reason is that the path integral over the dilaton imposes a delta function constraint for the Ricci scalar, and thus restricts the integration over metric configurations to those that have a constant negative curvature with an action governed only by a boundary term:
\begin{align}
	S_\del=-\int_\del \sqrt{h}\,\phi (K-1),\label{sdel}
\end{align}
where $h$ is the induced metric and we have included a counterterm as part of holographic renormalization. Noting that $\phi$ is constant on the boundary, the Gauss-Bonnet theorem implies
\begin{align}
	\int_\del \sqrt{h}\,K=2\pi \chi -\frac{1}{2}\int\,d^2x\, \sqrt{g}R,\nonumber\\
	=2\pi +\int d^2x\, \sqrt{g},\label{delk}
	\end{align}
where we used the fact that the Euler characteristic, $\chi$, is one for the disk and $R=-2$. With this, \eqref{sdel} becomes
\begin{align}
	S_{\del}=-\phi_B(2\pi+\int d^2x\,\sqrt{g}-\ell_\phi),\label{sding1}
\end{align} 
where $\ell_\phi$ is the length of the asymptotic boundary and $\phi_B$ is the value of the dilaton field at the boundary.
Furthermore, the second term above is just the integral of the volume form which, by using Stokes theorem, can be written as an integral of a 1-form
\begin{align}
	\int d^2x\, \sqrt{g}=\int_\del a.\label{sqgina}
\end{align} 
For instance, in Poincare coordinates for two-dimensional hyperbolic space $H_2$, the volume form is given by $\text{vol}=\frac{dx\wedge dy}{y^2}$, and so we can take $a = \frac{1}{y}dx$. Thus,
\begin{align}
	S_{\del}=\phi_B\left(-2\pi+l_\phi-\int_\del a\right).\label{sding}
\end{align} 
The last two terms above can be interpreted as the worldline action for a particle in an electric field with gauge field $a$. {Furthermore, given that our boundary is of fixed length, the relativistic point particle problem boils down to the analysis of a non-relativistic particle, now in the presence of an electric field. 
 The action becomes
\begin{align}
	S_{\text{particle}}=\int d{\tilde u} \left(\half \frac{x'^2+y'^2}{y^2}-q \frac{x'}{y}\right),\label{oartact}
\end{align}
where primes denote derivatives with respect to ${\tilde u}$. The quantization of the particle in an electric field in $H_2$ is outlined in \cite{Yang:2018gdb}.} The parameter $q$ is to be identified with $\phi_B$ for comparison with gravity,
\begin{align}
	q=\phi_B\label{qphib}.
\end{align}
The Green's function for the particle to propagate from ${\bf x_1} = (x_1,y_1)$ to ${\bf x_2} = (x_2,y_2)$ in a time ${\tilde u}$ is given by \cite{Yang:2018gdb,COMTET1987185}
\begin{align}
	G({\tilde u}, {\bf x_1},{\bf x_2})&=e^{i\varphi(\bf{x}_1,\bf{x}_2)}\tilde{K}({\tilde u}, {\bf x_1},{\bf x_2}),\nonumber\\
	\tilde{K}({\tilde u}, {\bf x_1},{\bf x_2})&=	\int_0^{\infty} ds\, s \frac{\sinh(2\pi s)}{2\pi (\cosh(2\pi s)+\cosh(2\pi q))}\frac{e^{-{{\tilde u} s^2\over 2}}}{d^{1+2is}}\,\,{}_2F_1\left(\half -i q+is, \half+iq+is,1, 1-\frac{1}{d^2}\right),\nonumber\\
	d&=\sqrt{\frac{(x_1-x_2)^2+(y_1+y_2)^2}{4y_1 y_2}},	\nonumber\\
	e^{i\varphi(\bf{x}_1,\bf{x}_2)}&=e^{2q\arctan(\frac{x_2-x_1}{y_1+y_2})}.\label{partge}
\end{align}
 As a check, we can reproduce the conventional JT gravity path integral answer for asymptotic boundary conditions  from the above Green's function by taking the limit of coincident points, $\bf{x}_1=\bf{x}_2$. It is easy to see that in this limit, $d=1$, the hypergeometric function evaluates to unity, and the phase factor vanishes. To get a match with JT answer, we take the limit of large $q$, for which we get
\begin{align}
	Z_{JT}(\tilde{u})=e^{2\pi q}G({\tilde u},{\bf{x},\bf{x}})=\frac{1}{\pi}\int_0^{\infty} ds \,s \sinh(2\pi s)e^{-{\frac{{\tilde u}s^2}{2}}},\label{coinc}
\end{align}
which indeed gives the JT gravity partition function after the identification\footnote{ In obtaining the above result for the partition function from the propagator, the coincident point $\bf{x}$ has to be integrated over the entire AdS space. This gives a factor of the volume of AdS, which is canceled by a factor of the volume of $SL(2,R)$ gauge group (in the denominator) in the definition of JT gravity path integral.}
\begin{align}
    {\tilde u}=\frac{\ell_\phi}{\phi_B}=\frac{u}{\phi_b}.\label{utilurel}
\end{align}
Thus, we see that JT gravity path integral can be recast as the problem of a non-relativistic  particle in an electric field. 

\subsubsection*{HH wavefunction from path integral}

\label{partmagf}
Let us now repeat the above analysis for the case of interest to us, i.e., the HH wavefunction with one asymptotic boundary and one extrinsic curvature boundary. Again, the dilaton integral will give a delta function constraint leaving us with just the boundary terms in the action. Let $\del_{A}$ and $\del_K$ stand for the asymptotic boundary and the boundary with constant extrinsic curvature respectively. We shall use the notation in eq.\eqref{bdcond} for various quantities along these boundaries. Then the boundary terms can be written as follows:
\begin{align}
	S_\del=-\int_{\del_A}\phi \,\,(K-1)+\sum_{j \in\text{corners}}\phi_j\theta_j,\nonumber\\
	=-q\left(-\ell_\phi+\int_{\del_A}K-\sum_{j\in \text{corners}}\theta_j\right).\label{sdeljt}
\end{align}
But now, we can use the Gauss-Bonnet theorem which relates the extrinsic curvature and  Ricci scalar to Euler character {\footnote{Note that it is the exterior angle at a corner $\tilde{\theta}=\pi-\theta$ that appears in the Gauss-Bonnet theorem.}} as:
\begin{align}
	\int_{\del_A}K+\int_{\del_{K}}K+\sum_i (\pi -\theta_i)\, =2\pi \chi-\frac{1}{2}\int d^2x \sqrt{g}\,R.\label{gbtheorem}
\end{align}
Using this in eq.\eqref{sdeljt}, we get
\begin{align}
	S_\del=&-q\left(-\ell_\phi-2\pi +2\pi \chi -\half\int\sqrt{g} R-\int_{\del_{K}}K \right),\nonumber\\
	=&-q\left(-\ell_\phi+\int\sqrt{g} -k \ell_k \right).\label{bdterm}
\end{align}
The next step is to write the bulk term $\int\sqrt{g}$ as a boundary term
\begin{align}
	\int d^2x\, \sqrt{g}=\int_\del a.\label{sqgasgf}
\end{align}
%where $f$ is the field strength and $a$ is the associated gauge field, $f=da$.
 In this way, again the path integral for gravity is converted to an integral for a particle in two-dimensional AdS in the presence of an electric field. So, 
\begin{align}
	S_\del&=q\left(\ell_\phi+k \ell_k-\int_{\del}a\right),\nonumber\\
	&=q\left(\ell_\phi+k \ell_k-\int_{\del_A}a-\int_{\del_K}a\right).\label{sdelingf}
\end{align}

  In terms of the Poincare coordinates that cover the full region of Euclidean $AdS_2$, let $(x_1,y_1)$ and $(x_2,y_2)$ be the coordinates of the points where the two boundaries meet (i.e., the corners).  We evaluate the wavefunction $\Psi$ for a fixed value of one of the points, say $(x_1,y_1)$, the location of which is to be integrated over the entire $AdS_2$ disk. This integration gives a divergent SL(2,R) volume of the hyperbolic disk which will be suitably cancelled by a similar factor appearing in the denominator, that is present in the definition of path integral. Given a fixed location of the first point, the second point is determined by the conditions that the boundaries have specified lengths and one of boundary should have a particular extrinsic curvature. So, we just compute the wavefunction for a fixed $(x_1,y_1)$ and $(x_2,y_2)$ and express the result in terms of gauge invariant quantities $\phi_B,\frac{\beta}{2}, k,\ell$. %Let us denote the contribution to the wavefunction from $\del_A$ and $\del_K$ as $\psi_A$ and $\psi_K$ respectively. Then, 
%\begin{align}
%	\Psi=\psi_A\psi_K.\label{psiak}
%\end{align} 
  
  Consider first the contribution from the asymptotic boundary. Along this boundary the action is that of a particle moving in the presence of an electric field in hyperbolic space as can be seen from eq.\eqref{sdelingf} with the only constraint that the length of the path traversed by the particle between the points $(x_1,y_1)$ and $(x_2,y_2)$ is fixed. This is nothing but the propagator of the particle between these points with the fixed proper distance. It is given by eq.\eqref{partge}, where $\tilde{u}$ is related to $\phi_B,\frac{\beta}{2}$ by eq.\eqref{utilurel}. %Thus, we have
  %\begin{align}
  %	\psi_A=G(\tilde{u}, {\bf x_1},{\bf x_2})\label{psiaval}
  %\end{align}
  %So, the full wavefunction with one boundary being $\del_K$ and another being $\del_A$ is given by 
%\begin{align}
%	\Psi=e^{- q k \ell_k+q\int_{\bf{x_2}}^{\bf{x_1}} a}\,\,G(\tilde{u},\bf{x}_1,\bf{x}_2)\label{psi2b1}
%\end{align}
%where $G(u,\bf{x}_1,\bf{x}_2)$ is given by eq.\eqref{partge}. 
%Note the minus sign in the exponent in eq.\eqref{psi2b1}. This is because the particle is assumed to traverse the boundary along $\del_A$, the asymptotic boundary from $\bf{x}_1$ to $\bf{x}_2$. For it to return back to $\bf{x}_1$, it should traverse from $\bf{x}_2$  to $\bf{x}_1$  along $\del_K$. So, with the definition of $\int_{\del_K}a $ in eq.\eqref{valwil}, it should come with an extra minus sign in the wavefunction.
%Since the bulk term in JT gravity action in eq.\eqref{sjt} vanishes in the full quantum theory as well, due to the dilaton integral. We can ignore this term.  
%Note that the particle action eq.\eqref{oartact} for which the propagator is eq.\eqref{partge} has only the gauge field turned on at the $\del_A$ boundary. But to compute the wavefunction in our problem we need to also turn on the gauge field at $\del_K$ as can be seen from eq.\eqref{sdelingf}. We now evaluate the on-shell value of this extra contribution. ({\color{red} does this extra term give any additional one-loop contribution?}). 
Next, the contribution along the constant extrinsic curvature boundary ${\del_{K}}$ can be evaluated by using the particle in a field approach, as again can be seen from eq.\eqref{sdelingf}. But this time, the path of the particle between the points $(x_1,y_1)$ and $(x_2,y_2)$ is not just constrained to have a fixed proper length but also should have a fixed extrinsic curvature. Thus, the contribution for the particle propagator with fixed proper time just gets the contribution from a single path corresponding to the specified extrinsic curvature.  So, we can obtain its contribution by simply evaluating the value of the action for this particular path. The relevant trajectory in AdS is given by 
\begin{align}
	&x(\lambda)=x_0+\frac{R \sinh (\lambda  \sqrt{1-k^2})}{k+\cosh (\lambda  \sqrt{1-k^2})},\nonumber\\
	&y(\lambda)=\frac{R\, \sqrt{1-k^2}}{k+\cosh (\lambda  \sqrt{1-k^2})}.\label{xlylcurp11}
\end{align}
%we again have to evaluate the  we need to evaluate the trajectory of the curve of constant extrinsic curvature. 
Let us first compute the distance between two points $(x_1, y_1), (x_2, y_2)$ along the  curve. Let
\begin{align}
	x(\lambda_i)=x_i, y(\lambda_i)=y_i,\label{xycond}
\end{align}
where $\lambda_1,\lambda_2$ are the values of the parameter labeling the curves at the endpoints. 
Solving these four equations gives $x_0, R, \lambda_1, \lambda_2 $ in terms of $(x_1, y_1), (x_2, y_2)$. The distance between these two points can then be obtained as
\begin{align}
	\ell_k&=\int_{\lambda_1}^{\lambda_2} d\lambda \frac{x'(\lambda)^2+y'(\lambda)^2}{y(\lambda)^2}=\lambda_2 -\lambda_1\nonumber\\
	&=\frac{1}{\sqrt{1-k^2}}\cosh^{-1}\left[1+(1-k^2)\left(\frac{(x_1-x_2)^2+(y_1-y_2)^2}{2y_1 y_2}\right)\right]\nonumber\\
	&=\frac{2}{\sqrt{1-k^2}}\sinh^{-1}(\sqrt{(d^2-1)(1-k^2)}),\label{lkval}
\end{align}
where $d$ is given in terms of $x_i, y_i$ in eq.\eqref{partge}. Inverting the above relation to obtain $d$, we get
\begin{align}
	d=\sqrt{1+\frac{1}{1-k^2}\sinh^2\left(\half \ell_k\sqrt{1-k^2}\right)}.\label{dinlk}
\end{align}
We can now compute the value of the Wilson line along the boundary $\del_K$. We get
\begin{align}
	\int_{\del_K}a&=\int_{\lambda_1}^{\lambda_2} d\lambda \frac{x'(\lambda)}{y(\lambda)}\nonumber\\
	&=\left[k\lambda +2 \arctan\left(\frac{1-k}{\sqrt{1-k^2}}\tanh(\frac{\sqrt{(1-k^2)}\lambda}{2})\right)\right]\bigg\vert_{\lambda_1}^{\lambda_2}\nonumber\\
	&=k\ell_k+\left[2 \arctan\left(\frac{1}{\sqrt{1+k}}\tanh(\frac{\sqrt{(1-k^2)}\lambda}{2})\right)\right]\bigg\vert_{\lambda_1}^{\lambda_2}	
	.\label{valwil}
\end{align}
%The total contribution from $\del_K$ in eq.\eqref{sdelingf1} then works out to be
%\begin{align}
%	\psi_K=\exp(-\left[2q \arctan\left(\frac{1}{\sqrt{1+k}}\tanh(\frac{\lambda\sqrt{1-k^2}}{2})\right)\right]\bigg\vert_{\lambda_1}^{\lambda_2}	).\label{delkpsik}
%\end{align}
Putting everything together, we find that the HH wavefunction is given by
\begin{align}
	\Psi&=e^{i\hat{\varphi}}\tilde{K}(\tilde{u},{\bf{x}_1,\bf{x}_2}),\nonumber\\
	e^{i\hat{\varphi}}&=e^{i\varphi({\bf{x}_1,\bf{x}_2})}e^{-2q\left[ \arctan\left(\frac{1}{\sqrt{1+k}}\tanh(\frac{\sqrt{1-k^2}}{2}\lambda)\right)\right]\bigg\vert_{\lambda_1}^{\lambda_2}	}.
	\label{psi2b2}
\end{align}
Using the explicit values of $\lambda_1, \lambda_2$ in terms of $(x_1, y_1),(x_2, y_2)$ it can easily be checked that 
\begin{align}
	\arctan(\frac{x_2-x_1}{y_1+y_2})&-\arctan\left(\frac{1}{\sqrt{1+k}}\tanh(\frac{\sqrt{(1-k^2)}\lambda_2}{2})\right)+\arctan\left(\frac{1}{\sqrt{1+k}}\tanh(\frac{\sqrt{(1-k^2)}\lambda_1}{2})\right)\nonumber\\
%	&=-\arctan({k}\sqrt{\frac{(x_1-x_2)^2+(y_1-y_2)^2}{(1-k^2)\left((x_1-x_2)^2+(y_1-y_2)^2\right)+4 y_1 y_2}})\nonumber\\
	&=-\arctan(\frac{k\sqrt{d^2-1}}{\sqrt{\left(d^2-1\right) \left(1-k^2\right)+1}}),\label{arctanid}
\end{align}
and so
\begin{align}
	e^{i\hat{\varphi}}&=\exp{-2q\arctan(\frac{k\sqrt{d^2-1}}{\sqrt{\left(d^2-1\right) \left(1-k^2\right)+1}})}\nonumber\\
	&=\exp{-2q\arctan(\frac{k}{\sqrt{1-k^2}}\tanh\left(\half \ell_k\sqrt{1-k^2}\right))},\label{varhatval}
\end{align}
where the last equality follows upon using eq.\eqref{dinlk} to replace $d$ in terms of $\ell_k$. 

\subsubsection*{Large $q$ limit}
We now wish to take the large $q$ limit in the above expression. { In this limit, the constant dilaton boundary becomes a boundary in the asymptotic AdS region. So, the appropriate boundary conditions to be taken for various quantities is given by eq.\eqref{bdcond}.  In terms of  the coordinates, in the limit of large $q$ as the points $(x_i, y_i)$ approach the boundary, the scaling of the quantities as in eq.\eqref{bdcond} corresponds to the limit $y_i\rightarrow 0$ as $y_i\sim \frac{1}{q}$}.  Thus, taking 
\begin{align}
	y_i=\frac{z_i}{q}\label{yizi}
\end{align}
we see that 
\begin{align}
	d\xrightarrow{q\rightarrow \infty}q\sqrt{\frac{(x_1-x_2)^2}{4 z_1 z_2}}\equiv q d_{\infty}.\label{dlargq}
\end{align}
As an aside, note that in the large $d$ limit, $\ell_k$ also becomes large as can be seen from eq.\eqref{lkval} for any fixed $k\neq 1$. Then from eq.\eqref{dinlk}, we get
\begin{align}
	d&\simeq \frac{1}{2\sqrt{1-k^2}}e^{\half \ell_k\sqrt{1-k^2}}\nonumber\\
	&\simeq \frac{L}{2\epsilon},\label{dinklk}
\end{align}
where we used $\ell_k$ in terms of $L$ as in eq.\eqref{bdcond} in obtaining the last equality. Noting $q$ in terms of $\phi_b$ from eq.\eqref{bdcond}, we get
\begin{align}
	d_\infty=\frac{d}{q}=\frac{L}{2\phi_b}.
\end{align}
Furthermore, using  the large $q$ limit of the hypergeometric function in eq.\eqref{partge} as
\begin{align}
	\frac{1}{d^{1+2is}}{}_2F_1\left(\half -i q+is, \half+iq+is,1, 1-\frac{1}{d^2}\right)\xrightarrow{q\rightarrow\infty}\frac{e^{\pi q}}{\pi d}K_{2is}\left(\frac{2}{d_\infty}\right),\label{hylargq}
\end{align}
where $K_\alpha(x)$ is the modified Bessel function of order $\alpha$. 
Also, the large $q$ limit of eq.\eqref{varhatval} gives
\begin{align}
	e^{i\hat{\varphi}}\xrightarrow{q\rightarrow\infty}e^{-2q\arctan(\frac{k}{\sqrt{1-k^2}})}=e^{-2q\arcsin(k)}.\label{varphlq}
\end{align}
Thus,  combining eq.\eqref{varphlq},\eqref{psi2b2},\eqref{partge} and eq.\eqref{hylargq}, we get
\begin{align}
	\Psi=\frac{1}{\pi^2 d_{\infty}}e^{-\pi q-2q\arcsin(k)}\int ds \,s \,{\sinh(2\pi s)}\,{e^{-{\tilde{u} s^2\over 2}}}\,\,K_{2is}\left(\frac{2}{d_\infty}\right).\label{psiinbek}
\end{align}
The factor of $d_{\infty}^{-1}$ in the prefactor in eq.\eqref{psiinbek} arises from the factor $d^{-1}$ in eq.\eqref{hylargq} with the factor of $q$ cancelled by the measure for integrating over the bulk point for the closed boundary, $\frac{dxdy}{y^2}\rightarrow q\frac{dxdz}{z^2}$.
This is the final form of the generalized Hartle-Hawking wavefunction, where recall that $d_{\infty}$ is given by 
{
\begin{align}
    d_\infty= \frac{1}{2\phi_B\sqrt{1-k^2}}e^{\half \ell_k\sqrt{1-k^2}}.\label{dinfinlkk}
\end{align}
From the above result eq.\eqref{psiinbek}, we can also read off the overlap between the energy eigenstate $|E\rangle$ and the length eigenstates $|\ell,k\rangle$ as
\begin{align}
    \Psi_{E,\text{PI}}(\ell)\equiv\langle \ell,k|E\rangle_{\text{PI}}=\frac{1}{d_\infty}
    e^{-2q\arcsin(k)}
    K_{2i\sqrt{E}}\left(\frac{2}{d_\infty}\right),\label{elkinpx1}
\end{align}
}
where the subscript PI is meant to indicate that this is the path integral results. Let us now compare this with the quantum wavefunction we obtained in eq.\eqref{psiek} from canonical quantization. Since, the wavefunction in eq.\eqref{elkinpx1} was computed in Euclidean signature,  we must first take $k\rightarrow i k$ in this formula before comparing it with eq.\eqref{psiek}. We then see that the Bessel function in eq.\eqref{elkinpx1} precisely matches with the Lorentzian answer in equation \eqref{psiek}. Note however that eq. \eqref{elkinpx1} also contains some additional factors: first, there is the overall phase factor (after sending $k \rightarrow ik$) $e^{-2iq \sinh^{-1}k}$  which is independent of $\ell$. There is also a second factor of $d_\infty$ in equation \eqref{elkinpx1}. This factor is present because the length states as defined in the path-integral are not exactly the same as the canonical approach. Indeed, it is easy to check that the correct completeness relation with respect to the path-integral length states is given by:
\begin{align}
\int d(d_\infty^2)\;| \ell\rangle\langle \ell|_{\text{PI}} = \mathds{1}.
\end{align}
This is forced upon us by the requirement that the norm of the TFD state should reproduce the thermal partition function of JT gravity, together with the following integral relation for the bessel functions $K_x(z)$:
\begin{align}
	\int_0^\infty\frac{dx}{x}K_{2is}\left(\frac{4}{x}\right)K_{2i\tau}\left(\frac{4}{x}\right)\propto \frac{\delta(\tau-s)}{{s}\sinh(2\pi s)}.\label{besselid}
\end{align}
All together, this implies that
\beq 
|\ell,k\rangle_{\text{PI}} = \frac{1}{d_{\infty}}|\ell,k\rangle_{\text{canonical}}.
\eeq 
While the path integral involves a Euclidean calculation together with analytic continuation, our previous calculation using canonical quantization was a purely Lorentzian calculation; since the answers agree, this gives a sanity check on the validity of the analytic continuation.
%it follows immediately that
%\begin{align}
%(\Psi(\ell_\phi,\phi_b,\ell_k,-k),\Psi(\ell_\phi,\phi_b,\ell_k,k))\propto %Z({2\ell_\phi,\phi_B})
%\end{align}
%where the proportionality constant is independent of $\ell_\phi, \phi_B$.
%}
\subsubsection*{Comparision with classical calculation}
\label{gravcaclcom}
As another simple check, we will now show that quantum HH wavefunction agrees with the classical calculation in the saddle-point approximation. For this, we need to use the following integral representation for the Bessel function:
\begin{align}
	K_{\alpha}(x)=\half\int_{-\infty}^{\infty}d\xi e^{-x \cosh\xi}\cosh(\alpha\xi).\label{intrepbsk}
\end{align}
Using this, the quantum wavefunction becomes
\begin{align}
	\Psi=\frac{1}{2\pi^2q d_{\infty}}e^{-\pi q-2q\arcsin(k)}\int_{-\infty}^{\infty}d\xi e^{-\frac{2}{d_\infty} \cosh\xi}\int ds \,s \,{\sinh(2\pi s)}\,{e^{-{\tilde{u} s^2\over 2}}}\,\,\cos(2s\xi).\label{psiinbek2}
\end{align}
We now have to  evaluate the integrals. First, the $s$ integral can be done exactly and gives
\begin{align}
	\int_{0}^{\infty} ds  s \,{\sinh(2\pi s)}\,{e^{-{\tilde{u} s^2\over 2}}}\,\,\cos(2s\xi)=\frac{1}{\tilde{u}^{3/2}}\sqrt{\frac{\pi }{2}} \left((\pi +i \xi ) e^{\frac{2 (\pi +i \xi )^2}{\tilde{u}}}+(\pi -i \xi ) e^{\frac{2 (\pi -i \xi )^2}{\tilde{u}}}\right).\label{sintval}
\end{align}
So, we are left to do the $\xi$ integral. Noticing that the above result is even in $\xi$, as is the rest of the integrand in eq.\eqref{psiinbek2}, we get
\begin{align}
	\Psi=\frac{1}{2\pi^2q d_{\infty}}\frac{\sqrt{2\pi }}{\tilde{u}^{3/2}} e^{-\pi q- 2q\arcsin(k)}\int_{-\infty}^{\infty}d\xi \, (\pi +i \xi ) e^{-\frac{2}{d_\infty} \cosh\xi}e^{\frac{2 (\pi +i \xi )^2}{\tilde{u}}}.\label{psixiint}
\end{align}
We evaluate the $\xi$ integral in the saddle point approximation. The saddle point is given by 
\begin{align}
	\frac{4i(\pi+i\xi_*)}{\tilde{u}}=\frac{2}{d_\infty}\sinh\xi_*,\label{sadxipt}
\end{align}
where the saddle point value of $\xi$ is denoted as $\xi_*$. Thus, the saddle point value of the integral is given by 
\begin{align}
	\Psi=\frac{(\pi +i \xi_* )}{2\pi^2q d_{\infty}}\frac{\sqrt{2\pi }}{\tilde{u}^{3/2}}  \exp{-\pi q- 2q\arcsin(k)+\frac{2}{\tilde{u}}(\pi+i\xi_*)^2-\frac{4i(\pi+i\xi_*)}{\tilde{u} \tanh(\xi_*)}}.\label{psixisad}
\end{align}
With the following change of variables:
\begin{align}
	\pi+i\xi_*=-\half r_s \beta,\label{xirsvar}
\end{align}
the saddle point equation eq.\eqref{sadxipt} becomes
\begin{align}
	\frac{2d_\infty\beta}{\tilde{u}}=\frac{2}{ r_s}\sin(\frac{\beta r_s}{2}),\label{rscsad}
\end{align}
which is similar to eq.\eqref{lrs}.  This gives the relation 
\begin{align}
	{L}=\frac{2d_\infty\beta}{\tilde{u}}.\label{ldinf}
\end{align}
Furthermore, comparing the large $q$ limit of eq.\eqref{lkval} with eq.\eqref{bdcond}, we get
\begin{align}
	\log(\frac{L\sqrt{1-k^2}}{\epsilon})=\log(2 \sqrt{1-k^2}qd_\infty)\implies {L}=2 \phi_b d_\infty.\label{lphibrel}
\end{align}
Using this in eq.\eqref{ldinf}, we get
\begin{align}
	\tilde{u}=\frac{\beta}{\phi_b}\label{uinbphib}
\end{align}
Comparing with eq.\eqref{utilurel}, we find that 
\begin{align}
    u=\beta\label{ubeta}
\end{align}
Using the result eq.\eqref{uinbphib} and eq.\eqref{xirsvar} in eq.\eqref{psixisad} and noting that $q=\phi_b r_c={\phi_b\over \epsilon}$, we get
\begin{align}
	\Psi=-\frac{r_su}{4\pi^2q d_{\infty}}\frac{\sqrt{2\pi }}{\tilde{u}^{3/2}} \exp{\frac{1}{2}\phi_br_s^2u+{2\phi_b r_s} \cot(\frac{u r_s}{2})-2\phi_b r_c\arccos(-k)}.\label{semicl}
\end{align}
%A plot of the above wavefunction as a function of the renormalised length, $\ell_{\text{ren}}$ is as below, 

%\begin{figure}
 %   \centering
 %   \includegraphics[width=0.5\linewidth]{Ver1/HH wfn.png}
 %   \caption{Above is a plot of the wavefunction as a function of $\ell_{\text{ren}}$ for various values of $k$ shown in the plot legend. }
 %   \label{fig:enter-label}
%\end{figure}
Comparing the exponent of the above result with eq.\eqref{psicl2}, we see that it agrees with the result of the classical calculation. Further, including the one-loop determinant around the saddle point gives the following one-loop corrected wavefunction:
\begin{align}
    \Psi &=-\frac{r_su}{4\pi^2q d_{\infty}}\frac{\sqrt{2\pi }}{\tilde{u}^{3/2}} \abs{{{\frac{2 }{{d_\infty }}\cos \left(\frac{u  r_s}{2}\right)+\frac{4}{(u r_s)^2}-\frac{4\phi_b}{{u}}}}}^{-\half}\nonumber\\
    &\times \exp{\frac{1}{2}\phi_br_s^2 u+{2\phi_b r_s} \cot(\frac{u r_s}{2})-2\phi_b r_c\arccos(-k)}.\label{semiclolpd}
\end{align}

%\begin{color}{red}(OP:Do we have anything more to say about this?)\end{color}

\subsection{York Hamiltonian}
\label{buphham}
In this section, we will show that equation \eqref{elkinpx1} for the overlap of an energy eigenstate with the bulk length state at different values of extrinsic curvature can be interpreted as a Schrodinger equation in York time. This will allow us to extract the York Hamiltonian corresponding to this bulk notion of time evolution.  

Our starting point will be equation \eqref{elkinpx1}, which we reproduce below:
\begin{align}
    \langle \ell,k|E\rangle=e^{2iq\sinh^{-1}k}K_{2i\sqrt{E}}\left(\frac{2}{\tilde{d}_\infty}\right), \tilde{d}_\infty= \frac{1}{\phi_B\sqrt{1+k^2}}e^{\half \ell_k\sqrt{1+k^2}}.
    \label{elkinpx2}
\end{align}
Under an infinitesimal shift in York time, we get
\begin{align}
    {\langle \ell,k+\delta k|E\rangle} = (1+\delta k \,\del_k)\langle {\ell,k|E\rangle} +O(\delta k^2).\label{elkdk}
\end{align}
%Now, a state at time $k+\delta k$ can be written as
%\begin{align}
 %   |\ell,k+\delta k\rangle =(1-i \delta k H_{\text{phys}})|\ell,k\rangle
%\end{align}
%Taking an innerproduct of the above equation with $\langle E|$, we get that
%\begin{align}
%   \langle E |\ell,k+\delta k\rangle =(1-i\, \delta k \,H_{\text{phys}})\langle E|\ell,k\rangle\label{hphdlk}
%\end{align}
Our goal is to compute the York time derivative above. Re-writing eq. \eqref{elkinpx2} as
\begin{align}
    {\langle \ell,k|E\rangle}=g(k)f(\tilde{d}_\infty),\quad f(\tilde{d}_\infty)=K_{2i\sqrt{E}}\left(\frac{2}{\tilde{d}_\infty}\right), \, g(k)=e^{2iq\sinh^{-1}k},
\end{align}
we notice that, 
\begin{align}
    \del_kf(\tilde{d}_\infty)&=(\del_k\tilde{d}_\infty) f'(\tilde{d}_\infty)=\frac{\del_k\tilde{d}_\infty }{\del_\ell\tilde{d}_\infty}\del_\ell f(\tilde{d}_\infty),\nonumber\\
    \del_kg(k)&=\frac{2i q}{\sqrt{1+k^2}}g(k).
    \label{dkfdlf}
\end{align}
This gives
\begin{align}
    \partial_k{\langle \ell,k|E\rangle} &= \left(\frac{\del_k\tilde{d}_\infty }{\del_\ell\tilde{d}_\infty}\del_\ell+\frac{2i q}{\sqrt{1+k^2}}\right) {\langle \ell,k|E\rangle}\nonumber\\
    &\simeq \left(\frac{k\ell }{1+k^2}\del_\ell+\frac{2i q}{\sqrt{1+k^2}}\right) {\langle \ell,k|E\rangle},
    \label{elkdkelk}
\end{align}
where we have used the expression for $\tilde{d}_\infty$ in eq.\eqref{dinfinlkk}, to find that at large $\ell$: %\begin{color}{red}(OP:renormalized $\ell$?)\end{color}
\begin{align}
{\del_k\tilde{d}_\infty}\simeq\frac{\ell\, k}{1+k^2}{\del_\ell \tilde{d}_\infty}.
\end{align}
Equation \eqref{elkdkelk} can be re-written as
\begin{equation}\label{wdw}
   - i\partial_k {\langle \ell, k|E\rangle} = \langle \ell,k | H_{\text{York}}|E\rangle .
\end{equation}
where 
\begin{equation}
    H_\text{York}(k)=\left(\frac{k }{1+k^2}\ell\,P+\frac{2  q}{\sqrt{1+k^2}}\right).\label{Hphyful}
\end{equation}
Together with the identifications $P=-i\del_\ell, \pi_k=-i\del_k$,\footnote{Throughout this paper, we have taken the viewpoint that $k$ is a parameter and the states $|\ell, k\rangle$ are physical, gauge-fixed states where the extrinsic curvature takes the value $k$. In discussing the WdW equation, we are implicitly moving to an enlarged kinematic phase space where $(k, \pi_k)$ are also dynamical variables. } equation \eqref{wdw} becomes 
\begin{align}
    \left(\pi_k-\frac{k\,\ell}{1+k^2}P-\frac{2q}{\sqrt{1+k^2}}\right)|\ell,k\rangle=0,\label{pikvsham}
\end{align}
which is nothing but the WDW equation for physical states (see eq.\eqref{hamlozer} for more details). Thus, we see that the WDW equation can be interpreted as a Schrodinger equation for unitary evolution in York time.  Note that the states defined in eq.\eqref{elkinpx3} and in eq.\eqref{elkinpx2} have an extra phase term compared to the states obtained in the canonical quantization approach, see eq.\eqref{psiek}. This extra phase term is important in getting the appropriate physical Hamiltonian eq.\eqref{Hphyful1} that is related to the WDW equation. More precisely, the last term in eq.\eqref{Hphyful} is also present in the WDW equation, as is carefully shown in appendix \ref{wdw}. 
%\begin{color}{red}(OP: Check sign and operator ordering ambiguity. Issue of measure. Is the Hamiltonian just $P$ with the right measure?)\end{color}

There is an important point worth highlighting about our result above. Naively, the Hamiltonian in equation \eqref{Hphyful} does not look like a Hermitian operator with respect to the inner product:
    \begin{align}
        \langle \ell_1,k|\ell_2,k\rangle
&=\frac{1}{\sqrt{1+k^2}}\delta(\ell_1-\ell_2).\label{l1l2inpc}
    \end{align}
    % The appropriate completeness relation for the length states that follows from eq.\eqref{l1l2inpc} is 
%\begin{align}
%        \int d\ell |\ell,k\rangle \langle \ell,k|=\frac{1}{\sqrt{1+k^2}}\mathds{1}\label{ellkinp1}
%\end{align}
    However, the cause of this apparent non-Hermiticity is the fact that the length states are not properly normalized, and in particular, they have a $k$-dependent normalization. The resolution to this is to define new basis states which have a time-independent normalization:
    \begin{align}
        \widetilde{|\ell,k\rangle}=(1+k^2)^{\frac{1}{4}}|\ell,k\rangle.
    \end{align}
    These new states above satisfy the standard completeness relation:
    \begin{align}
         \int d\ell \widetilde{|\ell,k\rangle}\widetilde
         {\langle \ell,k|}=\mathds{1}.\label{ellkinp1}
    \end{align}
    The overlap of these new states $\widetilde{|l,k\rangle}$ with energy eigenstates then becomes
    \begin{align}
        \widetilde{\langle \ell,k|}E\rangle=(1+k^2)^{\frac{1}{4}}e^{2iq\sinh^{-1}k}K_{2i\sqrt{E}}\left(\frac{2}{\tilde{d}_\infty}\right).
    \label{elkinpx3}
    \end{align}
    Because of the extra $k$-dependence in the prefactor, the new wavefunction satisfies a modified Schrodinger equation. Denoting the corresponding Hamiltonian as $\widetilde{H}_\text{York}$, we get
\begin{equation}
    \widetilde{H}_\text{York}(k)=\left(\half\frac{k }{1+k^2}(\ell\,P+ P\,\ell)+\frac{2  q}{\sqrt{1+k^2}}\right),\label{Hphyful1}
\end{equation}
which indeed looks manifestly Hermitian. From this point of view, the operator ordering ambiguity in the Hamiltonian is resolved simply by working with the correctly normalized length basis states. Such operator ordering ambiguities are routinely encountered in the study of the WdW equation, for e.g. in eq.\eqref{hamconkh} \cite{Isham:1992ms,dewittct,HAWKING1986185} (see also \cite{Nanda:2023wne} for discussion in the two dimensional de Sitter context). {In our case, the canonical quantization determines the wavefunction as in eq.\eqref{psiek}, upto a normalization factor which can be York time-dependent. This ambiguity is fixed by the path integral through the requirement that the states have a good normalization, as per the norm arising from path integral. The same presciption for the norm also leads to particular operator ordering for the York Hamiltonian that is Hermitian. }

%The corresponding WDW equation satisfied by these states would be given by
%\begin{align}
%    \left(\pi_k-\half\frac{k}{1+k^2}(\ell\,\pi_\ell+\pi_\ell\, \ell)-\frac{2q}{\sqrt{1+k^2}}\right)\widetilde{|\ell,k\rangle}=0,\label{pikvsham}
%\end{align}
%which is slightly different compared to eq.\eqref{pikvsham}, related to it by an operator ordering term. 

Importantly, we expect bulk time evolution (at fixed boundary time) to leave the physical state in the dual boundary theory unchanged. In other words, time evolution in $k$ is merely a change of coordinates (specifically, the time coordinate) from the bulk point of view and should not affect the physical state. Instead, our picture is that the York time dependence of the HH wavefunction comes from a $k$-dependent unitary transformation of the fixed-length basis states $\widetilde{|\ell,k\rangle}$:
\begin{equation}
    \widetilde{|\ell,k\rangle} = \mathcal{T}\exp\left[i\int_0^k dk'\widetilde{H}_{\text{York}}(k')\right]\widetilde{|\ell,0\rangle}\label{ltkuni}.
\end{equation}
Thus, the HH wavefunction in this basis satisfies a Schrodinger equation, not because the physical state $|E\rangle$ or $|\text{TFD}_{\beta}\rangle$ is evolving, but because the basis we use to compute the wavefunction is changing unitarily. We expect that our picture is very closely related to the notion of ``relational observables'' \cite{Page:1983uc, Hoehn:2019fsy, Fliss:2024don}, and it would be interesting to develop this connection further.\footnote{We thank Ronak Soni for discussions on this point.}
{The Unitary operator in eq.\eqref{ltkuni} can be explicitly written as
\begin{align}
    \hat{U}=e^{2i\sinh^{-1}k}e^{i\lambda(k)(\ell P+P\ell)},\qquad \lambda(k)=\frac{1}{4}\log(1+k^2),\label{Udef}
\end{align}
where now the length and momentum operators are defined at $k=0$. It can be seen from the above expression that the unitary operator acts as a dilatation operator in its action on the length operator and its conjugate momentum. Explicitly, we find

\begin{align}
    &U^\dagger{\ell}U=e^{-2\lambda}{\ell}=\frac{1}{\sqrt{1+k^2}}\,{\ell}\nonumber\\
    &U^\dagger{P}U=e^{-2\lambda}{P}={\sqrt{1+k^2}}\,{P}\label{Uactonlp}
\end{align}

The above transformations for the operators ${\ell},{P}$ show that the expectation value of ${\ell}$ decreases, while that of ${P}$ increases as a function of York time. Moreover, the variance of $\ell$ decreases as a function of York time,
\begin{align}
    \langle U^\dagger \ell^2 U\rangle - (\langle U^\dagger \ell U\rangle)^2=(1+k^2)^{-1}(\langle \ell^2\rangle -\langle \ell \rangle^2),\label{varlwik}
\end{align}
while the wavefunction spreads out in momentum:
\begin{align}
    \langle U^\dagger P^2 U\rangle - (\langle U^\dagger P U\rangle)^2=(1+k^2)(\langle P^2\rangle -\langle P \rangle^2).\label{varPwik}
\end{align}
From the above results, we expect that the expectation value and the spread of the ADM Hamiltonian (see eq.\eqref{hinpklk}) should increase with York time, indicating that the wavepacket delocalizes in energy as we move forward in York time. This suggests that the wavefunction becomes more non-classical after a large amount of York time evolution. The above discussion also shows that the overlap $\langle \ell, k| E\rangle$ can equivalently be written as $\langle \ell, 0 | U^{\dagger} | E\rangle$, where the unitary operator $U$ is given in equation \eqref{Udef}.

}

%\section{WDW as $T\bar{T}$ flow}
\section{Conclusion}
\label{concl}
%\tableofcontents
A natural issue that arises in the context of de Sitter holography is the problem of time -- in the absence of a timelike boundary, one must resort to finding an ``internal clock'', i.e., a degree of freedom in the gravitational theory internal to the system that can serve the role of a physical clock. Assuming the basic premise that spacetime and gravity emerge from the dynamics of an underlying quantum system, this notion of an internal clock must also emerge from some underlying evolution in the quantum system. In this paper, we have tried to emphasize that the better controlled setting of JT gravity in $AdS_2$ is a good playground to understand this issue of emergence of time. We sliced the bulk spacetime with constant extrinsic curvature slices, and took the value of the extrinsic curvature as a notion of internal time. This notion of time, also known as York time, is well-motivated and is indeed well-studied in higher dimensional theories, at least in the classical setting. We studied the York time evolution of the Hartle-Hawking wavefunction in JT gravity within the framework of canonical and path-integral quantization and showed that the York time dependence of the HH wavefunction emerges from a unitary transformation of the length basis states, not from a physical time evolution of the boundary quantum mechanical state. We identified the corresponding York Hamiltonian, i.e., the generator of the unitary rotation in the length basis, as being proportional to the squeezing operator $(\ell P + P \ell)$. 

Let us end with some future directions: 
\begin{enumerate}
\item JT gravity has a quantum mechanical dual in terms of an ensemble of random Hamiltonians \cite{Saad:2019lba}. It would be interesting if the York Hamiltonian we found above has a microscopic interpretation in the dual theory. For instance, the ADM Hamiltonian of JT gravity is the bulk avatar of the microscopic Hamiltonian of the dual quantum mechanical system. In a similar vein, it would be nice if the York Hamiltonian also had a natural interpretation in the microscopic theory. A starting point for this would be to understand the microscopic interpretation of the fixed-length basis states of JT gravity \cite{Iliesiu:2024cnh}. Recent work on Krylov (spread) complexity \cite{Parker:2018yvk,Balasubramanian:2022tpr} has also thrown up the possibility of the Krylov basis in the dual quantum theory as being dual to the length basis states (see, for instance, \cite{Rabinovici:2023yex, Xu:2024gfm, Nandy:2024zcd, Balasubramanian:2024lqk,Heller:2024ldz,Ambrosini:2024sre, Basu:2024tgg}). More precisely, the Krylov basis states are labelled by a discrete integer, and are analogous to the ``chord number'' states \cite{Lin:2022rbf} of the double-scaled SYK model \cite{Sachdev:1992fk,Kitaev-talks:2015,Maldacena:2016hyu,Berkooz:2018jqr,Cotler:2016fpe}. However, in the standard double-scaling limit for the matrix model (not the double-scaling limit of DSSYK), it is known that the chord-number states of DSSYK map to the length basis states of JT gravity\cite{Lin:2022rbf,Rabinovici:2023yex}. Given that the York time evolution of JT gravity can be written as the squeezing operator in phase space, it is natural to speculate that the microscopic version of the York evolution should be written similarly in terms of the Krylov position and momentum operators -- in the language of \cite{Basu:2024tgg}, this would be a Clifford unitary on the discrete Kylov phase space. It would be interesting to explore this possibility further. It would also be interesting to generalize our analysis to other two dimensional models of quantum gravity (see for instance \cite{Kuchar:1996zm}).

\item One of the reasons York time evolution is interesting is that in higher dimensions, it evolves the time-reflection symmetric Cauchy slice into the black hole interior, and in particular, closer to the black hole singularity. A microscopic understanding of the emergence of York time would thus also shed light on the black hole interior from a microscopic point of view. In this context, it would be particularly interesting to try and generalize our work to higher dimensions (see, for instance, \cite{York2, Carlip:1995zj, Coleman:2020jte, Hartnoll:2022snh, Hartnoll:2025hly, Blacker:2023oan, Blacker:2024rje} for results in this direction). We should note that while JT gravity is useful as an effective description of the near horizon physics of near-extremal black holes in higher dimensions, York type boundary conditions in higher dimensions do not translate simply to York boundary conditions in JT gravity. In this work, we have taken the point of view that JT gravity is a well-defined theory on its own and studied York time within this theory. The generalization to higher dimensions would therefore be non-trivially different.\footnote{We thank Edgar Shaghoulian for pointing this out.}

\item The York Hamiltonian that we obtained could equivalently have been derived from the WDW equation. A detailed discussion on this can be found in Appendix \ref{wdwmat}, but here we make a couple of remarks: firstly, discussions of the WdW equation in gravity are often plagued with operator ordering ambiguities. In our analysis in appendix \ref{wdwmat}, we encounter similar issues. One of the pleasant features of our analysis using canonical quantization and the JT gravity path integral is that such ambiguities are cleanly avoided. Secondly, at least in the context of Dirichlet boundary conditions, the WDW equation can be recast in terms of the $T\bar{T}$ flow \cite{Zamolodchikov:2004ce,Smirnov:2016lqw,Cavaglia:2016oda, McGough:2016lol, Hartman:2018tkw}. This is the perspective taken in Cauchy slice holography \cite{Caputa:2020fbc, Araujo-Regado:2022gvw, Soni:2024aop}. We have instead given a different interpretation of the WdW equation in terms of a unitary rotation of the length basis states. Nevertheless, our results should be related to Cauchy slice holography, loosely speaking, by a Legendre transformation. It would be interesting to build this connection further and to generalize it to higher dimensions.  
\end{enumerate}

\section*{Acknowledgements}
We thank Abhijit Gadde, Alok Laddha, Shiraz Minwalla, Ronak Soni and Sandip Trivedi for helpful discussions. We also thank Alok Laddha, Edgar Shaghoulian and Ronak Soni for helpful comments on the draft.  We acknowledge support from the Department of Atomic Energy, Government of India, under project identification number RTI 4002 and from the Infosys endowment for the study of the quantum structure of spacetime. The work of SKS is supported by MEXT KAKENHI Grant-in-Aid for Transformative Research Areas A “Extreme Universe” No. 21H05184. The authors also acknowledge
hospitality at the conference “Quantum Information, Quantum Field Theory and Gravity”
(ICTS/qftg2024/08) at ICTS-TIFR, India.

\vspace{1cm}	
	%----------------------------------------------------------------------------------------
	%	ADDRESSEE SECTION
	%----------------------------------------------------------------------------------------
	
	%\begin{letter}{%To \\ The Chairperson,\\
			%		%Selection Committe.	
			%} 
		
		%----------------------------------------------------------------------------------------
		%	YOUR NAME & ADDRESS SECTION
		%----------------------------------------------------------------------------------------
		\vspace{-0.6cm}

\appendix
\section{Variational principle for non-smooth boundaries}
\label{varpri}

The JT action, in units of $8\pi G=1$ is given by,
%\begin{equation}
	%	S= \frac{1}{16\pi G}\int \sqrt{g} \phi (R- \Lambda)
	%\end{equation}
%with equations of motion being,
%\begin{align}
	%	&  R- \Lambda =0  \label{EOMJT}\\
	%	& \nabla_{\mu} \nabla_{\nu} \phi - g_{\mu \nu} \nabla^2 \phi - g_{\mu \nu} \phi =0.
	%\end{align}
%In units $8 \pi G=1$, we have,
\begin{equation}
	S_{\text{bulk}}= \frac{1}{2} \int \sqrt{g} \phi (R- \Lambda)\label{jtin8g1}
\end{equation}
Taking the variation we obtain,
\begin{equation}
	\delta S_{\text{bulk}} = \frac{1}{2} \int \sqrt{g} \delta\phi (R- \Lambda) + \frac{1}{2}\int \phi \delta(\sqrt{g}) (R-\Lambda)+ \frac{1}{2} \int \sqrt{g}\phi \,\delta g^{\alpha \beta}\, R_{\alpha \beta}  + \frac{1}{2} \int \phi \sqrt{g} g^{\alpha \beta} \delta R_{\alpha \beta} \label{actionvar1}
\end{equation}
The first three terms do not give boundary terms and so we can ignore them.  Only the fourth term will give us the required boundary terms since it involves two derivatives acting on the variation of the metric.
%So,
%\begin{equation}
	%	g^{\alpha \beta} \delta R_{\alpha \beta} = g^{\alpha \beta} (\nabla_{\mu} \delta \Gamma ^{\mu}_{\alpha \beta} - \nabla_{\alpha} \delta \Gamma^{\mu}_{\mu \beta}) = \nabla_{\mu} (g^{\alpha \beta} \delta \Gamma ^{\mu}_{\alpha \beta} - g^{\mu \beta} \delta \Gamma ^{\alpha}_{\alpha \beta})
	%\end{equation}
The variation of the Ricci tensor is given by 
\begin{align}
	\delta R_{\alpha \beta}&=\nabla_{\mu} \delta \Gamma ^{\mu}_{\alpha \beta} - \nabla_{\alpha} \delta \Gamma^{\mu}_{\mu \beta},\nonumber\\
	\delta\Gamma^{\mu}_{\alpha\beta}&=\half g^{\mu \nu}(\nabla_\alpha \delta g_{\nu\beta}+\nabla_\beta \delta g_{\alpha\nu}-\nabla_\nu \delta g_{\alpha\beta})\label{delrindelg}
\end{align}
Using this, the last term in eq.\eqref{actionvar1} can be written as
\begin{align}
	\frac{1}{2} \int \phi \sqrt{g} g^{\alpha \beta} \delta R_{\alpha \beta}=\half \int \nabla_\mu \left(\phi \sqrt{g} \,(g^{\alpha\beta}\delta\Gamma^\mu_{\alpha\beta}-g^{\mu \beta}\delta\Gamma^{\nu}_{\nu\beta}) \right)-\half \int \sqrt{g}\,\nabla_\mu\phi \,(g^{\alpha\beta}\delta\Gamma^\mu_{\alpha\beta}-g^{\mu \beta}\delta\Gamma^{\nu}_{\nu\beta})\label{oninbpar}
\end{align}
We need to do one more integration by parts since the second term above involves derivatives acting on variations. Using the variation of Christoffel symbol in terms of variations of metric components, explicitly given in eq.\eqref{delrindelg}, it can be easily checked that,
\begin{align}
	g^{\alpha\beta}\delta\Gamma^\mu_{\alpha\beta}-g^{\mu \beta}\delta\Gamma^{\nu}_{\nu\beta}= \nabla_\nu (g^{\mu\alpha} g^{\nu\beta} \delta g_{\alpha\beta}- g^{\mu \nu} g^{\alpha \beta} \delta g_{\alpha \beta})\label{chrisid}
\end{align}
Using this in eq.\eqref{oninbpar}, and dropping the bulk term that we get after doing an integration by parts, we see that the total set of boundary terms are given by 
\begin{align}
	\delta S_{\text{bulk},\del}&=	\half \int \left[\nabla_\mu \left(\phi \sqrt{g} \,(g^{\alpha\beta}\delta\Gamma^\mu_{\alpha\beta}-g^{\mu \beta}\delta\Gamma^{\nu}_{\nu\beta}) \right)-\nabla_\nu\left(\sqrt{g}\nabla_\mu\phi\, (g^{\mu\alpha} g^{\nu\beta} \delta g_{\alpha\beta}- g^{\mu \nu} g^{\alpha \beta} \delta g_{\alpha \beta}) \right) \right]\nonumber\\
	&=	\half \int_\del  \sqrt{h} \,n_\mu\left[\phi  \,(g^{\alpha\beta}\delta\Gamma^\mu_{\alpha\beta}-g^{\mu \beta}\delta\Gamma^{\nu}_{\nu\beta}) -\nabla_\nu\phi\, (g^{\nu\alpha} g^{\mu\beta} \delta g_{\alpha\beta}- g^{\mu \nu} g^{\alpha \beta} \delta g_{\alpha \beta}) \right]
	\label{bdtdels}
\end{align}

We are interested in varying the boundary metric subject to the condition that the physical boundary is held fixed. The boundary is specified by some gauge invariant function of the bulk coordinates. In our case, it is either the condition that the dilaton or the extrinsic curvature takes a specific value.

Let us introduce some notation. Let $x^\alpha$ denote the bulk coordinates and $y^a$ denote the boundary coordinates.  Let $g_{\mu\nu}$ be the bulk metric and $h_{ab}$ be the boundary metric. The tangent vectors to the boundary hypersurface are given by 
\begin{align}
	e^\alpha_a\equiv \frac{\del x^\alpha}{\del y^a}\label{ealptv}
\end{align}
Let $e^a_\alpha$ be the inverse satisfying the relation
\begin{align}
	e^\alpha_a e^b_\alpha=\delta^b_a\label{eainv}
\end{align}
The induced metric on the boundary, $h_{ab}$ is related to $g_{\alpha\beta}$ by 
\begin{align}
	h_{ab}=g_{\alpha\beta}e^{\alpha}_ae^{\beta}_b\label{indbdmet}
\end{align}
Let $n^\alpha$ be the normal vector to the boundary. The completeness relation is then given by 
\begin{align}
	g^{\alpha\beta}=\epsilon\, n^\alpha n^\beta+h^{ab}e^\alpha_a e^\beta_b\label{compl}
\end{align}
where $\epsilon$ is the normalization factor for the normal $n^\alpha n_\alpha=\epsilon$. In the euclidean problem under consideration since the normal vector is spacelike we have $\epsilon=1$. 
Let us suppose the boundary is specified by the condition $\Phi(x)=\text{const}$. Here $\Phi$ is an arbitrary functional of the dynamical fields and should not be confused as just being a functional of dilaton field.  The normal vector is then given by 
\begin{align}
	n_\alpha = \epsilon \mu  \del_\alpha \Phi,\quad \mu^{-1}=\sqrt{g^{\alpha\beta}\del_\alpha\Phi\del_\beta\Phi}\label{normvec}
\end{align}
The variational problem that we are interested in is the one where the physical boundary is defined by the condition $\Phi = \text{const.}$ which is held fixed, but the induced boundary metric is being varied. 
Since, the physical boundary is not changing, the tangent vectors for the surface $x^\alpha=x^\alpha(y^a)$ are unchanged.  A variation of the metric however induces a variation in the normal due to the normalization of the normal vector which involves the metric. So, we have
\begin{align}
	&\delta e^\alpha_a=0, \nonumber\\
	&\delta n_\alpha=\frac{\delta \mu}{\mu}n_\alpha, \quad \frac{\delta\mu}{\mu}=-\frac{1}{2}\epsilon n_\alpha n_\beta \delta g^{\alpha\beta}\label{delnalpa}
\end{align}
Even though $e^\alpha_a$ is unchanged, the inverse components will have a variation. Noting the relation eq.\eqref{eainv}, we see that $e^\alpha_b\delta e^{a}_\alpha=0$ which means that the variation of $e^a_\alpha$ can only have a component along the normal direction. Let us denote this by $\deltabar A^a$. So, we have
\begin{align}
	\delta e^a_\alpha=\deltabar A^a n_\alpha\label{delealpah}
\end{align}	
Furthermore noting that $e^a_\alpha n^\alpha=0$ and $\epsilon=n^\alpha n_\alpha$, doing a variation  and using eq.\eqref{delealpah} and eq.\eqref{delnalpa} gives
\begin{align}
	e^a_\alpha \delta n^\alpha+\epsilon \deltabar A^a=0, \quad  n_\alpha \delta n^\alpha+\epsilon \frac{\delta \mu}{\mu}=0\label{delnuco}
\end{align}
which gives the component of $\delta n^\alpha$ along the tangent and normal direction 
\begin{align}
	\delta n^\alpha=-\frac{\delta\mu}{\mu} n^\alpha-\epsilon \deltabar A^a e^\alpha_a\label{delnuval}
\end{align}
The quantity $\deltabar A^a$ can be obtained either from eq.\eqref{delealpah} or eq.\eqref{delnuco} as
\begin{align}
	\deltabar A^a= \epsilon  n^\alpha \delta e^a_\alpha=- \epsilon  e^a_\alpha \delta n^\alpha\label{delAval}
\end{align}
It will also be useful to express the variation of the normal and $\deltabar A$ in terms of the variation of the bulk metric. For this, varying the condition $n^\alpha= g^{\alpha\beta}n_\beta$ gives 
\begin{align}
	\delta n^\alpha =n_\beta \delta g^{\alpha \beta}+g^{\alpha\beta}\delta n_\beta\Rightarrow \deltabar A^a=-\epsilon e^a_\alpha n_\beta \delta g^{\alpha\beta}\label{delindelg}
\end{align}
where we used eq.\eqref{delAval} in obtaining the second equality. 

We now have all the expressions required to simplify the boundary term in eq.\eqref{bdtdels}. The manipulations that are to follow are based on \cite{Lehner:2016vdi}. The extrinsic curvature is given in terms of the normal vectors as
\begin{align}
	K_{ab}=e^\alpha_a e^\beta_b \nabla_\alpha n_\beta\label{excurd}
\end{align}
Varying this and using eq.\eqref{delnalpa}, we get
\begin{align}
	\delta K_{ab}&=e^\alpha_a e^\beta_b( \nabla_\alpha \delta n_\beta-n_\mu \delta \Gamma^\mu_{\alpha\beta})\nonumber\\
	&=e^\alpha_a e^\beta_b(n_\beta \nabla_\alpha \delta\ln \mu+\ln \mu \nabla_\alpha \delta n_\beta -n_\gamma \delta \Gamma^\gamma_{\alpha\beta})\nonumber\\
	&=K_{ab}\delta\ln \mu -e^\alpha_a e^\beta_bn_\mu \delta \Gamma^\mu_{\alpha\beta}\label{deltaKab}
\end{align}
from which it follows that 
\begin{align}
	\delta K &=K_{ab} \delta h^{ab}+h^{ab}\delta K_{ab}\nonumber\\
	&=K_{ab} \delta h^{ab}+K \delta \ln \mu- h^{\alpha\beta}n_\mu \delta \Gamma^\mu_{\alpha\beta}\label{delk1st}
\end{align}
We now perform the variation of $\delta K$ in a slightly different way to obtain a different expression. Starting from the definition of the trace of extrinsic curvature given by 
\begin{align}
	K=h^\alpha_\beta\nabla_\alpha n^\beta\label{exk2def}
\end{align}
where $h^\alpha_\beta =e^\alpha_a e^a_\beta$ and doing a variation gives
\begin{align}
	\delta K= (\delta h^\alpha_\beta)\nabla_\alpha n^\beta+h^\alpha_\beta \nabla_\alpha \delta n^\beta+n^\alpha h^\beta_\mu \delta \Gamma^\mu_{\alpha\beta}\label{delK2int}
\end{align}
We can simplify the above expression as follows. Consider the first term
\begin{align}
\delta h^\alpha_\beta \,\nabla_\alpha n^\beta= e^\alpha_a \delta e^a_\beta\,\nabla_\alpha n^\beta=	e^\alpha_a \deltabar A^a n_\beta\,\nabla_\alpha n^\beta=0\label{firsdelk2}
\end{align}
The second term can be manipulated as follows
\begin{align}
	h^\alpha_\beta \nabla_\alpha \delta n^\beta &=e^\alpha_a e^a_\beta \nabla_\alpha \delta n^\beta\nonumber\\
	&=e^\alpha_a \nabla_\alpha (e^a_\beta\delta n^\beta)-e^\alpha_a(\nabla_\alpha e^a_\beta)\delta n^\beta\nonumber\\
	&=-\epsilon\, \del_a \deltabar A^a+\epsilon\, \deltabar A^be^\alpha_a e^\beta_b \nabla_\alpha e^a_\beta+\delta\ln \mu\,\, e^\alpha_an^\beta \nabla_\alpha e^a_\beta\label{secdelk2}
\end{align}
The second and third terms in the last equality can be manipulated as follows.
\begin{align}
	&e^\alpha_a e^\beta_b \nabla_\alpha e^a_\beta=e^\alpha_a \nabla_\alpha (e^\beta_b e^a_\beta)-e^\alpha_a e^a_\beta \nabla_\alpha e^\beta_b=e^\alpha_a \nabla_\alpha(\delta^a_b)-e^a_\beta \Gamma^c_{ab}e^\beta_c=-\Gamma^c_{cb}\nonumber\\
	&e^\alpha_a n^\beta\nabla_\alpha e^a_\beta=e^\alpha_a \nabla_\alpha (n^\beta e^a_\alpha)-e^\alpha_a e^a_\beta \nabla_\alpha n^\beta=-K\label{inteakgki}
\end{align}
Using eq.\eqref{inteakgki} in eq.\eqref{secdelk2}, we get
\begin{align}
		h^\alpha_\beta \nabla_\alpha \delta n^\beta= -\epsilon\, \del_a \deltabar A^a-\epsilon\, \deltabar A^b\, \Gamma^a_{ab}-\delta\ln \mu\,\, K= -\epsilon D_a \deltabar A^a-K \delta \ln \mu  \label{secdelk3}
\end{align}
where $D_a$ is the boundary covariant derivative. Using eq.\eqref{firsdelk2} and eq.\eqref{secdelk3} in eq.\eqref{delK2int}, we get that
\begin{align}
	\delta K = -\epsilon D_a \deltabar A^a-K \delta \ln \mu+n^\alpha h^\beta_\mu \delta \Gamma^\mu_{\alpha\beta}\label{delK2int2}
\end{align}
Combining eq.\eqref{delk1st} and \eqref{delK2int2}, we find 
\begin{align}
 h^{\alpha\beta}n_\mu \delta \Gamma^\mu_{\alpha\beta}-n^\alpha h^\beta_\mu \delta \Gamma^\mu_{\alpha\beta}=-\epsilon D_a \deltabar A^a+K_{ab} \delta h^{ab}-	2 \delta K 
 \label{hgamarel}
\end{align}
But the combination appearing in eq.\eqref{bdtdels} involves the bulk metric. However, it is the same as eq.\eqref{hgamarel} which can be easily checked as
\begin{align}
	n_\mu \,(g^{\alpha\beta}\delta\Gamma^\mu_{\alpha\beta}-g^{\mu \beta}\delta\Gamma^{\nu}_{\nu\beta})&=(n_\mu g^{\alpha\beta}-n^\alpha g^\beta_\mu)\delta \Gamma^\mu_{\alpha\beta}\nonumber\\
	&=(n_\mu (\epsilon n^\alpha n^\beta +h^{\alpha \beta})-n^\alpha (\epsilon n^\beta n_\mu + h^\beta_\mu))\delta \Gamma^\mu_{\alpha\beta}\nonumber\\
	&=(n_\mu h^{\alpha \beta}-n^\alpha  h^\beta_\mu)\delta \Gamma^\mu_{\alpha\beta}\label{nggid1}
\end{align}
Using this result in eq.\eqref{bdtdels}, we get the boundary terms in the variation of the bulk action to be
\begin{align}
	\delta S_{\text{bulk},\del}&=	\half \int_\del \sqrt{h}\,n_\mu\left[\phi  \,(g^{\alpha\beta}\delta\Gamma^\mu_{\alpha\beta}-g^{\mu \beta}\delta\Gamma^{\nu}_{\nu\beta}) -\nabla_\nu\phi\, (g^{\nu\alpha} g^{\mu\beta} \delta g_{\alpha\beta}- g^{\mu \nu} g^{\alpha \beta} \delta g_{\alpha \beta}) \right]\nonumber\\
	&=\half \int_\del \sqrt{h}\,\left[\phi  \,(-\epsilon D_a \deltabar A^a+K_{ab} \delta h^{ab}-	2 \delta K ) -n_\mu\nabla_\nu\phi\, (g^{\nu\alpha} g^{\mu\beta} \delta g_{\alpha\beta}- g^{\mu \nu} g^{\alpha \beta} \delta g_{\alpha \beta}) \right]\nonumber\\
	&=\half \int_\del \sqrt{h}\,\left[ \,(\epsilon \,\deltabar A^a  \, D_a \phi +\phi K_{ab} \delta h^{ab}-	2 \phi \delta K ) -n_\mu\nabla_\nu\phi\, (g^{\nu\alpha} g^{\mu\beta} \delta g_{\alpha\beta}- g^{\mu \nu} g^{\alpha \beta} \delta g_{\alpha \beta}) \right]\nonumber\\
	&\quad -\int_{\del\del} \sqrt{\gamma}\epsilon r_m \phi \deltabar A^m\label{sbulkbb}
\end{align}
where in obtaining the last line from the second, we have done an integration by parts of the first term in the second line, since the derivative is along the directions tangent to the boundary. The last term in the eq.\eqref{sbulkbb} is the boundary term that is obtained due to the integration by parts. It is a codim-2 surface and hence in 2D spacetime, a point. The vector $r_m$ is the normal to the codim-2 surface along the codim-1 boundary. If the codim-1 boundary is smooth, {\it i.e } it does not have boundaries, there is no such boundary term. For our problem in which we have part of the boundary with $\phi$ fixed and part of it with $K$ fixed, there will be such codim-2 surface terms. On the other hand, if the full boundary is a codim-1 boundary with $\phi$ fixed, it will be a smooth boundary and so will not have such codim-2 boundary terms.  $\gamma_{ab}$ is the induced metric on the codim-2 surface from the metric $h_{ab}$ on the codim-1 surface. 

Let us first simplify the terms boundary terms on codim-1 boundary in eq.\eqref{sbulkbb} as follows. We can write the derivative of dilaton along a general direction in terms of its components along the codim-1 boundary and along the normal to the codim-1 boundary as
\begin{align}
	\nabla_\alpha \phi =e^a_\alpha D_a\phi+\epsilon n_\alpha (n\cdot \nabla \phi)\label{deinbdn}
\end{align}
Using this, we have
\begin{align}
	\epsilon \deltabar A^a D_a\phi- n_\mu g^{\nu\alpha}g^{\mu\beta}\delta g_{\alpha\beta}\nabla_\nu \phi &= \epsilon\, \deltabar A^a D_a \phi + n_\mu \delta g^{\mu\nu}\nabla_\nu \phi\nonumber\\
	&=-e^a_\alpha n_\beta \delta g^{\alpha\beta}D_a\phi + (e^a_\nu D_a\phi +\epsilon n_\nu (n\cdot \nabla\phi))n_\mu \delta g^{\mu\nu}\nonumber\\
	&=\epsilon n_\mu n_\nu \delta g^{\mu\nu}(n\cdot \nabla \phi)\label{simpl1}
\end{align}

Using eq.\eqref{simpl1} in eq.\eqref{sbulkbb} gives
\begin{align}
	\delta S_{\text{bulk},\del}&=\half \int_\del \sqrt{h}\,\left[ \epsilon n_\mu n_\nu \delta g^{\mu\nu}n\cdot \nabla \phi+\phi K_{ab} \delta h^{ab}-	2\phi \delta K  +\,  g^{\alpha \beta} \delta g_{\alpha \beta}\, n\cdot \nabla\phi \right]-\int_{\del\del} \sqrt{\gamma}\epsilon r_m \phi \deltabar A^m\label{delsbulsim3}
\end{align}
Further noting that 
\begin{align}
	&g^{\alpha \beta}\delta g_{\alpha \beta}=-g_{\alpha \beta}\delta g^{\alpha \beta}=-(\epsilon n_\alpha n_\beta +h_{\alpha \beta})\delta g^{\alpha \beta},\nonumber\\
	&h_{\alpha\beta}\delta g^{\alpha \beta}=h_{\alpha \beta}\delta h^{\alpha\beta}= h_{ab}\delta h^{ab}\label{ggdelide}
\end{align}
and that for a 1D boundary in 2D, 
\begin{align}
	K_{ab}= K h_{ab}\label{2dexk}
\end{align}
the boundary term in eq.\eqref{delsbulsim3} simplifies to 
\begin{align}
		\delta S_{\text{bulk},\del}&=\half \int_\del \sqrt{h}\,\left[ (\phi K-n\cdot \nabla\phi) h_{ab} \delta h^{ab}-	2\phi  \delta K \right]-\int_{\del\del} \sqrt{\gamma}\epsilon r_m \phi \deltabar A^m\label{delsbulsim4}
\end{align}
Thus, for a single smooth boundary with no codim-2 boundaries with $\phi, h_{ab}$ fixed, we need to add a GH term to cancel the variation of $\delta K$ above. On the other hand for a single smooth boundary with no codim-2 boundaries and with $K, h_{ab}$ fixed, we do not need to add any boundary terms. For a non-smooth boundary, we need to add additional local terms on codim-2 boundaries to cancel the last term in eq.\eqref{delsbulsim4}. Let us now simplify this last term. 

Let $\del_1, \del_2$ be the two boundaries on which the boundary conditions are different but have a codim-2 boundaries in common. Let these codim-2 boundary for $\del_1$ be labelled  $\del\del_1$ and for $\del\del_2$  for $\del_2$.  In our problem, these would be
 $\del_\phi$ and $\del_K$ where $\phi$ and $K$ are fixed respectively. The corresponding codim-2 boundaries would be $\del\del_\phi$ and $\del\del_K$. Since, the codim-2 boundaries are points for 2 bulk dimensions, we have, using eq.\eqref{delindelg}
 \begin{align}
 	\int_{\del\del}\sqrt{\gamma}\epsilon r_a \phi \deltabar A^a=-\epsilon r_a \phi e^ a_\alpha n_\beta \delta g^{\alpha\beta}=\epsilon\phi r^a e^\alpha_a n^\beta \delta g_{\alpha\beta}=\epsilon\phi r^\alpha n^\beta \delta g_{\alpha\beta}\label{cod2delsa}
 \end{align}
Consider a codim-2 boundary at the intersection of the two codim-1 boundaries $\del_1,\del_2$.  Since in the variation of the total action, each boundary $\del_i$ gives a $\del\del$ term, the net contribution for a single codim-2 boundary at their intersection would be
\begin{align}
		\int_{\del\del_{1,2}}\sqrt{\gamma}\epsilon r_a \phi \deltabar A^a=\phi (r^\alpha_1 n^\beta_1+r^\alpha_2 n^\beta_2) \delta g_{\alpha\beta}\label{deldel2}
\end{align}
where the subscripts $1,2$ on $n,r$ denote their origin from $\del_1,\del_2$ respectively. For example, $n_1^\alpha$ is the normal vector to the $\del_1$ in the 2D bulk and $r_1^\alpha$ is the normal to $\del\del_1$ along $\del_1$. These can be visualized in a figure as shown below, fig.\ref{conerfig}.

 \begin{figure}[h!]
 	
 	\centering
 	\tikzset{every picture/.style={line width=0.75pt}} %set default line width to 0.75pt        
 	
 	\begin{tikzpicture}[x=0.75pt,y=0.75pt,yscale=-1,xscale=1]
 		
 		%Curve Lines [id:da11999882930164207] 
 		\draw    (266,113) .. controls (279,192) and (392,214) .. (420,112) ;
 		%Curve Lines [id:da7662151615428864] 
 		\draw    (266,113) .. controls (293,121) and (349,156) .. (420,112) ;
 		%Straight Lines [id:da7826825395566887] 
 		\draw    (310,173) -- (300.77,195.15) ;
 		\draw [shift={(300,197)}, rotate = 292.62] [color={rgb, 255:red, 0; green, 0; blue, 0 }  ][line width=0.75]    (10.93,-3.29) .. controls (6.95,-1.4) and (3.31,-0.3) .. (0,0) .. controls (3.31,0.3) and (6.95,1.4) .. (10.93,3.29)   ;
 		%Straight Lines [id:da16767504944746114] 
 		\draw    (391,126) -- (387.29,100.98) ;
 		\draw [shift={(387,99)}, rotate = 81.57] [color={rgb, 255:red, 0; green, 0; blue, 0 }  ][line width=0.75]    (10.93,-3.29) .. controls (6.95,-1.4) and (3.31,-0.3) .. (0,0) .. controls (3.31,0.3) and (6.95,1.4) .. (10.93,3.29)   ;
 		%Straight Lines [id:da783538013718958] 
 		\draw    (266,113) -- (231.81,96.85) ;
 		\draw [shift={(230,96)}, rotate = 25.28] [color={rgb, 255:red, 0; green, 0; blue, 0 }  ][line width=0.75]    (10.93,-3.29) .. controls (6.95,-1.4) and (3.31,-0.3) .. (0,0) .. controls (3.31,0.3) and (6.95,1.4) .. (10.93,3.29)   ;
 		%Straight Lines [id:da4944703780119153] 
 		\draw    (266,113) -- (260.33,78.97) ;
 		\draw [shift={(260,77)}, rotate = 80.54] [color={rgb, 255:red, 0; green, 0; blue, 0 }  ][line width=0.75]    (10.93,-3.29) .. controls (6.95,-1.4) and (3.31,-0.3) .. (0,0) .. controls (3.31,0.3) and (6.95,1.4) .. (10.93,3.29)   ;
 		%Straight Lines [id:da7267269138720739] 
 		\draw    (266,113) -- (229.98,118.69) ;
 		\draw [shift={(228,119)}, rotate = 351.03] [color={rgb, 255:red, 0; green, 0; blue, 0 }  ][line width=0.75]    (10.93,-3.29) .. controls (6.95,-1.4) and (3.31,-0.3) .. (0,0) .. controls (3.31,0.3) and (6.95,1.4) .. (10.93,3.29)   ;
 		%Straight Lines [id:da8890733742200576] 
 		\draw    (266,113) -- (280.15,82.81) ;
 		\draw [shift={(281,81)}, rotate = 115.11] [color={rgb, 255:red, 0; green, 0; blue, 0 }  ][line width=0.75]    (10.93,-3.29) .. controls (6.95,-1.4) and (3.31,-0.3) .. (0,0) .. controls (3.31,0.3) and (6.95,1.4) .. (10.93,3.29)   ;
 		%Shape: Arc [id:dp3294475411882213] 
 		\draw  [draw opacity=0] (252.48,106.25) .. controls (253.8,103.91) and (255.44,101.69) .. (257.41,99.65) .. controls (259.13,97.85) and (261.01,96.3) .. (263,95) -- (277.32,118.82) -- cycle ; \draw   (252.48,106.25) .. controls (253.8,103.91) and (255.44,101.69) .. (257.41,99.65) .. controls (259.13,97.85) and (261.01,96.3) .. (263,95) ;  
 		
 		% Text Node
 		\draw (356,180.4) node [anchor=north west][inner sep=0.75pt]  [font=\small]  {$\partial _{1}$};
 		% Text Node
 		\draw (315,114.4) node [anchor=north west][inner sep=0.75pt]  [font=\small]  {$\partial _{2}$};
 		% Text Node
 		\draw (289,194.4) node [anchor=north west][inner sep=0.75pt]    {$n_{1}$};
 		% Text Node
 		\draw (379,75.4) node [anchor=north west][inner sep=0.75pt]    {$n_{2}$};
 		% Text Node
 		\draw (207,113.4) node [anchor=north west][inner sep=0.75pt]    {$n_{1}$};
 		% Text Node
 		\draw (281,63.4) node [anchor=north west][inner sep=0.75pt]    {$n_{2}$};
 		% Text Node
 		\draw (249,52.4) node [anchor=north west][inner sep=0.75pt]    {$r_{1}$};
 		% Text Node
 		\draw (210,80.4) node [anchor=north west][inner sep=0.75pt]    {$r_{2}$};
 		% Text Node
 		\draw (245,84.4) node [anchor=north west][inner sep=0.75pt]  [font=\small]  {$\theta $};
 		% Text Node
 		\draw (251,113.4) node [anchor=north west][inner sep=0.75pt]    {$A$};
 		% Text Node
 		\draw (422,115.4) node [anchor=north west][inner sep=0.75pt]    {$B$};

 	\end{tikzpicture}
 \label{conerfig}
 \caption{Corner angle, normal and tangent vectors for various boundaries.}
 \end{figure}
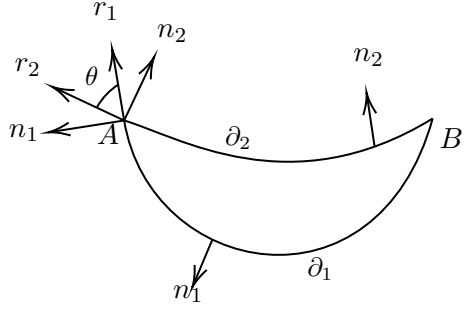

Since, $r_i$ is along $\del_i$ and $n_i$ normal to $\del_i$, it immediately follows that $r_1\cdot n_1=0=r_2\cdot n_2$. So $(r_1,n_1)$ and $(r_2, n_2)$ are a pair of orthogonal bases for the two dimensional vector space at codim-2 boundary. So, either one can be expressed in terms of the other. Let $\theta$ be the angle between $r_1, r_2$ as shown in the figure. Expressing $(n_2,r_2)$ in terms of $(n_1, r_1)$ we have
\begin{align}
	n_2&=r_1\sin\theta -n_1\cos\theta \nonumber\\
	r_2&=r_1\cos\theta + n_1\sin\theta\label{nr2innr1}
\end{align}
and so we have
\begin{align}
	r^\alpha_1 n^\beta_1+r^\alpha_2 n^\beta_2 =(n_1^\alpha n_1^\beta+n_2^\alpha n_2^\beta )\cot\theta + (n_1^\alpha n_2^\beta +n_1^\beta  n_2^\alpha) \csc \theta \label{r1n2r12n2in}
\end{align}
Using this result in eq.\eqref{deldel2} we get
\begin{align}
		\int_{\del\del_{1,2}}\sqrt{\gamma}\epsilon r_a \phi \deltabar A^a=\epsilon\phi \left( (n_1^\alpha n_1^\beta+n_2^\alpha n_2^\beta )\cot\theta + 2n_1^\alpha n_2^\beta \csc \theta\right) \delta  g_{\alpha\beta}\label{corinn1n2}
\end{align}
Let $\tilde{\theta}$ be the angle between the normal vectors $n_1, n_2$. Varying this we get
\begin{align}
	\cos\tilde{\theta}= g^{\alpha\beta}n_{1\alpha}  n_{2\beta}\Rightarrow -\sin\tilde{\theta} \delta \tilde{\theta}&= n_{1\alpha}n_{2\beta}\delta g^{\alpha\beta}+n_{1\alpha}g^{\alpha\beta} \delta n_{2\beta}+g^{\alpha\beta} n_{2\beta}\delta n_{1\alpha}\nonumber\\
	&=n_{1\alpha}n_{2\beta}\left(\delta g^{\alpha\beta}+g^{\alpha\beta} \frac{\delta \mu_1}{\mu_1}+g^{\alpha\beta} \frac{\delta \mu_2}{\mu_2}\right)\nonumber\\
	&=\left(n_{1\alpha}n_{2\beta}- \half (\epsilon_1 n_{1\alpha }n_{1\beta}+\epsilon_2 n_{2\alpha}n_{2\beta})\cos\tilde{\theta}\right)\delta g^{\alpha\beta}\nonumber\\
	\Rightarrow \delta\tilde{\theta}&=-\half \left(2n_{1\alpha}n_{2\beta}\csc\tilde{\theta}-  (\epsilon_1 n_{1\alpha }n_{1\beta}+\epsilon_2 n_{2\alpha}n_{2\beta})\cot\tilde{\theta}\right)\delta g^{\alpha\beta}\label{varannorm}
\end{align}
From the figure it is easy to see that $\tilde{\theta}$, the angle between $n_1,n_2$ is  $\pi-\theta$. Thus, 
\begin{align}
	\tilde{\theta}=\pi-\theta\Rightarrow \delta \theta=-\half \left(2n_{1}^\alpha n_{2}^\beta\csc{\theta}+  (\epsilon_1 n_{1 }^\alpha n_{1}^\beta+\epsilon_2 n_{2}^\alpha n_{2}^\beta)\cot{\theta}\right)\delta g_{\alpha\beta}\label{varthintht}
\end{align}
Thus, the corner term in eq.\eqref{corinn1n2} becomes
\begin{align}
		\int_{\del\del_{1,2}}\sqrt{\gamma}\epsilon r_a \phi \deltabar A^a=-\phi \delta\theta\label{bdcindelt}
\end{align}
So, the  variation of the bulk action gives the net boundary term
\begin{align}
		\delta S_{\text{bulk},\del}&=\half \int_\del \sqrt{h}\,\left[ (\phi K-n\cdot \nabla\phi) h_{ab} \delta h^{ab}-	2\phi  \delta K \right]+\sum_{\mathcal{J}\in\, \text{Corners}}\phi_J \delta\theta_J\label{totvarbuc}
\end{align}
where $\phi_J$ is the value of the dilaton field at the corner and $\theta$ is the angle between the boundaries at the corner. So, the appropriate action with the correct boundary terms for various types of boundaries are as follows
\begin{align}
	S_{\phi,l} &=\half \int d^2x \sqrt{g}\, \phi \,(R-2)+\int_\del \sqrt{h} \,\phi\, K \nonumber\\
	S_{K,l}&=\half \int d^2x \sqrt{g}\, \phi\, (R-2)\nonumber\\
	S_{(\phi, l_\phi),(K,l_K)}&=\half \int d^2x \sqrt{g} \,\phi \,(R-2)+\int_{\del_\phi} \sqrt{h}\, \phi \,K - \sum_{{J}\in\, \text{Corners}}\phi_J \theta_J
	\label{fullactjt}
\end{align}
The net variation of the action with one asymptotic boundary and one extrinsic curvature boundary is given by 
\begin{align}
	\delta S_{(\phi, l_\phi),(K,l_K)}=&\half \int_{\del_K} \sqrt{h}\,\left[ (\phi K-n_K\cdot \nabla\phi) h_{ab} \delta h^{ab}-	2\phi  \delta K \right]+\half \int_{\del_\phi} \sqrt{h}\,\left[ -n_\phi\cdot \nabla\phi\, h_{ab} \delta h^{ab}+	2K\delta\phi   \right]\nonumber\\
    &-\sum_J\theta_J\delta\phi_J\label{askbvar}
\end{align}

\section{Transition amplitude}
\label{twoyorkclass}
\subsection{Classical amplitude}
In this section, we discuss the computation of the on-shell value of the JT gravity action with two York boundary segments where we fix the extrinsic curvatures to be some constants $k_1$ and $k_2$ respectively. We also fix the renormalized lengths of the individual boundary segments to be $\ell_{\text{ren},1}$ and $\ell_{\text{ren},2}$ respectively (see figure \ref{fig:bcfig}(b)). We can interpret this gravitational calculation as the transition amplitude
$ \langle \ell_1,k_1|\ell_2,k_2\rangle.$ 

Let $\del_{1,2}$ be the two boundaries with extrinsic curvatures $k_{1,2}$ and lengths $\ell_{1,2}$ respectively. To be more precise, the boundary conditions are specified as:
\begin{align}
	K_{\del_i}=k_i, \ell_{\del_i}\sim \ell_{\text{ren},i}+\frac{2}{\sqrt{1-k_i^2}}\log\frac{1}{\epsilon} .\label{bdcondkk}
\end{align}
where $\ell_{\text{ren,i}}$ stands for the renormalized length of the $i$-th boundary. As before, it is convenient to introduce the quantities $L_i$ as
\begin{align}
\ell_{\text{ren,i}}=\frac{2}{\sqrt{1-k^2}}\log(L_i\sqrt{1-k^2}).\label{lreniLi}
\end{align}
Also, the two corner points where these boundaries meet will be taken to have the same value of the dilaton $\phi_B$, and we will be interested in the limit $\phi_B =\frac{\phi_b}{\epsilon}$ with $\epsilon \to \infty$.
 The action for such a configuration is given by 
\begin{align}
	S_{JT}=-\half \int \phi (R+2)+\sum_{j\in \text{corners}} \phi_j\theta_j,\label{jtactk1k2}
\end{align}
where $\theta_j$ are the angles of intersection between the two constant-$K$ boundaries at the corners, and $\phi_j=\phi_B$. Since the bulk term vanishes on-shell, we are left to just compute the corner angles. For this computation, we find it convenient to use the Poincare coordinate system eq.\eqref{poinco}. Let us parametrize a constant-$K$ curve as $(x(\lambda),y(\lambda))$ on $H_2$, where $\lambda$ is the proper length coordinate such that,
\begin{align}
	\frac{x'^2+y'^2}{y^2}=1,\label{lambdprot}
\end{align}
with primes denoting derivative with respect to $\lambda$. 
One can now explicitly compute the extrinsic curvature and then use the above equation to solve for $(x(\lambda),y(\lambda))$. However, since we have already computed constant-$K$ curves in terms of polar coordinates $(r,t)$ in eq.\eqref{rtslasolkzer}, we can use the coordinate transformation between $r,t$ and $x,y$ to obtain these curves in Poincare coordinates.  Curves of constant extrinsic curvature computed in the previous subsection in eq.\eqref{rtslasol} when written in terms of Poincare coordinates (using the transformation eq.\eqref{poltopoin}) take the form: 
\begin{align}
	x(\lambda)&= \frac{\left(\sqrt{J^2+1}-J\right) \sinh \left(\sqrt{1-k^2} \lambda \right)}{\cosh \left(\sqrt{1-k^2} \lambda \right)+k},\nonumber\\
	y(\lambda)&=\frac{\left(\sqrt{J^2+1}-J\right) \sqrt{1-k^2}}{\cosh \left(\sqrt{1-k^2} \lambda \right)+k}.\label{xlylcur}
\end{align}
Actually, $x(\lambda)$ appears in eq.\eqref{lambdprot} and in the formula for extrinsic curvature with derivatives, so it is only determined upto an additive constant. With this additional constant, and denoting $R = (\sqrt{J^2+1}-J)$, we can parametrize the curves as
\begin{align}
	&x(\lambda)=x_0+\frac{R \sinh (\lambda  \sqrt{1-k^2})}{k+\cosh (\lambda  \sqrt{1-k^2})},\nonumber\\
	&y(\lambda)=\frac{R\, \sqrt{1-k^2}}{k+\cosh (\lambda  \sqrt{1-k^2})}.\label{xlylcurp1}
\end{align}
We consider $0<k<1$ in what follows. Further since we require $y>0$, we take $R>0$.

From the above formulas, it is straightforward to show that the trajectory is a circle in the Euclidean plane:
\begin{align}
	(x-x_0)^2+\left(y+\frac{Rk}{\sqrt{1-k^2}}\right)^2=\frac{R^2}{1-k^2}.\label{circle}
\end{align}
Thus the two curves with extrinsic curvatures $k_1$ and $k_2$ respectively can be parametrized as 
\begin{align}
	&x_1(\lambda)=x_{c,1}+\frac{R_1 \sinh (\lambda  \sqrt{1-k_1^2})}{k+\cosh (\lambda  \sqrt{1-k_1^2})},\quad y_1(\lambda)=\frac{R_1\, \sqrt{1-k_1^2}}{k_1+\cosh (\lambda  \sqrt{1-k_1^2})},\nonumber\\
	&x_2(\tilde{\lambda})=x_{c,2}+\frac{R_2 \sinh (\tilde{\lambda}  \sqrt{1-k_2^2})}{k_2+\cosh (\tilde{\lambda}  \sqrt{1-k_2^2})},\quad y_2(\tilde{\lambda})=\frac{R_2\, \sqrt{1-k_2^2}}{k_2+\cosh (\tilde{\lambda}  \sqrt{1-k_2^2})}.\label{xlylcurp}
\end{align}

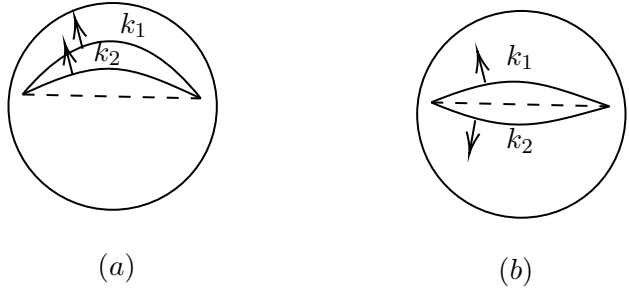
\begin{figure}
	
	\centering
	
	\tikzset{every picture/.style={line width=0.75pt}} %set default line width to 0.75pt        

\tikzset{every picture/.style={line width=0.75pt}} %set default line width to 0.75pt        

\begin{tikzpicture}[x=0.75pt,y=0.75pt,yscale=-1,xscale=1]

%Shape: Circle [id:dp8430119796564053] 
\draw   (390,150) .. controls (390,121.28) and (413.28,98) .. (442,98) .. controls (470.72,98) and (494,121.28) .. (494,150) .. controls (494,178.72) and (470.72,202) .. (442,202) .. controls (413.28,202) and (390,178.72) .. (390,150) -- cycle ;
%Curve Lines [id:da0034984030400806443] 
\draw    (397,144) .. controls (438,127) and (449,133) .. (486,146) ;
%Curve Lines [id:da11165324626291762] 
\draw    (397,144) .. controls (437,161) and (450,156) .. (486,146) ;
%Shape: Circle [id:dp8222387163637678] 
\draw   (186,146) .. controls (186,117.28) and (209.28,94) .. (238,94) .. controls (266.72,94) and (290,117.28) .. (290,146) .. controls (290,174.72) and (266.72,198) .. (238,198) .. controls (209.28,198) and (186,174.72) .. (186,146) -- cycle ;
%Curve Lines [id:da8648897489397624] 
\draw    (193,140) .. controls (222,105) and (251,103) .. (282,142) ;
%Curve Lines [id:da6072052132414357] 
\draw    (193,140) .. controls (233,121) and (244,124) .. (282,142) ;
%Straight Lines [id:da26587936771178977] 
\draw    (223,117) -- (219.49,102.94) ;
\draw [shift={(219,101)}, rotate = 75.96] [color={rgb, 255:red, 0; green, 0; blue, 0 }  ][line width=0.75]    (10.93,-3.29) .. controls (6.95,-1.4) and (3.31,-0.3) .. (0,0) .. controls (3.31,0.3) and (6.95,1.4) .. (10.93,3.29)   ;
%Straight Lines [id:da6342278061864451] 
\draw    (218,130) -- (214.49,115.94) ;
\draw [shift={(214,114)}, rotate = 75.96] [color={rgb, 255:red, 0; green, 0; blue, 0 }  ][line width=0.75]    (10.93,-3.29) .. controls (6.95,-1.4) and (3.31,-0.3) .. (0,0) .. controls (3.31,0.3) and (6.95,1.4) .. (10.93,3.29)   ;
%Straight Lines [id:da7490841388417411] 
\draw    (424,134) -- (420.49,119.94) ;
\draw [shift={(420,118)}, rotate = 75.96] [color={rgb, 255:red, 0; green, 0; blue, 0 }  ][line width=0.75]    (10.93,-3.29) .. controls (6.95,-1.4) and (3.31,-0.3) .. (0,0) .. controls (3.31,0.3) and (6.95,1.4) .. (10.93,3.29)   ;
%Straight Lines [id:da37849819820047625] 
\draw    (419,153) -- (416.35,168.03) ;
\draw [shift={(416,170)}, rotate = 280.01] [color={rgb, 255:red, 0; green, 0; blue, 0 }  ][line width=0.75]    (10.93,-3.29) .. controls (6.95,-1.4) and (3.31,-0.3) .. (0,0) .. controls (3.31,0.3) and (6.95,1.4) .. (10.93,3.29)   ;
%Straight Lines [id:da11842705174581036] 
\draw  [dash pattern={on 4.5pt off 4.5pt}]  (193,140) -- (282,142) ;
%Straight Lines [id:da014871430954026321] 
\draw  [dash pattern={on 4.5pt off 4.5pt}]  (397,144) -- (486,146) ;

% Text Node
\draw (229,218.4) node [anchor=north west][inner sep=0.75pt]    {$( a)$};
% Text Node
\draw (429,220.4) node [anchor=north west][inner sep=0.75pt]    {$( b)$};
% Text Node
\draw (240,97.4) node [anchor=north west][inner sep=0.75pt]    {$k_{1}$};
% Text Node
\draw (227,111.4) node [anchor=north west][inner sep=0.75pt]    {$k_{2}$};
% Text Node
\draw (433,113.4) node [anchor=north west][inner sep=0.75pt]    {$k_{1}$};
% Text Node
\draw (433,155.4) node [anchor=north west][inner sep=0.75pt]    {$k_{2}$};

\end{tikzpicture}
\caption{Convention for extrinsic curvature}
\label{cur}
\end{figure}

Let the two curves intersect at $(x_A, y_A)$ and $(x_B,y_B)$. Let $\theta_A,\theta_B$ be the corner angles at these points. We will be interested in the special case when the value of the dilaton is the same at both these points, and scales as $\frac{\phi_b}{\epsilon}$, i.e., the intersection points lie on the asymptotic boundary. Then, by symmetry, we expect that $\theta_A=\theta_B$. So, we focus on the evaluation at one of the corners, say $A$. The expression for the angle is given by 
\begin{align}
	\cos\theta_A= \frac{x_1'(\lambda_A)x_2'(\tilde{\lambda}_A)+y_1'(\lambda_A)y_2'(\tilde{\lambda}_A)}{y_A^2}.\label{cosatAex}
\end{align}
To compute the above angle, we first need to evaluate the quantities $x_{c,i},R_i$ for each of the curves $(x_i(\lambda),y_i(\lambda))$. The equations to determine them are, say for the curve $(x_1,y_1)$:
\begin{align}
	x_1(\lambda_A)=x_A, 	y_1(\lambda_A)=y_A, \nonumber\\
	x_1(\lambda_B)=x_B,		y_1(\lambda_B)=y_B. \label{solcons}
\end{align}
Note that there are four variables $(x_{c,1},R_1, \lambda_A, \lambda_B)$ and four equations, and hence this system is solvable. An alternative way to directly obtain $(x_{c,1},R_1)$ without solving for $\lambda_A, \lambda_B$ explicitly is to notice that the components $(x_1(\lambda), y_1(\lambda))$ solve the equation of a circle:
\begin{align}
	(	x_1-x_{c,1})^2+\left(y_1+\frac{R_1k}{\sqrt{1-k^2}}\right)^2=\frac{R_1^2}{1-k^2}.\label{x1y1}
\end{align}
Since both the points $(x_A, y_A)$ and $(x_B,y_B)$ satisfy these equations, we have two equations in two variables, which can be solved. The expressions for $x_{c,1}, R_1$ are given by 
{\footnotesize
	\begin{align}
		&x^\pm _{AB}=x_A\pm x_B,\quad y^\pm _{AB}=y_A\pm y_B,\quad \nonumber\\
		&x_{c,1}=
		\frac{k_1 | y^-_{AB}|  \sqrt{-\left(\left(x^-_{AB}{}^2+y^-_{AB}{}^2\right) \left(\left(k_1^2-1\right) x^-_{AB}{}^2+k_1^2 y^-_{AB}{}^2-y^+_{AB}{}^2\right)\right)}+k_1^2 x^+_{AB} \left(x^-_{AB}{}^2+y^-_{AB}{}^2\right)-x^-_{AB} (x^-_{AB} x^+_{AB}+y^-_{AB} y^+_{AB})}{2 \left(k_1^2-1\right) x^-_{AB}{}^2+2 k_1^2 y^-_{AB}{}^2},\nonumber\\
		&R_1=\frac{\sqrt{1-k_1^2} \left(x^-_{AB} \sqrt{-\left(\left(x^-_{AB}{}^2+y^-_{AB}{}^2\right) \left(\left(k_1^2-1\right) x^-_{AB}{}^2+k_1^2 y^-_{AB}{}^2-y^+_{AB}{}^2\right)\right)}-k_1 y^+_{AB} \text{sgn}(y^-_{AB}) \left(x^-_{AB}{}^2+y^-_{AB}{}^2\right)\right)}{2 \text{sgn}(y^-_{AB}) \left(\left(k_1^2-1\right) x^-_{AB}{}^2+k_1^2 y^-_{AB}{}^2\right)}.\label{xc1R1val}
	\end{align}
}
We can similarly obtain $x_{c,2},R_2$ in terms of $(x_A,y_A), (x_B, y_B)$ and $k_2$; the expressions are similar to the ones above with $k_1$ replaced by $k_2$. Having obtained the values of $x_{c,i},R_i$, we can now evaluate eq.\eqref{cosatAex} by noting that 
\begin{align}
	x_i'(\lambda)=\frac{ y(\lambda ) }{R}\left({k R}+y(\lambda )\sqrt{1-k^2}\right),\nonumber\\
	y_i'(\lambda)=-\frac{ y(\lambda ) }{R}(x(\lambda )-x_{c,i})\sqrt{1-k^2}.\label{yplxpl}
\end{align}
Using these expressions, the angle in eq.\eqref{cosatAex} becomes
\begin{align}
	\cos\theta_A= \frac{(k_1 R_1+y_A\sqrt{1-k_1^2} ) (k_2 R_2+y_A\sqrt{1-k_2^2}  )+\sqrt{(1-k_1^2)(1-k_2^2)}  (x_{c,1}-x_A) (x_{c,2}-x_A)}{R_1 R_2}.\label{costhe}
\end{align}
Using the explicit expressions for $x_{c,1},R_1$ shown in eq.\eqref{xc1R1val} and the corresponding expressions for $x_{c,2},R_2$ with $k_1\rightarrow k_2$, we get the following explicit expression for the angle: 
\begin{align}
	\cos\theta_A =\frac{\sqrt{\left(k_1^2+d^2 \left(1-k_1^2\right)\right) \left(k_2^2+d^2 \left(1-k_2^2\right)\right)}+\left(d^2-1\right) k_1 k_2}{d^2},\label{costhefin}
\end{align}
where $d$ is the following function of $x_A, x_B, y_A$ and $y_B$:
\begin{align}
	d&=\sqrt{\frac{(x_A-x_B)^2+(y_A+y_B)^2}{4y_A y_B}}	.\label{dxaxbyayb}
\end{align}
{The quantity $d$ above is related to the geodesic distance $\ell_0$ as
\begin{align}
    d=\cosh(\frac{\ell_0}{2})\label{dingeodis}
\end{align}
Note that the formula for the angle between two curves computed is if both the curves are on the same side of the geodesic curve (shown in dotted lines in fig.(\ref{cur}))  }

For the situation where the two curves are on the opposite sides of the geodesic curve, as in fig.\ref{cur}(b), the angle between the curves at the corners is given by 
\begin{align}
	\cos\theta_A =\frac{\sqrt{\left(k_1^2+d^2 \left(1-k_1^2\right)\right) \left(k_2^2+d^2 \left(1-k_2^2\right)\right)}-\left(d^2-1\right) k_1 k_2}{d^2}.\label{costhefin2}	
\end{align}
The results eq.\eqref{costhefin},\eqref{costhefin2} can be written more compactly by defining $\theta_1, \theta_2$ as
\begin{align}
	\cos\theta_1&=\frac{\sqrt{\left(k_1^2+d^2 \left(1-k_1^2\right)\right) }}{d},\nonumber\\
	\cos\theta_2&=\frac{\sqrt{\left(k_2^2+d^2 \left(1-k_2^2\right)\right) }}{d},
	\label{costhefin3}
\end{align}
so that
\begin{align}
	\cos\theta_A=\cos(\theta_1\pm\theta_2).\label{thin12}
\end{align}
or equivalently
\begin{align}
	\theta_A=\arctan(\frac{k_1\sqrt{d^2-1}}{\sqrt{\left(d^2-1\right) \left(1-k_1^2\right)+1}})\pm \arctan(\frac{k_2\sqrt{d^2-1}}{\sqrt{\left(d^2-1\right) \left(1-k_2^2\right)+1}})\label{prtangl}
\end{align}
{Essentially, $\theta_1,\theta_2$ are the angles between the geodesic curve and the extrinsic curvature curves $k_1,k_2$ respectively. The above signs in eq.\eqref{thin12} then is just a refection of the fact that if both the curves are on the same side of geodesic, the angle between them is the difference of the angles between each of the curve and geodesic and if they are on the opposite side, the total angle is the sum of individual angles with the geodesic curve. }

So, the transition amplitude in the saddle point limit corresponding to the on-shell action becomes
\begin{align}
	\langle \ell_1,k_1|\ell_2,k_2\rangle =e^{-\phi_A \theta_A-\phi_B\theta_B},\label{psial}
\end{align}
where $\theta_A$ is given by eq.\eqref{costhefin} and a similar expression for $\theta_B$, $\theta_A=\theta_B$. In the limit where the two corners go to infinity and so $d$ becomes large, eq.\eqref{costhefin} simplifies to 
\begin{align}
	\theta_A\simeq \arccos(\sqrt{(1-k_1^2)(1-k_2^2)}-k_1k_2)=\arcsin(k_1)+\arcsin(k_2)+\order{(\epsilon^2)},\label{theainld}
\end{align}
so the classical amplitude becomes:
\begin{align}
	\langle \ell_1,k_1|\ell_2,k_2\rangle\simeq e^{-2\phi_B((\arcsin(k_1)+\arcsin(k_2)))}.\label{lklapsi}
\end{align}
We shall see later, when we compute the full quantum wavefunction, that not all four quantities, the two lengths of the boundaries $l_1,l_2$ and the associated extrinsic curvature $k_1,k_2$, are independent. There is a single relation between these quantities which leads to a delta function condition in the wavefunction, apart from the exponential factor shown above eq.\eqref{lklapsi}.

\subsection{Quantum amplitude}
\label{qtwybd}

Before we turn to performing some checks of our result, let us also evaluate the quantum wavefunction for bulk geometry with two boundaries $\del_{1,2}$ each of which are York boundaries  with the boundary conditions specified in eq.\eqref{bdcondkk}. Denoted as $\Psi_Y$, this will be proportional to $\langle l_1,k_1|l_2,k_2\rangle$. As before, let the points where the two boundaries meet be labelled by $(x_1,y_1)$ and $(x_2,y_2)$.  In terms of the particle picture, the action is given by 
\begin{align}
		S_{\del}=q\left( k_1\ell_{1}-\int_{\del_{1}}a\right) +\left(k_2\ell_{2}-\int_{\del_{2}}a  \right).\label{skink12}
\end{align}
As in the previous subsection, the contribution from $\del_i$ can be obtained by evaluating the contribution from the single path between the end points $(x_1,y_1)$ and $(x_2,y_2)$ with the extrinsic curvature $k_i$. The value of the Wilson line along this path is the same as in eq.\eqref{valwil} with $k$ replaced by $k_i$ appropriate to the path. Thus, the wavefunction is given by 
\begin{align}
\Psi_Y=	&\exp(-2q\left[ \tan^{-1}\left(\frac{1}{\sqrt{1+k_1}}\tanh(\frac{\sqrt{(1-k_1^2)}\lambda}{2})\right)\right]\bigg\vert_{\lambda_1}^{\lambda_2})\nonumber\\
&\times\exp(2q\left[\tan^{-1}\left(\frac{1}{\sqrt{1+k_2}}\tanh(\frac{\sqrt{(1-k_2^2)}\lambda}{2})\right)\right]\bigg\vert_{\lambda_1}^{\lambda_2}	).
\label{psiink1k2}
\end{align}
This expression can be further simplified by adding and subtracting $	\arctan(\frac{x_2-x_1}{y_1+y_2})$ in the exponent and making use of the result in eq.\eqref{varhatval} which gives
\begin{align}
	\Psi_Y=\exp[-2q\tan^{-1}\left(\frac{k_1}{\sqrt{1-k_1^2}}\tanh\left(\frac{\ell_1}{2} \sqrt{1-k_1^2}\right)\right)+2q\tan^{-1}\left(\frac{k_2}{\sqrt{1-k_2^2}}\tanh\left(\frac{\ell_2}{2} \sqrt{1-k_2^2}\right)\right)].
	\label{psifk1k2}
\end{align}
Further the quantity $d$, related to the end points of the curve by eq.\eqref{dxaxbyayb}, can be expressed in terms of either $\ell_1, k_1$ or $\ell_2,k_2$ as in eq.\eqref{dinlk}, 
\begin{align}
	d=\sqrt{1+\frac{1}{1-k_1^2}\sinh^2\left(\half \ell_1\sqrt{1-k_1^2}\right)}=\sqrt{1+\frac{1}{1-k_2^2}\sinh^2\left(\half \ell_2\sqrt{1-k_2^2}\right)}.\label{dl1l12k1k2}
\end{align} 
We expect that this condition should be enforced in the wavefunction by an explicit delta function; we will give further evidence for this below. Defining
\begin{align}
	d_1=\sqrt{1+\frac{1}{1-k_1^2}\sinh^2\left(\half \ell_1\sqrt{1-k_1^2}\right)},\nonumber\\
	d_2=\sqrt{1+\frac{1}{1-k_2^2}\sinh^2\left(\half \ell_2\sqrt{1-k_2^2}\right)},\label{d1d2}
\end{align}
the wavefunction is given by 
{\small
\begin{align}
		\Psi_Y=\delta(d_{\infty,1}-d_{\infty,2})\exp[-2q\tan^{-1}\left(\frac{k_1}{\sqrt{1-k_1^2}}\tanh\left(\frac{\ell_1}{2} \sqrt{1-k_1^2}\right)\right)+2q\tan^{-1}\left(\frac{k_2}{\sqrt{1-k_2^2}}\tanh\left(\frac{\ell_2}{2} \sqrt{1-k_2^2}\right)\right)],
	\label{psifk1k2de}
\end{align}}
where 
\begin{align}
	d_{\infty,i}=\frac{d_i}{q}.\label{dinfd}
\end{align}
{Note that the above result eq.\eqref{psifk1k2de} has a relative minus sign between the two terms in the exponent and corresponds to the configuration in fig.\ref{cur}(a). The result above also matches with that obtained directly by evaluating the corner angle in the on-shell calculation, see eq.\eqref{prtangl}, noting the relation eq.\eqref{dinlk}.

A particular special case is worth mentioning explicity. For the case $k_1=k_2=k$, the above simplifies to
\begin{align}
    \Psi_Y=%\frac{q^2\sqrt{1-k^2}}{\sinh(\ell_1\sqrt{1-k^2})}
    2\sech(\half\ell_1\sqrt{1-k^2})\delta(\ell_1-\ell_2)\label{delk1k2k}
\end{align}
It is more convenient to write the above result just as
\begin{align}
    \Psi_Y(\ell_1,k;\ell_2,-k)=\delta(d_{\infty,1}-d_{\infty,2})\label{psiyl1l2}
\end{align}
The delta function above can be justified alternatively as follows. We can use the result in eq.\eqref{elkinpx1} for the overlap of an energy eigenstate with the length state and the completeness relation of the energy eigenstates to derive the delta function in eq.\eqref{psiyl1l2}. The completeness relation for the energy eigenstates is given by 
\begin{align}
\int_0^\infty dE\rho(E)|E\rangle\langle E|=\mathds{1},\qquad \rho(E)=\frac{2}{\pi^2}\sinh(2\pi\sqrt{E})\label{ecomprel}
\end{align}
Using this, we have
\begin{align}
\langle \ell_1,k|\ell_2,k\rangle_{\text{PI}}&=\int dE \, \rho(E)\,\langle \ell_1,k|E\rangle_{\text{PI}} \langle E|\ell_2,k\rangle_{\text{PI}} \nonumber\\
&=\frac{2}{\pi^2d_{\infty,1}d_{\infty,2} }\int dE {\,\sinh(2\pi\sqrt{E})}e^{-2q\sin^{-1}(k)-2q\sin^{-1}(-k)}K_{2i\sqrt{E}}\left(\frac{2}{d_{\infty,1}}\right)K_{2i\sqrt{E}}\left(\frac{2}{d_{\infty,2}}\right)\nonumber\\
&=\frac{2}{\pi^2d_{\infty,1}d_{\infty,2} }\left(\frac{\pi^2}{4d_{\infty,1}}\delta\left(\frac{1}{d_{\infty,1}}-\frac{1}{d_{\infty,2}}\right)\right)\nonumber\\
&=\frac{1}{2 d_{\infty,1}} \delta(d_{\infty,1}-d_{\infty,2})  \label{l1l2inp}
\end{align}
where the subscript $PI$ is to denote that the states are defined by the path integral result eq.\eqref{elkinpx1}. We used the following identity for Bessel functions in obtaining the last line from the penultimate one, 
\begin{align}
    \int_0^\infty dy \, y\, K_{iy}(z)K_{iy}(x)\sinh{\pi y}=\pi^2 x\, \delta(x-z)\label{besfid}
\end{align}
It follows from eq.\eqref{besfid} that the resolution of identity in terms of the length states, at any value of physical clock $k$, can be written as
\begin{align}
    \int d(d_{\infty}^2)|\ell,k\rangle\langle \ell,k|_{\text{PI}}=\mathds{1}\label{lkpimes}
\end{align}
}

The wavefunction in eq.\eqref{psifk1k2de} should be interpreted as an overlap between the length eigenstates at times $k_1$ and $k_2$. More precisely, analogous to eq.\eqref{elkinpx1}, we can write, 
\begin{align}
    \langle \ell_1,k_1|\ell_2,k_2\rangle &=\half(\sqrt{d_{\infty,1}d_{\infty,2}})^{-1}\Psi_Y\nonumber\\
    &\simeq \half(\sqrt{d_{\infty,1}d_{\infty,2}})^{-1}\delta(d_{\infty,1}-d_{\infty,2})e^{-2q\arcsin{k_1}+2q\arcsin{k_2}}
\end{align}
where the second inequality is obtained from the first in the limit of large $\ell_1,\ell_2$.
From this information, we can now extract the bulk Hamiltonian which acts as a physical evolution operator, as was done in subsection \ref{buphham}. We need to do the analytic continuation from $k_i\rightarrow i k_i$ to compute the physical Bulk Hamiltonian that generates $k$ evolution in the Lorentzian spacetime. A computation analogous to the one presented earlier after eq.\eqref{elkinpx2} shows that the physical Hamiltonian corresponding to the York time evolution will again be given by eq.\eqref{Hphyful}.

\subsection{Consistency checks}
\label{conv2wf}
%{\color{red}{check the conventions in this section}}

In this subsection we shall consider the inner product of various wavefunctions obtained  in this paper. As a part of this, we show how to obtain the partition function of JT gravity in AdS with a single asymptotic boundary from the inner product of two wavefunctions each of which is made up of one asymptotic boundary and one extrinsic curvature boundary. This serves as a strong check of our result in eq.\eqref{psiinbek}. Let the two wavefunctions be given by 
\begin{align}
	\Psi_A(\phi_B, \ell_\phi, k,\ell_k)\equiv \frac{\hat{\Psi}_A(\phi_B, \ell_\phi, k,\ell_k)}{d_\infty}=\frac{1}{d_{\infty}}e^{-2\phi_B\sin^{-1}(k)}\int ds \,\rho(s)\,{e^{-\tilde{u} s^2}}\,\,K_{2is}\left(\frac{2}{d_{\infty}}\right)
%	\Psi_2(\phi, l_\phi, k_2,l_{2	})=\frac{1}{\pi^2 d_{2,\infty}}e^{-2\phi\cos^{-1}k}\int ds \,\rho(s)\,{e^{-u_2s^2}}\,\,K_{2is}\left(\frac{2}{d_{2,\infty}}\right)
\label{twowfss}
\end{align}
where
\begin{align}
	\rho(s)&=s\sinh(2\pi s),\nonumber\\
	\tilde{u}&=\frac{\ell_{\phi}}{\phi_B}\nonumber\\
	d_{\infty}&=\frac{d}{\phi_B}=	\frac{L}{\phi_b\sqrt{1-k^2}}
	\label{rhos}
\end{align}
and where $L$ is the renormalized length of the extrinsic curvature slice eq.\eqref{bdcond}. Note that we have included the counterterm for corners, eq.\eqref{ctforcor}, in obtaining in eq.\eqref{twowfss}. This will be needed to obtain correct results for overlaps which involve the wavefunction $\Psi_A$ as we shall see below. 
%The change in the sign of $\phi\sin^{-1}k$ term in the exponent of $\Psi_2^*$ can be thought as due to the change in the direction of the normal of the extrinsic curvature slice as compared to $\Psi_2$. 

We glue the two wavefunctions along the extrinsic curvature boundary. The length of the boundary of the extrinsic curvature slice is related to $d_\infty$. The appropriate measure for gluing of the two wavefunctions $\hat{\Psi}_A$ is given by 
\begin{align}
    \int \frac{d(d_\infty)}{d_\infty}=\half \int d\ell_k \sqrt{1-k^2}
\end{align}

\begin{align}
&	\int \frac{d(d_\infty)}{d_\infty}\hat{\Psi}_A(\phi_B, \ell_{\phi,1}, k,\ell)\hat{\Psi}_A(\phi_B, \ell_{\phi,2}, -k,\ell)\nonumber\\ &\propto\int ds_1 ds_2\rho(s_1)\rho(s_2)e^{-u_1s_1^2-u_2s_2^2-2\phi\sin^{-1}k-2\phi\sin^{-1}(-k)}\int_0^\infty \frac{d(d_\infty)}{d_\infty}K_{2is_1}\left(\frac{2}{d_\infty}\right)K_{2is_2}\left(\frac{2}{d_\infty}\right)\label{psi12s}
\end{align}

Using the identity
\begin{align}
	\int_0^\infty\frac{dx}{x}K_{2is}\left(\frac{4}{x}\right)K_{2i\tau}\left(\frac{4}{x}\right)\propto \frac{\delta(\tau-s)}{\rho(s)}\label{besselid}
\end{align}
we find that eq.\eqref{psi12s} becomes
\begin{align}
		\int \frac{d(d_\infty)}{d_\infty}\hat{\Psi}_A(\phi_B, \ell_{\phi,1}, k,\ell)\hat{\Psi}_A(\phi_B, \ell_{\phi,2}, -k,\ell)\propto& \int ds_1 ds_2\rho(s_1)\rho(s_2)\frac{\delta(s_1-s_2)}{\rho(s_1)}e^{-\tilde{u}_1s_1^2-\tilde{u}_2s_2^2}\nonumber\\
	=&	\int ds\rho(s)e^{-(\tilde{u}_1+\tilde{u}_2)s^2}\nonumber\\
	=&Z_{\text{AdS}}(\ell_{\phi,1}+\ell_{\phi,2},\phi_B)\label{convadspf}
\end{align}
which is indeed the AdS partition function for the total length of the boundary $\ell_{\phi,1}+\ell_{\phi,2}$ and the value of the dilaton on the boundary.

We can also consider the innerproduct of two wavefunctions, with one or both of them being ones with two extrinsic curvature boundaries, 
\small{
\begin{align}
		\Psi_Y(k_1,\ell_1,k_2,\ell_2)&\equiv\frac{\hat{\Psi}_Y(k_1,\ell_1,k_2,\ell_2)}{\sqrt{d_{\infty,1}d_{\infty,2}}}=\delta(d_{\infty,1}-d_{\infty,2})\exp[2\phi_B\left(w(k_1,\ell_1)+w(k_2,\ell_2)\right)]\nonumber\\
		\Psi_Y(-k_2,\ell_2,k_3,\ell_3)&\equiv\frac{\hat{\Psi}_Y(k_2,\ell_2,k_3,\ell_3)}{\sqrt{d_{\infty,2}d_{\infty,3}}}=\delta(d_{\infty,2}-d_{\infty,3})\exp[2\phi_B\left(w(k_3,\ell_3)-w(k_2,\ell_2)\right)]\nonumber\\
		w(k,\ell)&=\tan^{-1}\left(\frac{k}{\sqrt{1-k^2}}\tanh\left(\frac{\ell}{2} \sqrt{1-k^2}\right)\right)\simeq \arcsin(k)
		\label{k1k2k3}
\end{align}
}
It is easy to see that 
\begin{align}
	\hat{\Psi}_Y(k_1,\ell_1,k_3,\ell_3)=    \int \frac{d(d_{\infty,2})}{d_{\infty,2}}\hat{\Psi}(k_1,\ell_1,k_2,\ell_2)\hat{\Psi}(-k_2,\ell_2,k_3,\ell_3)\label{k1k2k3inp}
\end{align}

Similary, the innerproduct between an asymptotic wavefunction and an extrinsic curvature wavefunction gives

\begin{align}
	\hat{\Psi}_Y(k_1,\ell_1,-k,\ell)&=\delta(d_{\infty,1}-d_\infty)\exp[2\phi_B\left(w(k_1,\ell_1)-w(k,\ell)\right)]\nonumber\\
	\hat{\Psi}_A(\phi_B,\ell_\phi, k,\ell)&=e^{-2\phi_B\sin^{-1}(k)}\int ds \,\rho(s)\,{e^{-\tilde{u} s^2}}\,\,K_{2is}\left(\frac{2}{d_\infty}\right)
		\label{k1k2ph}
\end{align}
Their innerproduct leads to
\begin{align}
	\hat{\Psi}_A(\phi_B,\ell_\phi, k_1,\ell_1)=\int \frac{d(d_{\infty})}{d_{\infty}}	\hat{\Psi}_Y(k_1,\ell_1,-k,\ell) \hat{\Psi}_A(\phi_B,\ell_\phi, k,\ell)
\end{align}

\section{The Wheeler-DeWitt equation}
\label{wdwmat}
In this appendix, we present the derivation of the Wheeler-DeWitt (WDW) constraint for various types of boundaries. In section \ref{wdwconder}, we derive the quantum WDW equation for different types of boundaries, namely asymptotic boundary with Dirichlet boundary condition on dilaton and induced metric, York boundary with extrinsic curvature and induced metrix fixed and more general mixed boundary with partly asymptotic boundary and partly York boundary. In particular, since we only discuss the dependence of the wavefunction on the zero modes of $\phi,\sqrt{h},K$ (dilaton, induced metric and extrinsic curvature respectively) in this paper, we shall be mainly concerned with the corresponding zero mode term of the WDW equation. 

Following this, in subsection \ref{wdwsec}, we shall show that the wavefunctions obtained in section \ref{qsols} satisfy the appropriate WDW equation derived in \ref{wdwconder}. See also \cite{Chowdhury:2021nxw} for interesting recent work on the role of the WdW equation in the canonical approach to gravity. 
%Noting that x$K$ also has a dependence on $N$ given by 
%\begin{align}
%	K=\nabla_\mu n^\mu\label{kh}
%\end{align}
\subsection{The WDW constraint}
\label{wdwconder}
The local WDW equation obtained by doing a variation of the action, eq.\eqref{buljt} with respect to $N$ gives
\begin{align}
	\frac{\delta S}{\delta N}\equiv\mathcal{H}=\sqrt{h}(-\phi -D^2\phi +K \mathfrak{t}^\alpha \nabla_\alpha\phi)\label{delsn}
\end{align}
where $D^2=D^iD_i$ and $D_i$ is the boundary covariant derivative, see eq.\eqref{delsn1}. To express it as a differential equation acting on quantum states which is the standard Wheeler De-Witt equation, we need to express the time derivatives in terms of the conjugate momenta of the physical variables in the theory. This further depends on the boundary conditions on the variables. Let us now analyze various boundary conditions. 
\begin{itemize}
	\item {\bf Asymptotic Boundary}: For the case where $\phi, g_{ab}$ is fixed on the boundary, as mentioned earlier, we need to add the GHY boundary term which leads to the action, 
	\begin{align}
			S_{\phi, h}&=\int dt\int dx\sqrt{h}(h^{\beta\alpha}\nabla_\beta N\nabla_\alpha\phi-NK\mathfrak{t}^\alpha\nabla_\alpha\phi -N\phi )\label{phhcon}
	\end{align}
from which we can read off the conjugate momenta as
\begin{align}
	\pi_\phi&\equiv \frac{\del S_{\phi, h}}{\del \dot{\phi}}=-\sqrt{h}K\nonumber\\
	\pi_{h}^{ij}&\equiv \frac{\del S_{\phi, h}}{\del \dot{h}_{ij}}=-\frac{\sqrt{h}}{2}h^{ij}\mathfrak{t}^\alpha\nabla_\alpha\phi\label{dilconmo}
\end{align}
Noting that $h_{ij}$ has only one component since boundary is 1D, we find that the Hamiltonian constraint eq.\eqref{delsn} is given by 
\begin{align}
	\mathcal{H}=\frac{2}{\sqrt{h}}h_{ij}\pi_{h}^{ij}\pi_\phi -\sqrt{h}D^2\phi-\sqrt{h}\phi\label{phhhamc}
\end{align}
The zero mode part of the above can be obtained by integrating it along the boundary of constant $\phi$. This gives, upto operator ordering ambiguity, 
\begin{align}
		\mathcal{H}_0=\pi_{\ell_\phi}\pi_\phi -\ell_\phi \phi\label{phhhamcz}
\end{align}
where $\ell_\phi$ is the length of the boundary, given by $\ell_\phi=\int dx\sqrt{h}$.
To quantize the theory, we replace{\footnote{In Euclidean signature, the commutation relations are of the form $[x,p]=1$}}
\begin{align}
	\pi_\phi\rightarrow -\del_\phi,\nonumber\\
	\pi_{\ell_\phi}\rightarrow -\del_{\ell_\phi}\label{piqt}
\end{align}
The path integral result for the configuration where dilaton and metric are fixed on the boundary, in the asymptotic limit where both the length of the boundary and dilaton diverge at the boundary with their ratio held fixed,
\begin{align}
	\phi,\ell_\phi\rightarrow \infty ,\frac{\ell_\phi}{\phi}=\text{fixed}\label{asyc}
\end{align}
 is given by 
\begin{align}
	\Psi(\ell_\phi,\phi)\propto \int ds s \sinh(2\pi s)e^{\ell_\phi\phi-\frac{\ell_\phi}{2\phi}s^2	}\label{psilph}
\end{align}
which satisfies the WDW equation in the asymptotic limit (WKB limit)
\begin{align}
	\left(\del_{\ell_\phi}\del_\phi-\frac{1}{\ell_\phi}\del_\phi -\ell_\phi\phi\right)\Psi(\ell_\phi,\phi)=0\label{wkbwdwlph}
\end{align}
More general solutions to the above differential equation are of the form 
\begin{align}
	\Psi(\ell_\phi,\phi)=\int \rho_+(M)e^{\ell_\phi\sqrt{\phi^2-M}}+\int \rho_-(M)e^{-\ell_\phi\sqrt{\phi^2-M}}\label{psilphi}
\end{align}
The path integral result eq.\eqref{psilph} is a particular solution of the form above with $\rho_+(M)=\sinh(2\pi\sqrt{M}), \rho_-=0$ and the identification $M=s^2$.
\item {\bf York boundary}: Now turning to the case where extrinsic curvature $K$ and $g_{ab}$ are fixed on the boundary which requires no additional boundary terms, we find from eq.\eqref{sbul1} that
\begin{align}
		S_{K, h}&=\int dt\int dx\sqrt{h}\left(h^{\beta\alpha}\nabla_\beta N\nabla_\alpha\phi+\frac{\phi}{\sqrt{h}}\del_\mu(KN\sqrt{h}\mathfrak{t}^\mu)-N\phi\right)\label{phhcon2}
\end{align}
From this, the conjugate momenta can be read off as
\begin{align}
	\pi_K&\equiv \frac{\del S_{K, h}}{\del \dot{K}}=\sqrt{h}\phi\nonumber\\
	\pi_{h}^{ij}&\equiv \frac{\del S_{K, h}}{\del \dot{h}_{ij}}=-\frac{\sqrt{h}h^{ij}}{2}(\mathfrak{t}^\alpha\nabla_\alpha\phi-\phi K)\label{pikh}
	\end{align}
The Hamiltonian constraint in terms of conjugate momenta then becomes
\begin{align}
\mathcal{H}=	-\pi_K-\sqrt{h}D^2\phi-{K}(2h_{ij}\pi_h^{ij}- K\pi_K)\label{hamconkh}
\end{align}
 The zero mode part of the WDW constraint can be obtained by integrating the above local equation over the entire boundary. This gives, again upto operator ordering ambiguity, 
\begin{align}
	\mathcal{H}_0=	-(1-k^2)\pi_k-k\ell_k\pi_{\ell_k}\label{hamconkhz}
\end{align}
where $\ell_k$ is the length of the boundary and $k$ is the zero mode of $K$,
\begin{align}
	 \ell_k&=\int dx\sqrt{h}\nonumber\\
	 k&=\int dx K(x).\label{lkdef}
\end{align}

As an aside let us also mention an alternate way to derive the conjugate momenta as follows. The extrinsic curvature and metric can be thought of as an alternate set of canonical variables to the dilaton and the boundary metric. The classical phase space is two dimensional which can be labelled by either of the variables. Since the symplectic form on the phase space is an invariant, we can use it to obtain it the conjugate momenta of one set if we know them for the other set and the transformation between these sets of variables. From eq.\eqref{dilconmo}, we can write the symplectic form on the phase space in terms of the coordinates $\phi,h_{ij}$ and their conjugate momenta as
\begin{align}
		\omega&=d\phi\wedge d\pi_\phi+dh_{ij}\wedge d{\pi}_h^{ij}\nonumber\\
		&=-d\phi\wedge d(K\sqrt{h})-dh_{ij}\wedge d\left(\frac{\sqrt{h}}{2}h^{ij}\mathfrak{t}^\alpha\nabla_\alpha\phi\right)\nonumber\\
		&=-d\phi\wedge \sqrt{h}dK-d\phi\wedge \frac{K\sqrt{h}}{2}h^{ij}dh_{ij}-dh_{ij}\wedge d\left(\frac{\sqrt{h}}{2}h^{ij}\mathfrak{t}^\alpha\nabla_\alpha\phi\right)\nonumber\\
	&=dK\wedge d(\phi\sqrt{h})+dh_{ij}\wedge d\left(-\frac{\sqrt{h}}{2}h^{ij}(\mathfrak{t}^\alpha\nabla_\alpha\phi-\phi K)\right)\label{sympform}
\end{align}
From the above, we can immediately read off the conjugate momenta when $K, h_{ij}$ are the coordinate variables on the phase space and they match with what we have already obtained in eq.\eqref{pikh}.
The path integral result for the boundary with extrinsic curvature fixed to be $K=k$ and length $\ell_k$ is given by 
\begin{align}
	\Psi({k,\ell_k})\propto\delta(\ell_k\sqrt{k^2-1}-2\pi)\label{psiklk}
\end{align}
which again can be shown to satisfy the WDW equation
\begin{align}
((1-k^2)	\del_k+k\ell_k\del_{\ell_k})\Psi(k, \ell_k)=0\label{wdwexk1}
\end{align}
A more general set of solutions to the above differential equation can be obtained as follows. Defining the coordinates $\rho, \sigma$ as
\begin{align}
	\rho=\ell_k\sqrt{1-k^2},\sigma=\frac{\ell_k}{\sqrt{1-k^2}}\label{lksq}
\end{align}
the WDW can be expressed in terms of these variables using the chain rule for derivatives
\begin{align}
	\del_{\ell_k}&=\frac{\del\rho}{\del \ell_k}\del_\rho+\frac{\del\sigma}{\del \ell_k}\del_\sigma=\sqrt{1-k^2}\del_\rho+\frac{1}{\sqrt{1-k^2}}\del_\sigma\nonumber\\
	\del_{k}&=\frac{\del\rho}{\del k}\del_\rho+\frac{\del\sigma}{\del k}\del_\sigma=-\frac{k\,\ell_k}{\sqrt{1-k^2}}\del_\rho+\frac{k \, \ell_k}{(\sqrt{1-k^2})^3}\del_\sigma\label{pardefch}
\end{align}
Using these we find
\begin{align}
	((1-k^2)\del_{k}+ k \ell_k\del_{\ell_k})\Psi=2k\sigma\del_\sigma\Psi=0\label{wdweqpsi}
\end{align}
This means that 
\begin{align}
	\Psi=g(\rho)\label{grhpsi}
\end{align}
where $g(\rho)$ is an arbitrary function of $\rho$. 

Let us also quickly mention the result for a similar calculation in the Lorentzian signature. Manipulations similar to those that lead to eq.\eqref{phhcon}, when carried out in Lorentzian signature leads to 
\begin{align}
S_{K,h}=\int d^2x \sqrt{h}\left(NK \del_t\phi-h^{ij}\nabla_i\phi\nabla_jN-N\phi\right)\label{Slk}
\end{align}
where $\kappa$ is an overall normalization factor and is related to the 2D Newton's constant.
The corresponding conjugate momenta are found to be
\begin{align}
    \pi_K=-{\sqrt{h}\phi},\,\,\pi_h^{ij}=\frac{\sqrt{h}h^{ij}}{2}\left(\frak{t}\cdot\nabla\phi-K\phi\right)\label{piklo}
\end{align}
The Hamiltonian constraint turns out to be
\begin{align}
    \mathcal{H}^{(K)}&=-{\sqrt{h}}\left(K\frak{t}^\alpha\nabla_\alpha\phi+\phi-D^2\phi\right)\nonumber\\
    &=(1+K^2)\pi_K-2Kh_{ij}\pi^{ij}+{\sqrt{h}}D^2\phi\label{hamlo}
\end{align}
where $D_i$ is the covariant derivative compatible with the spatial metric $h_{ij}$.
The zero mode part, expressed in terms of the length of the spatial slice, $\ell_k$, reads
\begin{align}
    \mathcal{H}^{(K)}_0=(1+k^2)\pi_k-k\ell_k\pi_{\ell_k}+\sum_{j\in \text{corner}}{r_j\cdot\nabla\phi}\label{hamlozer}
\end{align}
where $r_j$ is the outward tangent vector to the slice at the endpoints.
%The solution we constructed earlier eq.\eqref{psisol} is indeed of this form.

	\item {\bf Mixed boundary}: For the case where the boundary is non-smooth with part of it being asymptotic boundary and part being extrinsic curvature boundary, the WDW is more non-trivial. To avoid confusion, let us denote the induced metric on the asymptotic boundary as $\gamma_{ij}$ and on the York boundary as ${h}_{ij}$. First consider the extrinsic curvature boundary. 
	
The conjugate momenta are still given by eq.\eqref{pikh}. 
%	\begin{align}
%		\pi_\phi&\equiv \frac{\del S_{\phi, h}}{\del \dot{\phi}}=-\sqrt{h}K\nonumber\\
%	\pi_{h}^{ij}&\equiv \frac{\del S_{\phi, h}}{\del \dot{h}_{ij}}=-\frac{\sqrt{h}}{2}h^{ij}n^\alpha\nabla_\alpha\phi
%		\pi_K&=-\sqrt{\tilde{h}}\phi\nonumber\\
%	\pi_{\tilde{h}}^{ij}&=\frac{\sqrt{\tilde{h}}\tilde{h}^{ij}}{2}(Nn^\alpha\nabla_\alpha\phi-\phi K)\label{pikh}
%	\end{align}
The local Hamiltonian constraint on the asymptotic boundary is given by 
\begin{align}
\mathcal{H}^{(K)}=	-\pi_K-\sqrt{h}D^2\phi-{K}(2h_{ij}\pi_h^{ij}- K\pi_K)\label{hlcph}
\end{align}
The zero mode part of it is obtained by integrating over this along the asymptotic boundary, 
\begin{align}
	\mathcal{H}_0^{(K)}&=	-(1-k^2)\pi_k-k\ell_k\pi_{\ell_k}-\sum_{{j}\in\, \text{corners}}r_j\cdot\nabla\phi\nonumber\\
	&=	-(1-k^2)\pi_k-k\ell_k\pi_{\ell_k}-\sum_{{j}\in\, \text{corners}}r_j\cdot\nabla\left(\frac{\pi_K}{\sqrt{h}}\right)
	\label{hzeras}
\end{align}
where $r^\mu$ is the tangent vector to the asymptotic boundary at the corners.  As can be seen from the above, the Hamiltonian constraint now has extra corner terms. The  action of this term on the wavefunction can be evaluated by knowing the dependence of the wavefunction on the non-zero mode part of the extrinsic curvature. To this end, we would need to evaluate the dependence on the non-zero mode part of $K$. This is harder and so we shall evaluate this dependence perturbatively by considering the non-zero modes as a small perturbation around the zero-mode part.  The details of this calculation are relegated to the appendix. 

Now, consider the asymptotic boundary. Again, the conjugate momenta are the same as in eq.\eqref{dilconmo}.  Similarly, the local Hamiltonian constraint on the asymptotic boundary is given by 
\begin{align}
	\mathcal{H}^{(\phi)}=\frac{2}{\sqrt{h}}h_{ij}\pi_{h}^{ij}\pi_\phi -\sqrt{h}D^2\phi-\sqrt{h}\phi\label{hlock}
\end{align}
The zero mode part of it is obtained by integrating it along the York boundary, 
\begin{align}
	\mathcal{H}_0^{(\phi)}&=\pi_{\ell_\phi}\pi_\phi -\ell_\phi \phi-\sum_{{j}\in\, \text{corners}}r_j\cdot\nabla\phi \nonumber\\
	&=\pi_{\ell_\phi}\pi_\phi -\ell_\phi \phi
	\label{hzerk}
\end{align}
where the second line is obtained from the first by noting that along the asymptotic boundary, $\phi$ is held fixed so that $r\cdot\nabla\phi=0$.
	
%To verify that this is indeed the correct WDW, we need to evaluate the corresponding path integral and obtain the resulting wavefunction. Directly evaluating the path integral is a bit non-trivial. So, we resort to an alternate way of evaluating the path integral in this case by noting an analogy between the JT gravity path integral and a relativistic particle in a background electric field. We elucidate this in the next subsection.
\end{itemize}

%Let us end this section with one comment. An alternative way to derive the Hamiltonian and momentum constraints following the Hamilton-Jacobi method is discussed in appendix \ref{constrainsvar}. In this method, the Hamiltonian and momentum constraints can be obtained by considering the most general variation of the action, including a change in the boundary itself. The Hamiltonian and momentum constraint can then be read off from the coefficient of the change of the boundary, in the directions tangent and normal to the boundary itself. We considered the mixed boundary condition in appendix \ref{constrainsvar} and the final result is in eq.\eqref{DelSjt2}. The Hamiltonian and momentum constraint for smooth boundaries can be obtained as a special case of the mixed boundary by ignoring the boundary which is not needed. 

\subsection{WDW equation and York time}
\label{wdwsec}
We now show how the wavefunctions obtained in section \ref{qsols} satisfy the WDW equation. The WDW constraint is a local constraint and should be satisfied on every codim-1 slice. However, we have only computed the wavefunction for a slice of constant extrinsic curvature and constant dilaton on the asymptotic boundary, which correspond to the zero modes of the corresponding local quantities. So, to check how the wavefunctions satisfy the WDW constaints, we have to extract the zero mode part of the constraints. These are given by eq.\eqref{hzerk} and eq.\eqref{hzeras} respectively. Reinstating factors of $G$ we get the constraint equations as
\begin{align}
	\mathcal{H}^{(\phi)}_0&=G^2\pi_{\ell_\phi}\pi_\phi+\frac{G^2}{\ell_\phi}\pi_\phi -\ell_\phi \phi\label{gaswde}\\
	\mathcal{H}_0^{(K)}&=-G	\left((1-k^2)\pi_k+k\ell_k\pi_{\ell_k}+\sum_{{j}\in\, \text{corners}}r_j\cdot\nabla\left(\frac{\pi_K}{\sqrt{h}}\right)\right)\label{gexkwde}
\end{align}

Note that eq.\eqref{gaswde} has an extra term compared to eq.\eqref{hzerk}. This corresponds to an operator ordering ambiguity term that is fixed by the path integral as mentioned in eq.\eqref{wkbwdwlph}.

We note that eq.\eqref{gexkwde} has a corner term which depends on the tangential derivative of $\pi_K$. Thus, to evaluate the action of this constraint on the wavefunction, we need the dependence of the wavefunction on the non-zero modes of the extrinsic curvature, at least at the corner. However computing this dependence is highly non-trivial. So, we resort to a perturbation theory by considering a curve which has a small non-zero mode part around a constant zero-mode value for extrinsic curvature. We then evaluate the value of the on-shell action for such a configuration. We then argue that to leading order in $G$, this contribution from on-shell action is good enough to compute the action of the last term in eq.\eqref{gexkwde}.

Let us now find the parametrization of the curve for an extrinsic curvature of the form 
\begin{align}
	K(\lambda)=k+\delta k(\lambda)\label{klam}
\end{align}
where $k$ is the zero mode part with $\delta k$ being the non-zero mode but small in magnitude so that we can account for it perturbatively. Working in Poincare coordinates, the extrinsic curvature for a curve $(x(\lambda),y(\lambda))$ is given by 
\begin{align}
    K(\lambda)=\frac{-y(\lambda ) x''(\lambda ) y'(\lambda )+x'(\lambda )^3+x'(\lambda ) \left(y(\lambda ) y''(\lambda )+y'(\lambda )^2\right)}{\left(x'(\lambda )^2+y'(\lambda )^2\right)^{3/2}}
    \label{exktoqd}
\end{align}
Let the curve corresponding to eq.\eqref{klam} be denoted as $(\tilde{x}_0(\lambda)+\delta x(\lambda),\tilde{y}_0(\lambda)+\delta y(\lambda))$, where $(\tilde{x}_0(\lambda), \tilde{y}_0(\lambda))$ correspond to the zero mode part in eq.\eqref{klam} and are given by eq.\eqref{xlylcurp1}, 
\begin{align}
	\tilde{x}_0(\lambda)&=x_0+\frac{R \sinh (\lambda  \sqrt{1-k^2})}{k+\cosh (\lambda  \sqrt{1-k^2})}\nonumber\\
	\tilde{y}_0(\lambda)&=\frac{R\, \sqrt{1-k^2}}{k+\cosh (\lambda  \sqrt{1-k^2})}\label{zermx0y0}
\end{align}

The perturbations $\delta x$ and $\delta y$ are related, as a result of eq.\eqref{lambdprot}, as
\begin{align}
	\delta x'(\lambda)=-\frac{\tilde{y}_0(\lambda)}{\tilde{x}_0(\lambda)}\delta y'(\lambda)\label{delxdely}
\end{align}
Expanding eq.\eqref{exktoqd} to linear order on both sides, we  get
\begin{align}
	\delta k=-\frac{1}{\sqrt{\tilde{y}_0^2-\tilde{y}_0'^2}}\delta y''+\frac{\tilde{y}_0'(3\tilde{y}_0^2-2\tilde{y}_0'^2-\tilde{y}_0\tilde{y}_0'')}{\tilde{y}_0 (\tilde{y}_0-\tilde{y}_0'^2)^\frac{3}{2}}\delta y'+\frac{\tilde{y}_0^2\tilde{y}_0''+2\tilde{y}_0'^4-3\tilde{y}_0^2\tilde{y}_0'^2}{\tilde{y}_0^2(\tilde{y}_0^2-\tilde{y}_0'^2)^\frac{3}{2}}\delta y
\end{align}

Solving this equation to obtain $\delta y$ for a specified $\delta k$ gives
\begin{align}
	\delta y(\lambda) &= h_1(\lambda)\left(c_1+c_2 h_2(\lambda)-g_1(\lambda)-g_2(\lambda)\right)\nonumber\\
	h_1(\lambda)&=\frac{\left(k \cosh \left(\sqrt{1-k^2} \lambda \right)+1\right) }{2 \left(\cosh \left(\sqrt{1-k^2} \lambda \right)+k\right)^2}\exp{2 \tanh ^{-1}\left(\frac{k \,e^{\sqrt{1-k^2} \lambda }+1}{\sqrt{1-k^2}}\right)}\nonumber\\
	h_2(\lambda)&=  \frac{ \left(\left(k^2+2 \sqrt{1-k^2}-1\right) \sinh \left(\sqrt{1-k^2} \lambda \right)-\cosh \left(\sqrt{1-k^2} \lambda \right)-k\right)}{\sqrt{1-k^2} \left(k e^{\sqrt{1-k^2} \lambda }+\sqrt{1-k^2}+1\right)^2}	e^{\sqrt{1-k^2} \lambda } \nonumber\\
	h_3(\lambda)&= 2R \left(\frac{ \left(2-\sqrt{1-k^2}\right) \sqrt{1-k^2} \sinh \left(\sqrt{1-k^2} \lambda \right)- \cosh \left(\sqrt{1-k^2} \lambda \right)- k}{k \sqrt{1-k^2}}\right)	\nonumber\\
	h_5(\lambda)&=	   -4R \left(k \cosh \left(\sqrt{1-k^2} \lambda \right)+1\right) \exp{2 \tanh ^{-1}\left(\frac{k e^{\sqrt{1-k^2} \lambda }+1}{\sqrt{1-k^2}}\right)} \nonumber\\
	g_2(\lambda)&=\int_0^{\lambda}\delta k(\rho)h_3(\rho)d\rho\nonumber\\
	g_1(\lambda)&=-h_2(\lambda)\int_0^\lambda \delta k(\rho)h_5(\rho)d\rho
	\label{delkdely}
\end{align}
where $c_1, c_2$ are integration constants. These constants can be fixed by the requirement that $\delta y=0$ at the end points of the curve. Let $\lambda_1,\lambda_2$ be the values of the parameter $\lambda$ near the endpoints.  Doing this leads to $c_1,c_2$ as
\begin{align}
	c_1=& \frac{(g_1(\lambda_2)+g_2(\lambda_2))h_2(\lambda_1)-(g_1(\lambda_1)+g_2(\lambda_1))h_2(\lambda_2)}{h_2(\lambda_1)-h_2(\lambda_2)}\nonumber\\
	c_2=&\frac{g_1(\lambda_2)-g_1(\lambda_1)+g_2(\lambda_2)-g_2(\lambda_1)}{h_2(\lambda_2)-h_2(\lambda_1)}\label{c2c1}
\end{align}

Having obtained the solution for $\delta y$, we can now compute the modified value of the on-shell action due to this change. The corner angle between a geodesic curve parametrized as $(x_g,y_g)$, and the curve with perturbed extrinsic curvature discussed above, at the corner where $\lambda=\lambda_1$ so that $x=x_1,y=y_1$ denoted by $\theta_{1}$, gives
\begin{align}
	\cos\theta_1&=\frac{x_g'(\hat{\lambda}_{1})(\tilde{x}_0'(\lambda_1)+\delta x'(\lambda_1))+y_g'(\hat{\lambda}_{1})(\tilde{y}_0'(\lambda_1)+\delta y'(\lambda_1))}{y_1^2}\nonumber\\
	&\simeq\cos\theta_{1,0}+\left( y_g'(\hat{\lambda}_1) -  x_g'(\hat{\lambda}_1)  \frac{\tilde{y}_0'(\lambda_1)}{\tilde{x}_0'(\lambda_1)} \right)\frac{\delta y'(\lambda_1)}{y_1^2}\label{costhl1}
\end{align}
where $\hat{\lambda}$ is the parameter labeling the geodesic curve, $\theta_{1,0}$ is the angle between the unperturbed curve and the geodesic curve at this corner. 
 We also used  eq.\eqref{delxdely} in obtaining the second line in eq.\eqref{costhl1} from the first. From eq.\eqref{costhl1}, we get that 
\begin{align}
	\theta_1\equiv\theta_{1,0}+\delta\theta_1\simeq\theta_{1,0}-\left( y_g'(\hat{\lambda}_1) -  x_g'(\hat{\lambda}_1)  \frac{\tilde{y}_0'(\lambda_1)}{\tilde{x}_0'(\lambda_1)} \right)\frac{\delta y'(\lambda_1)}{y_1^2\sin\theta_{1,0}}\label{cornex1}
\end{align}

 We can compute the value of $\delta y'(\lambda_1)$, which will also be useful later on,  from eq.\eqref{delkdely}, eq.\eqref{c2c1} and we get
{
\small{
\begin{align}
	\delta y'(\lambda_1)=h_1(\lambda_1)\left(\delta k(\lambda_1)\left(h_5(\lambda_1)h_2(\lambda_1)-h_3(\lambda_1)\right)-\frac{h_2'(\lambda_1)}{h_2(\lambda_2)-h_2(\lambda_1)}\int_{\lambda_1}^{\lambda_2}d\rho\,\delta k(\rho)(h_5(\rho)h_2(\lambda_2) -h_3(\rho))\right)\label{delkyp}
\end{align}}
}
An identical computation gives a similar result for the other corner with $\lambda_1\rightarrow\lambda_2$,
{
	\small{
		\begin{align}
			\delta y'(\lambda_2)&=h_1(\lambda_2)\left(\delta k(\lambda_2)\left(h_5(\lambda_2)h_2(\lambda_2)-h_3(\lambda_2)\right)-\frac{h_2'(\lambda_2)}{h_2(\lambda_2)-h_2(\lambda_1)}\int_{\lambda_1}^{\lambda_2}d\rho\,\delta k(\rho)(h_5(\rho)h_2(\lambda_1) -h_3(\rho))\right)			\label{delkyp2}
	\end{align}}
}
The action of the corner term in eq.\eqref{gexkwde} can be evaluated as follows. The function $\delta k$ can be expanded in a complete basis of functions, say the eigenfunctions of the Laplacian operator satisfying Dirichlet boundary conditions at the corner. Let this expansion be given by 
\begin{align}
	\delta k(\lambda)=\sum_n f_n(\lambda) \delta k_n\label{kmod}
\end{align} 
The mode functions $f_n$ are assumed to satisfy  completeness relation of the form, along with the Dirichlet boundary conditions at the corner
\begin{align}
f_n(\lambda)\bigg\vert_{\text{corner}}=0 ,\qquad 	\sum_n f_n(\lambda)f_n(\tilde{\lambda})=\delta({\lambda-\tilde{\lambda}})\label{fnfm}
\end{align}
Let the corresponding expansion for the conjugate momentum  in the above basis be given by 
\begin{align}
	\pi_K=\sum_n f_n(\lambda)\pi_{k,n}\label{pikmo}
\end{align}
Using the basis expansion above, the corner term in the WDW equation in eq.\eqref{gexkwde} can be written as
\begin{align}
	r.\nabla\left(\frac{\pi_K}{\sqrt{h}}\right)=\sum_n\del_\lambda f_n(\lambda)\pi_{k,n}\label{rdelkl1}
\end{align}

We now have all the results at our disposal to check the WDW constraint. For concreteness, let us first compute the action of this term on the wavefunction with two York boundaries. We remind the reader that this wavefunction is given by

\begin{align}
		\Psi_Y(k_1,\ell_{k_1},k_2,\ell_{k_2})&=\delta(d_{\infty,1}^2-d_{\infty,2}^2)\exp[2q\left(w(k_1,\ell_{k_1})+w(k_2,\ell_{k_2})\right)]\nonumber\\
	w(k,\ell_k)&=\tan^{-1}\left(\frac{k}{\sqrt{1-k^2}}\tanh\left(\frac{\ell_k}{2} \sqrt{1-k^2}\right)\right)\label{wkl}
\end{align}

Note that this wavefunction has been obtained by evaluating the on-shell action, either directly in terms of gravity variables, eq.\eqref{psial}, or by rewriting as a particle in a field, eq.\eqref{psifk1k2de}. For a curve with non-zero mode turned on as in eq.\eqref{klam}, the on-shell contribution changes due to the extra contribution to the corner angle given by eq.\eqref{cornex1}. The wavefunction, with the extra contribution to the corner angle eq.\eqref{cornex1}, reads
\begin{align}
	\Psi_Y=e^{-q(\theta_{1,0}+\theta_{2,0}+\delta\theta_1+\delta\theta_2)}\label{psiyex}
\end{align}
where $\theta_{2,0}$ has the same value as $\theta_{1,0}$ given by 
\begin{align}
	\theta_{1,0}=\arctan(\frac{k_1}{\sqrt{1-k_1^2}}\tanh(\frac{\ell_{k_1}}{2}\sqrt{1-k_1^2}))+ \arctan(\frac{k_2}{\sqrt{1-k_2^2}}\tanh(\frac{\ell_{k_2}}{2}\sqrt{1-k_2^2}))\label{the10kl}
\end{align}
Action of the operator in eq.\eqref{rdelkl1} on the wavefunction eq.\eqref{psiyex} at corner $i$ is given by 
\begin{align}
	\sum_n\del_\lambda f_n(\lambda)\pi_{k,n}\Psi_Y&=-\sum_n\del_\lambda f_n(\lambda)\frac{\del \Psi_Y}{\del (\delta k_n)}\nonumber\\
	&=-\sum_n f_n'(\lambda_i)\left( y_g'(\hat{\lambda_i}) -  x_g'(\hat{\lambda_i})  \frac{\tilde{y}_0'(\lambda_i)}{\tilde{x}_0'(\lambda_i)} \right)\frac{q}{y_i^2\sin\theta_{i,0}}\frac{\del \delta y'(\lambda_i)}{\del (\delta k_n)}\Psi_Y\label{cornacw}
\end{align}
where the second line is obtained from first by noting that the change in action comes entirely due to the change in the angle at the corner and using eq.\eqref{cornex1}. 
Using eq.\eqref{delkyp} and noting that $\delta k$ vanishes at the corner, we find that 
\begin{align}
	\frac{\del \delta y'(\lambda_1)}{\del (\delta k_n)}&=-\frac{h_1(\lambda_1)h_2'(\lambda_1)}{h_2(\lambda_2)-h_2(\lambda_1)}\int_{\lambda_1}^{\lambda_2}d\rho\,f_n(\rho)(h_5(\rho)h_2(\lambda_2) -h_3(\rho))\nonumber\\
	\Rightarrow\sum f_n'(\lambda_1)\frac{\del \delta y'(\lambda_1)}{\del (\delta k_n)}&=-\frac{h_1(\lambda_1)h_2'(\lambda_1)}{h_2(\lambda_2)-h_2(\lambda_1)}\int_{\lambda_1}^{\lambda_2}d\rho\sum_n f_n'(\lambda_1)f_n(\rho)(h_5(\rho)h_2(\lambda_2) -h_3(\rho))\nonumber\\
	&=-\frac{h_1(\lambda_1)h_2'(\lambda_1)}{h_2(\lambda_2)-h_2(\lambda_1)}\int_{\lambda_1}^{\lambda_2}d\rho\sum_n f_n'(\lambda_1)f_n(\rho)(h_5(\rho)h_2(\lambda_2) -h_3(\rho))\nonumber\\	&=-\frac{h_1(\lambda_1)h_2'(\lambda_1)}{h_2(\lambda_2)-h_2(\lambda_1)}\int_{\lambda_1}^{\lambda_2}d\rho\,\del_\lambda\delta(\lambda_1-\rho)(h_5(\rho)h_2(\lambda_2) -h_3(\rho))\nonumber\\
	&=-\frac{h_1(\lambda_1)h_2'(\lambda_1)}{h_2(\lambda_2)-h_2(\lambda_1)}(h_5'(\lambda_1)h_2(\lambda_2) -h_3(\lambda_1))
	\label{delkypfn1}
\end{align}
Similar calculation for the other corner $\lambda=\lambda_2$ gives
\begin{align}
		\frac{\del \delta y'(\lambda_2)}{\del (\delta k_n)}&=-\frac{h_1(\lambda_2)h_2'(\lambda_2)}{h_2(\lambda_2)-h_2(\lambda_1)}\int_{\lambda_1}^{\lambda_2}d\rho\,f_n(\rho)(h_5(\rho)h_2(\lambda_1) -h_3(\rho))\nonumber\\
		\Rightarrow\sum f_n'(\lambda_2)\frac{\del \delta y'(\lambda_2)}{\del (\delta k_n)}&=-\frac{h_1(\lambda_2)h_2'(\lambda_2)}{h_2(\lambda_2)-h_2(\lambda_1)}(h_5'(\lambda_2)h_2(\lambda_1) -h_3'(\lambda_2))\label{delkypfn2}
\end{align}

Thus, the action of the corner term in eq.\eqref{gexkwde} on the wavefunction gives
\begin{align}
	\sum_{\text{corners}}r_j\cdot\nabla\left(\frac{\pi_K}{\sqrt{h}}\right)\Psi_Y=&\left( y_g'(\hat{\lambda}_1) -  x_g'(\hat{\lambda}_1)  \frac{\tilde{y}_0'(\lambda_1)}{\tilde{x}_0'(\lambda_1)} \right)\frac{q}{y_1^2\sin\theta_{1,0}}\frac{h_1(\lambda_1)h_2'(\lambda_1)}{h_2(\lambda_2)-h_2(\lambda_1)}(h_5'(\lambda_1)h_2(\lambda_2) -h_3'(\lambda_1))\Psi_Y\nonumber\\
	+&\left( y_g'(\hat{\lambda}_2) -  x_g'(\hat{\lambda}_2)  \frac{\tilde{y}_0'(\lambda_2)}{\tilde{x}_0'(\lambda_2)} \right)\frac{q}{y_2^2\sin\theta_{2,0}}\frac{h_1(\lambda_2)h_2'(\lambda_2)}{h_2(\lambda_2)-h_2(\lambda_1)}(h_5'(\lambda_2)h_2(\lambda_1) -h_3'(\lambda_2))\,\Psi_Y\label{cornercon}
\end{align}
Using the form of the functions $h_i, \theta_{i,0}, \tilde{x}_0,\tilde{y}_0$ in eq.\eqref{zermx0y0},\eqref{delkdely} and \eqref{the10kl} ($\theta_{2,0}=\theta_{1,0}$) and the corresponding expression for the geodesic curve that can be obtained from eq.\eqref{zermx0y0} by setting $k=0$, we find that the contribution from the corner term, to leading order in large $\ell_i$, is given by 
\begin{align}
	\sum_{\text{corners}}r_j\cdot\nabla\left(\frac{\pi_K}{\sqrt{h}}\right)\Psi_Y\simeq -2q \sqrt{1-k^2}\,\,\Psi_Y\label{psiy}
\end{align}
The action of the full WDW constraint on this wavefunction can be understood as follows. Since the wavefunction is calculated perturbatively in $\delta k$, the action of the WDW constraint also has to be evaluated perturbatively in $\delta k$. We shall only show that the WDW constraint is satisfied to $\order{(\delta k)^0}$.
The action of the zero mode part in eq.\eqref{gexkwde} coming from the bulk part of the slice to $\order{(\delta k)^0}$ gives
\begin{align}
-(	(1-k^2)\del_k+k\ell_k\del_{\ell_k})\Psi_Y&=\left(2q\frac{\sqrt{1-k^2} \tanh \left(\frac{1}{2} \sqrt{1-k^2} \ell_k\right)}{1-k^2 \text{sech}^2\left(\frac{1}{2} \sqrt{1-k^2} \ell_k\right)}+\order{(\delta k)}\right)\Psi_Y\nonumber\\
&\simeq 2q\sqrt{1-k^2}\,\,\Psi_Y
\label{zeodk}
\end{align}
Combining this result with eq.\eqref{zeodk}, we find that the wavefunction $\Psi_Y$ in eq.\eqref{wkl} satisfies the constraint equation
\begin{align}
	\mathcal{H}_0^{(K_{1(2)})}\Psi_Y=\left((1-k^2)\pi_k+k\ell_k\pi_{\ell_k}-2q\sqrt{1-k^2}\right)\Psi_Y=0\label{hkps}
\end{align}

An exactly similar analysis can be done for the wavefunction with mixed boundary conditions, eq.\eqref{psiasyk}. This involves computing the change in the corner angle contribution due to a perturbation in the extrinsic curvature eq.\eqref{klam} following the series of steps from eq.\eqref{exktoqd} through eq.\eqref{psiy}. Doing this will again lead to a WDW equation of the form eq.\eqref{hkps} and it can be checked that the wavefunction in eq.\eqref{psiasyk} satisfies the WDW equation. 
Now let us turn to the WDW constraint on the asymptotic boundary.

 We will check that eq.\eqref{gaswde} is the correct equation that will be satisfied  by the wavefunctions we obtain in this paper, eq.\eqref{psiinbek}. Note that the result in eq.\eqref{psiinbek} is obtained with an action which has an additional counterterm, see eq.\eqref{sdel}. Thus, for checking the action of WDW constraint  eq.\eqref{gaswde}, we need to add back the contribution of the counterterm to eq.\eqref{psiinbek}, doing which we get
\begin{align}
	\Psi_A(\phi,\ell_\phi, k, \ell_k,\phi_c)=\frac{\phi_c}{ d}e^{-\pi q-2\phi_c\arcsin(k)}\int ds \,s \,{\sinh(2\pi s)}\,{e^{{\ell_\phi\phi\over G}-{G \ell_\phi s^2\over 2\phi}}}\,\,K_{2is}\left(\frac{2\phi_c}{Gd}\right)\label{psiasyk}
\end{align}
where a distinction is made between the value of the zero mode part of the dilaton on the asymptotic boundary, denoted $\phi$, and the value of the dilaton at the corner, denoted $\phi_c$. Even though their value is the same, it is important that this distinction be made because, the zero mode of local WDW constraint acts only on the dilaton configuration on the asymptotic boundary and not at the corner. It can be easily checked that the constraint eq.\eqref{gaswde} is satisfied by eq.\eqref{psiasyk} to $\order{(G^3)}$. To obtain the action of $\mathcal{H}_{0}^{(K)}$, an analysis similar to the one carried out earlier for $\Psi_Y$ has to be done.

 	\bibliographystyle{JHEP}
 
\bibliography{refs}
	\end{document}